\documentclass[
    aps,
    prb,
    showpacs,
    reprint,
    superscriptaddress,
    floatfix,
    amsmath
]{revtex4-1}

\pdfoutput=1
\usepackage{braket}
\usepackage{graphics}
\usepackage{graphicx}
\usepackage{multirow}
\usepackage{tasks}
\usepackage{dcolumn}
\usepackage{scrextend}
\usepackage{longtable}
\usepackage{dblfnote}
\usepackage[multiple]{footmisc}
\newcolumntype{.}{D{.}{.}{-1}}

\usepackage{epsfig}
\usepackage{epstopdf}
\usepackage{hyperref}
\usepackage{verbatim}
\usepackage[english]{babel}

\newcommand\Bstrut{\rule[-0.9ex]{0pt}{0pt}}

\newcommand{\NMaterials}{71 }
\newcommand{\NMaterialsNoSpace}{71}

\begin{document}

    \title{Accessible computational materials design with high fidelity and high throughput}
    
    \author{Protik Das}
    \thanks{
        mail-to: info@exabyte.io; \\
        PD was a student at the University of California, Riverside, California 92507, USA, at the moment of this writing
    }
    \author{Mohammad Mohammadi}
    \author{Timur Bazhirov}
    \thanks{PD and TB contributed equally to this work}
    \affiliation{Exabyte Inc., San Francisco, California 94103, USA}

    \begin{abstract}

    Despite multiple successful applications of high-throughput computational materials design from first principles, there is a number of factors that inhibit its future adoption. Of particular importance are limited ability to provide high fidelity in a reliable manner and limited accessibility to non-expert users. We present example applications of a novel approach, where high-fidelity first-principles simulation techniques, Density Functional Theory with Hybrid Screened Exchange (HSE) and GW approximation, are standardized and made available online in an accessible and repeatable setting. We apply this approach to extract electronic band gaps and band structures for a diverse set of \NMaterials materials ranging from pure elements to III-V and II-VI compounds, ternary oxides and alloys. We find that for HSE and G$_0$W$_0$, the average relative error fits within 20$\%$, whereas for conventional Generalized Gradient Approximation the error is 55$\%$. For HSE we find the average calculation time on an up-to-date server centrally available from a public cloud provider to fit within 48 hours. This work provides a cost-effective, accessible and repeatable practical recipe for performing high-fidelity first-principles calculations of electronic materials in a high-throughput manner.

\end{abstract}

    \maketitle
    
    \section{Introduction}
\label{sec:introduction}

    Materials design and discovery based on first-principles modeling is an inter-disciplinary research area that recently received much attention with multiple success stories reported in the field of catalysis, hydrogen storage materials, Li-ion batteries, photovoltaics, topological insulators, carbon capture, piezoelectrics, and thermoelectrics \cite{jain2013materialsproject, curtarolo2012aflowlib, saal2013openQMD, pizzi2016aiida, nomad}. These efforts enabled the integration of computational materials science with information technology (e.g., web-based dissemination, databases and data-mining), expanded access to properties computed by first-principles modeling approaches to new communities and promoted new collaborative work. Nevertheless, when compared with the more established computer-aided design and engineering sector, there is still much room for improvement in the way first-principles modeling is performed with respect to the accessibility and repeatability of high fidelity calculations.
    
    High-throughput virtual screening produced large repositories of data for its further consumption by other scientists, notably the Materials Project \cite{jain2013materialsproject}, AFLOW\cite{curtarolo2012aflowlib}, and the open quantum materials database\cite{saal2013openQMD}. Other initiatives, like AIIDA\cite{pizzi2016aiida}, also provided a set of building blocks for the construction of the simulation workflows. Recently, other approaches like NOMAD\cite{nomad}, emerged with the idea of an open access data repository aimed to allow for advanced data analytics and the creation of machine learning models. Other notable example includes the Computational 2D Materials Database\cite{haastrup2018computational, rasmussen2015computational} targeted at the applications in semiconductor area. The efforts above include a significant computer science aspect and created software tools that facilitate the execution of simulations in a high-throughput way, such as Pymatgen\cite{ong2013python}, Atomic Simulations Environment\cite{larsen2017atomic}, AIIDA stack\cite{pizzi2016aiida} and similar. These tools facilitate the adoption of the original techniques by other computational materials scientists and help organize and standardize the community efforts. Naturally, the data-centric approaches to the development of new materials followed after\cite{citrine, tilde, villars2004pauling, isayev2017ml-descriptors, ward2016ml-framework}. Additionally, the area of cloud computing as applied to the first-principles materials modeling emerged in the last few years\cite{yang2018matcloud, 2016-exabyte-aps-abstract}.
    
    We present the approach conceived and implemented by Exabyte Inc. inside its web-based modeling platform since 2014. The approach is focused on the accessibility and repeatability of modeling workflows, is designed to support creation and execution of multiscale models online, and is reminiscent to NanoHUB.org\cite{klimeck2008nanohub}. Compared to standalone software tools, such an approach allows users to focus on the physical essence of the problem and removes any obstacles related to the computational complexity, such as installation and parallelization concerns. Our approach enables access for (eg. experimental) scientists without direct knowledge of modeling techniques, promotes the exchange of ideas, and extends creative breadth of the resulting research. By relying on centrally-available cloud-based high performance computing the platform yields the computational power to facilitate high fidelity in a reproducible way\cite{exabyte2018hp3c}, and its data-centric nature eliminates unnecessary repetition, facilitates collaboration, and embraces traceability, version control and other computer science paradigms\cite{pizzi2016aiida}.
    
    In this manuscript we report the example application of the above platform\cite{exabytePlatform} to the electronic structural properties of semiconducting materials. We use Density Functional Theory in the plane-wave pseudopotential approximation\cite{hohenberg-kohn1964DFT, mlcohen1979pseudopotentialDFT} and obtain the electronic band structures and band gaps for a diverse set of \NMaterials compounds ranging from pure elements to ternary oxides and further referred to as ESC-\NMaterialsNoSpace. We provide the results for the Generalized Gradient Approximation \cite{perdew1996generalized}, the Hybrid Screened Exchange\cite{heyd2003hybrid} and the GW approximation\cite{hybertsen-louie1985GW}. We compare the results with the available experimental data and present the assessment of the accuracy levels for each model. For the first time ever this work presents all the following combined together: the results, the tools that generated the results, the simulations with all associated data, and an easy-to-access way to reproduce, improve and contribute results for other materials into a centralized ever-growing repository.\cite{exabytePlatformBGPhaseIURL}.
    
    \onecolumngrid
     \begin{figure}[t]
     \centering
        \label{figure:execution-flow}
        \includegraphics[width = 1.0\textwidth]{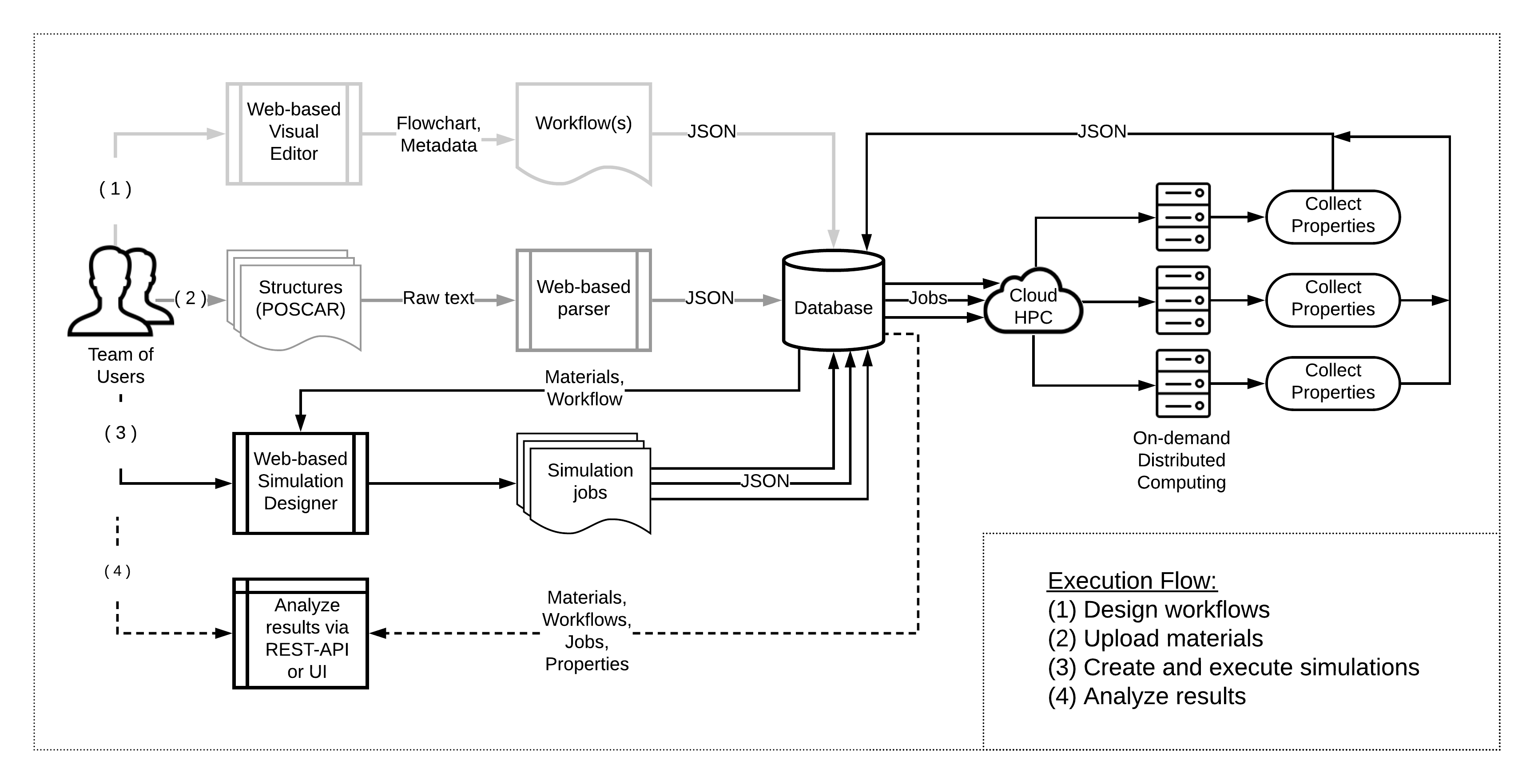}
        \parbox{1.0\textwidth}{\caption{
            Flowchart with  the execution logic of the simulations. Branch (1), shown in light gray, represents the initial design of the simulation workflow with its subsequent storage as JSON object in database. Branch (2), dark gray, illustrates the upload and conversion to database entries of the structural materials information. (3), in black, demonstrates the main execution logic for the creation and execution of simulation jobs. Finally, (4) denotes further analysis and is show using dashed black lines.
        }}
    \end{figure}
    \twocolumngrid

    \begin{figure}[h!]
	  \centering
	  \label{figure:example-unit}
	  \includegraphics[width = 0.4825\textwidth]{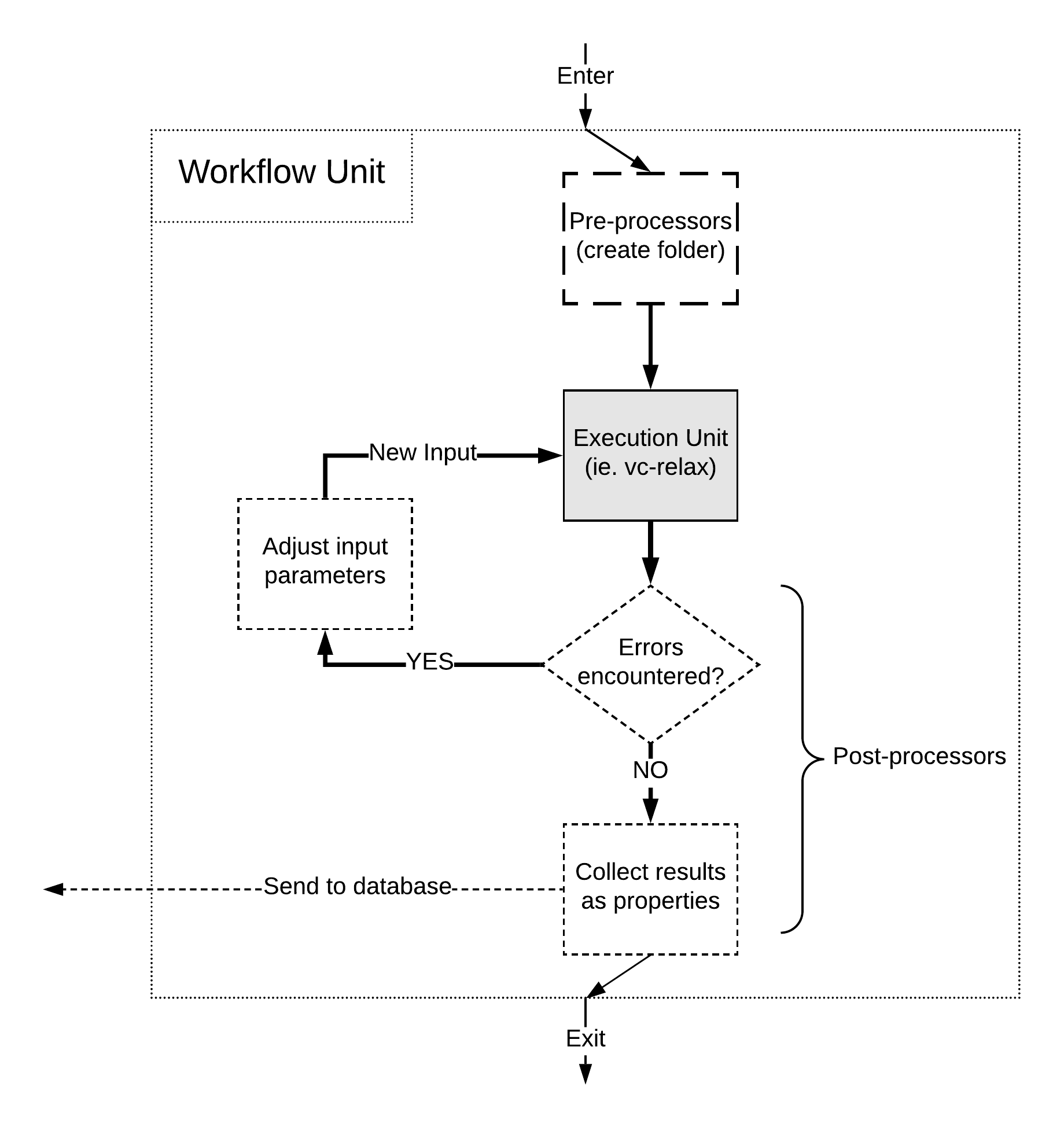}
	  \caption{
	      Example unit of a simulation workflow with a pre-processor, main execution part, and post-processors.
	  }
    \end{figure}
    \begin{figure}[ht!]
	  \centering
	  \label{figure:example-workflow}
	  \includegraphics[width = 0.45\textwidth]{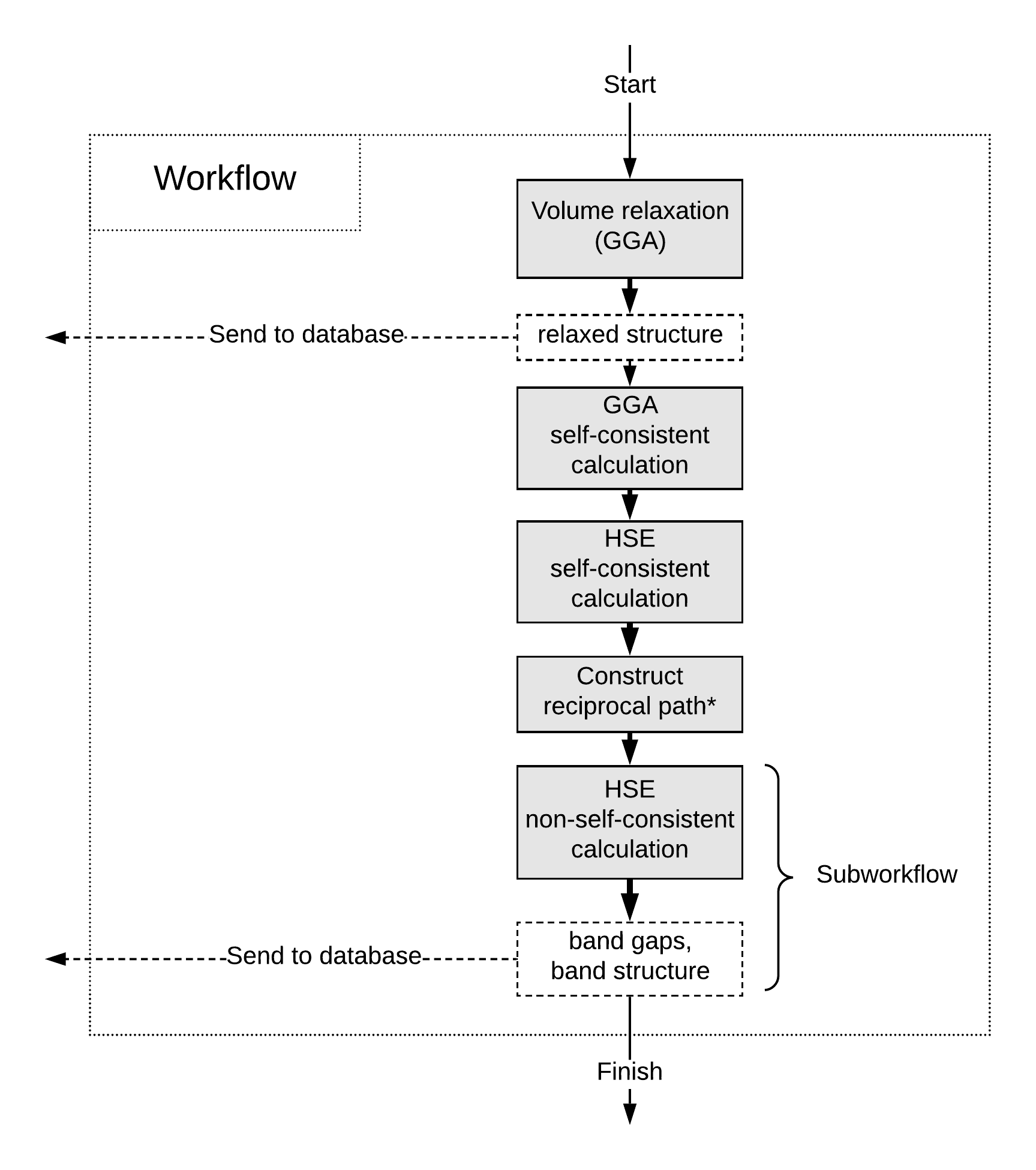}
	  \caption{
	      Simulation workflow for an HSE calculation. Post-processors (dashed) used to extract materials properties. ($^*$) denotes an auxiliary intermediate step.
	  }
    \end{figure}

\section{Methodology}
\label{sec:methodology}

\subsection{General logic}

    \subsubsection{Execution flow}

    We demonstrate the general execution flow employed in this work in Fig. \ref{figure:execution-flow}. We start from the design of the simulation workflows (designated by (1) in the figure). We represent the logic of the workflows through a data structure encoded using a JavaScript Object Notation (JSON) using a data convention further referred to as Exabyte Data Convention (EDC). Next, we upload the initial structures for the materials to be studied (branch (2) in the figure). After that, we create and execute the simulation jobs using the cloud-based high-performance computing infrastructure assembled on-demand by our software (corresponding to branch (3) in the figure), and collect the resulting properties in a database. Finally, we analyze the results either through a graphical user interface or by means of the RESTful application programming interface (API). The general execution flow, including all the above components and the associated entities are freely available online. The users of Exabyte platform can clone the associated entities (eg. materials, workflows) - and re-create our calculations in order to reproduce or further improve the results.
    
    \subsubsection{Workflow units}
    
    Within the EDC each workflow contains multiple units and each unit contains the corresponding input parameters for the simulation engine(s) used within. We logically separate each individual unit into pre-processors, main execution, and post-processors parts as demonstrated in Fig. \ref{figure:example-unit}. The pre-processors are ran before the main part and are used for auxiliary tasks, such as creating the required system folders for data on disk. The main execution part is where the main simulation is done. Post-processors are used to assert the completion of the simulation or attempt the main simulation again with a set of adjusted parameters. For the work described in this manuscript we used the error correction logic implemented in \cite{jain2013materialsproject}. After asserting the validity of the simulation a set of material properties is extracted, organized into JSON data structures, and stored in the database.
    
    \subsubsection{Workflows}
    
    An example workflow for HSE calculations utilized in this work in shown in Fig. \ref{figure:example-workflow}. We start with obtaining the relaxed structures, and then self-consistently pre-calculate the electronic wavefunctions and charge density. These steps are done within the GGA. Next, we repeat the self-consistent calculation, this time including the exchange interaction within HSE. Next, we run an auxiliary step to assist with the construction of the reciprocal path for the final part of the calculation - the non-self-consistent HSE calculation. The latter produces the resulting band gap and band structure properties.

\subsection{Materials}

    All materials studied in this work constitute the E-\NMaterialsNoSpace set and are divided in 7 categories according to their stoichiometric composition.  The categories together with their shorthand names are listed in Table \ref{table:mat_category}. We attempted to cover a diverse set of semiconductor stoichiometries accessible to the modeling from first principles. We prioritized compounds with smaller number of atoms within the crystal unit cell, however did not impose a hard limit on the unit cell size. Most of the structures studied have 4 or less atoms inside the unit cell, the largest unit cell has 32 atoms. We further sub-categorized materials into groups by the associated difficulty levels for the simulation workflows involved, as explained below. The details about the materials studied, including the corresponding categories and the results are given in Table \ref{table:materials-data}. Our approach is similar to that of \cite{heyd2005energy}, with an attempt to improve the range of compounds studied and include materials with potential industrial applications.

    \begin{table}[h!]
    	\centering
    	\begin{tabular}{l |  c |  c c c}
    		\hline
    		\hline
            Material category   & Symbol & Count & $N_{at}^{min}$ & $N_{at}^{max}$ \\
            \hline
            Elemental           & EL    & 10     & 2               & 12 \\
            III-V               & 35    & 10     & 2               & 4  \\
            II-VI               & 26    & 11     & 2               & 4  \\
            Binary oxides       & BO    & 15     & 2               & 12 \\
            Ternary oxides      & TO    & 10     & 5               & 32 \\
            Dichalcogenides     & DC    & 5      & 3               & 6  \\
            Alloys              & AL    & 10     & 2               & 8  \\
    		\hline
    		\hline
    	\end{tabular}
    	\caption{
    	    Summary of the material categorization employed in this work with counts. $N_{at}$ - number of sites (atoms) in the crystal unit cell.
    	    }
    	\label{table:mat_category}
    \end{table}

\subsection{Workflows}

\begin{figure}[ht!]
        \includegraphics[width = 0.48\textwidth]{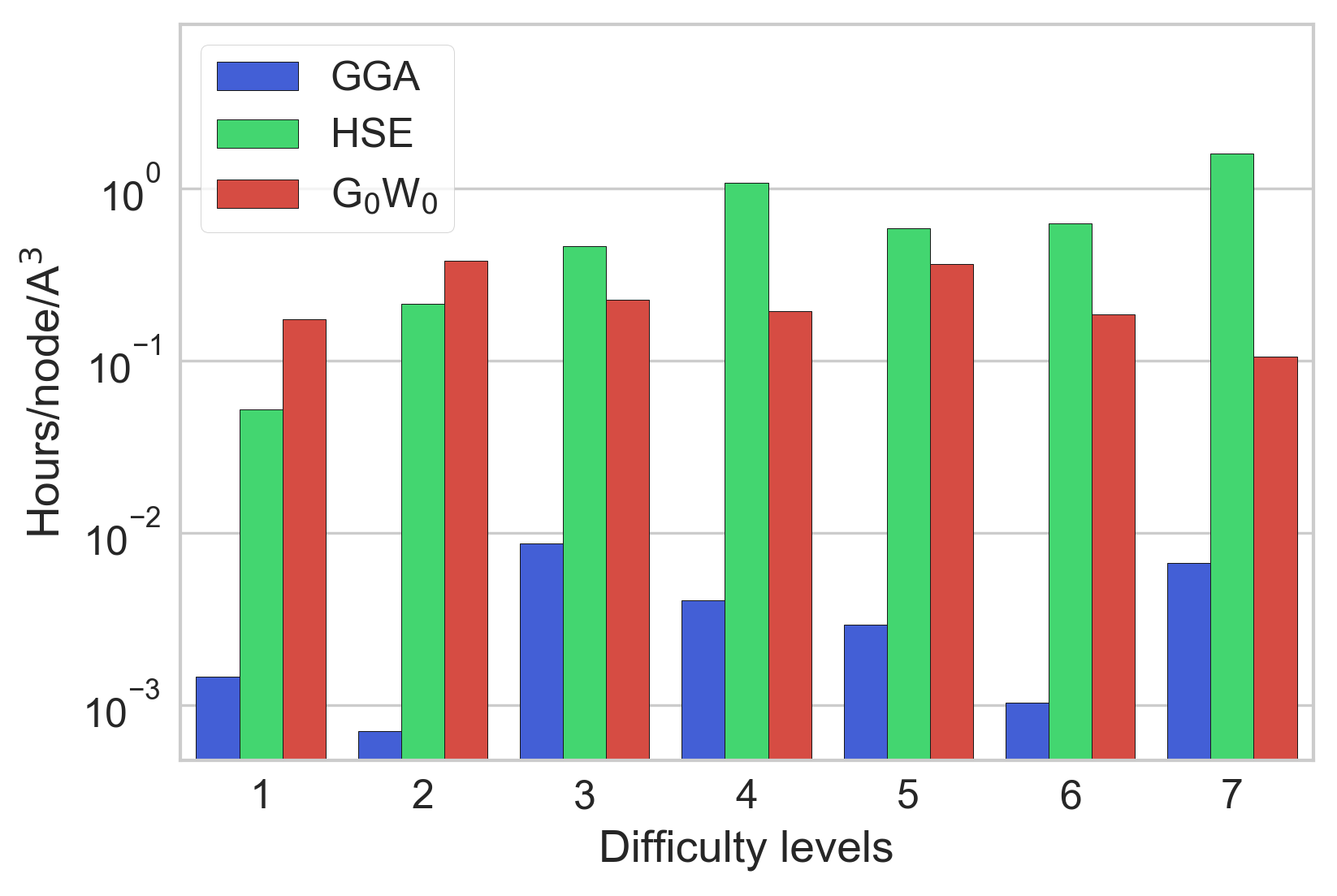}
        \caption{
            Calculation time per each difficulty level (as defined in section \ref{sec:methodology}). The time is normalized per one compute node and unit cell volume (\AA$^3$). The electronic band structures are calculated in full for GGA and HSE only.
        }
        \label{figure:difficulty-wise-runtime}
    \end{figure}

    In order to organize the information about the simulation workflows we employ the categorization illustrated in Table \ref{table:difficulty}. The categorization depends on: (a) whether the semi-core electronic states are included in the pseudopotentials, (b) whether the treatment of spin-orbit coupling is considered within the calculation, and (c) whether the treatment of magnetic interactions is included. Larger numbers, as included in Table \ref{table:difficulty} do not necessarily correspond to the higher computational difficulty (see Fig.\ref{figure:difficulty-wise-runtime}, for example).  Difficulty 1 (D1) workflows are GGA calculations with default set of pseudopotentials, as implemented in VASP 5.4.4\cite{kresse1996software}. In terms of theory, the even difficulty numbers, (eg. 2) workflows are similar to the nearest odd (eg. 1), except for the inclusion of semi-core states in the pseudopotential for the following elements: Ga, Ge, In, Sn, Ti, Pb, Bi, Li, Na, Ca, K, Rb, Sr, Cs, Ba. Readers may consult the data online for further detailed information about the types of semi-core states included\cite{exabytePlatformBGPhaseIURL}. We prioritize the most comprehensive set whenever available, such that if a pseudopotential with only $p$ and both $p$ and $s$ states are present, we use the latter.
    
    The difficulty 3 and 4 workflows incorporate spin-orbit coupling (SOC). We treated all materials that contain elements with atomic number Z $>$ 45 (Rh) as the ones that require spin-orbit coupling to be included in the calculations. This is, notably, a somewhat "loose" approach, as there exists a well known spin-orbit splitting effect for GaAs, for example\cite{louie1991gaas-soc}. We argue, however, that since the latter effect is of the order of 100 meV it would not be critical to the results of this study. This statement is further supported by the band gap value for GaAs found in this work. For the difficulty levels 5 and 6 we incorporate collinear magnetism as follows: we switch on the magnetic interactions and set the initial magnetic moments to a pre-defined value for all ferromagnetic atoms (V, Cr, Mn, Fe, Co, Ni). When more than one atom is present in the unit cell, we alternate the signs for the magnetic moment effectively creating an anti-ferromagnetic arrangement in this case. Lastly, the difficulty 7 workflows have all three, and, due to the nature of the computational implementation, resolve the non-collinear magnetic interactions.

    \begin{table}[h!]
    	\centering
    	\begin{tabular}{c |  c   c  c | c}
    		\hline
    		\hline
    		Difficulty level   & Semi-core & SOC   & Magnetism & materials \\
    	    \hline
    	    1                  & no        & no    & no   & 23 \\
      	    2                  & yes       & no    & no   & 16 \\
      	    3                  & no        & yes   & no   & 8 \\
      	    4                  & yes       & yes   & no   & 9 \\
      	    5                  & no        & no    & yes  & 4 \\
      	    6                  & no        & no    & yes  & 5 \\
      	    7                  & yes       & yes   & yes  & 6 \\
    		\hline
    		\hline
    	\end{tabular}
    	\caption{
    	    Summary of the simulation workflows categorization employed in this work. "Semi-core" indicates that the pseudopotentials with semi-core states were used, "SOC" stands for the inclusion of the spin-orbit coupling, and "Magnetism" is used to denote the inclusion of collinear magnetic moments, except for the difficulty 7 when spin-orbit coupling and magnetism are included both, which lead to the treatment of non-collinear magnetic interactions.
    	}
    	\label{table:difficulty}
    \end{table}

\subsection{Computational setup}

    \subsubsection{Software/Theory}

    All Density Functional Theory\cite{kresse1996efficient, kohn1965self} calculations were performed within the pseudopotential projector augmented wave (PAW)\cite{blochl1994projector} formalism using the Vienna Ab initio Simulation Package (VASP)\cite{kresse1996software, hacene2012accelerating}. Within the generalized gradient approximation the exchange-correlation effects were modeled using the Perdew-Berke-Ernzerhof (PBE)\cite{perdew1996generalized} functional. All calculations were performed with the largest default plane wave cutoff energy of the pseudopotentials involved. The energies of all calculations were converged to within $10^{-4}$ eV. The Gaussian method was chosen as the smearing algorithm, the blocked Davidson iteration scheme\cite{johnson2001block} was chosen as the electron minimization algorithm, and ions were updated using the conjugated gradient algorithm. A smearing value of $50$ meV was chosen for all the calculations. The semi-empirical Grimme-D2 correction to the Kohn-Sham energies were incorporated in all of our calculations\cite{grimme2006vdWcorrection}. The Heyd-Scuseria-Ernzerhof (HSE) calculations incorporate a 25$\%$ short-range Hartree-Fock exchange\cite{heyd2003hybrid}. The screening parameter $\mu$ is set to 0.2 \AA$^{-1}$. GW calculations were performed at the non-self-consistent G$_0$W$_0$ level. The number of unoccupied bands for the band gap calculation step was set to the total number of plane waves in the SCF step.
    
    We implemented sampling in the reciprocal cell based on k-points per reciprocal atom (KPPRA) with a uniform unshifted grid. In our calculations, KPPRA of 2,000 were used unless specified otherwise. The density of states (DOS) calculations were performed within a denser grid with KPPRA of 16,000 using tetrahedron interpolation as implemented in VASP\cite{kresse1996software}. We ran most of the calculations within a single compute node described in the next subsection. For some $G_0W_0$ calculations in particular, the memory requirements were larger than the resources available on a single node, however and efficient parallelization scheme for memory distribution is yet to be implemented for GW calculations in VASP at the moment of this writing. To accommodate the calculations within the available memory, we reduced the precision in a controlled way as follows: we limited the number of bands to 1000 at most, instead of using all available, then we reduced the KPPRA value to reduce memory requirements. The details about the cases with reduced precision are summarized in the footnotes of Table\ref{table:materials-data}.

    \subsubsection{Hardware}
    
    All calculations were performed using the hardware available from Microsoft Azure cloud computing service\cite{azure-instance-types}. We utilized the "H16r" and "H16mr" instances specifically designed to handle high performance computing workloads. The instances are based on the Intel Xeon E5-2667 v3 Haswell 3.2 GHz (3.6 GHz with turbo) with 16 cores per node, and 112 and 224 GB of memory respectively. The instances carry a low latency, high-throughput network interface optimized and tuned for remote direct memory access. Computational resources were provisioned and assembled on-demand by software implemented and available within the Exabyte platform\cite{exabytePlatform}. Most of the calculations were executed within a two-week period with a few requiring further work beyond that time frame. The peak size of the computational infrastructure used during this work was administratively limited to 125 nodes or 2000 total computing cores.

\subsection{Data extraction}

    The relevant data for each workflow unit is extracted from the calculation output, parsed and stored in the database in the JSON format according to EDC. For instance, the forces on each atom after the volume relaxation are extracted and shown in the results page for each material. The band structure and the density of states (DOS) are also extracted from the band structure and density of states calculations, respectively. Thus, results for each material can be viewed online on Exabyte platform \cite{exabytePlatformBGPhaseIURL}. The platform also support a programmatic way of extraction of the data associated with materials and simulations through a RESTful application programming interface\cite{exabyteRESTAPIClient}, partly used in this work as well.

\subsection{Repeatability}
    
    The materials, workflows, batch jobs for each material with the associated properties, and files for each step of the simulation workflows are all made readily available online\cite{exabytePlatformBGPhaseIURL}. The Exabyte platform now contains all materials and workflows mentioned in this work, so readers may create an account, copy one or more materials to their account collection, copy a workflow similarly, and use the simulations designer as mentioned in Fig. \ref{figure:execution-flow}  to recreate the simulation for this material. Furthermore, users can introduce modifications to our workflow and further improve the results.
    

    \begin{table*}[t!]
\centering
\begin{ruledtabular}
\begin{tabular}{@{}cccccccccccccccccc@{}}
    Formula & Diff. & \multicolumn{3}{c}{Calculated} & \multicolumn{3}{c}{Experiments}   & \multicolumn{6}{c}{Band gaps (eV)}\Bstrut& \multicolumn{4}{c}{References}  \\ \cline{3-5} \cline{6-8} \cline{9-14} \cline{15-18}
    & & a & b & c & a & b & c & GGA & HSE & G$_0$W$_0$ & HSE$^\prime$ & G$_0$W$_0^\prime$ & Expt. & Lat. & Gap & HSE$^\prime$ & G$_0$W$_0^\prime$ \\
     \hline
    \multicolumn{18}{c}{Elemental}  \\
    \hline
    Si & 1 & 3.75 & 3.75 & 3.75 & 3.82 & 3.82 & 3.82 & 0.56 & 1.14 & 1.09 & 1.28\footnotemark[1] & 1.12 & 1.17 & \cite{jette1935precision} & \cite{kittel1996introduction} & \cite{heyd2005energy} & \cite{shishkin2007self} \\
    Ge & 2 & 3.98 & 3.98 & 3.98 & 4.00 & 4.00 & 4.00 & 0.16 & 0.86 & 0.84 & 0.56\footnotemark[1] & 0.66 & 0.75 & \cite{madelung2012semiconductors} &  & \cite{heyd2005energy} & \cite{van2006adequacy} \\
    Te & 3 & 4.32 & 4.32 & 6.02 & 4.45 & 4.45 & 4.45 & 0.00 & 0.42\footnotemark[2] & 0.00 & 0.32 & - & 0.32 & \cite{Se_Te_abc} & \cite{anzin1977measurement} & \cite{yi2018nature} &  \\
	B & 1 & 4.85 & 4.85 & 5.00 & 5.06 & 5.06 & 5.06 & 1.20 & 1.70 & 1.58 & - & - & 1.49 & \cite{decker1959crystal} & \cite{madelung2012semiconductors} &  &  \\
	Bi & 4 & 4.50 & 4.50 & 4.50 & 4.54 & 4.54 & 4.54 & 0.06 & 0.00 & 0.00\footnotemark[4]\footnotemark[3] & - & - & 0.00 & \cite{cucka1962crystal} & \cite{madelung2012semiconductors} &  &  \\
	P & 1 & 3.33 & 4.37 & 5.45 & 3.31 & 5.92 & 4.38 & 0.14 & 0.14 & 0.16 & 0.39 & 0.30 & 0.35 & \cite{brown1965refinement} & \cite{asahina1984band} & \cite{gomes2015phosphorene} & \cite{tran2014layer} \\
	As & 1 & 3.79 & 3.79 & 3.99 & 3.65 & 3.65 & 4.47 & 0.03 & 0.29 & 0.15 & 0.00 & - & 0.30 & \cite{smith1975structures} & \cite{madelung2012semiconductors} & \cite{kecik2016stability} &  \\
	Sb & 3 & 4.31 & 4.31 & 4.46 & 4.30 & 4.30 & 4.30 & 0.00 & 0.00 & 0.00 & - & - & 0.00 & \cite{barrett1963crystal} & \cite{madelung2012semiconductors} &  &  \\
	Se & 1 & 4.21 & 4.21 & 5.10 & 4.37 & 4.37 & 4.96 & 0.64 & 1.34 & 1.38 & - & - & 1.85 & \cite{Se_Te_abc} & \cite{madelung2012semiconductors} &  &  \\
	grey-Sn & 3 & 4.59 & 4.59 & 4.59 & 4.57 & 4.57 & 4.57 & 0.00 & 0.00 & 0.00 & 0.00 & - & 0.00 & \cite{brownlee1950lattice} & \cite{madelung2012semiconductors} & \cite{hummer2009heyd} &  \\
    \hline
    \multicolumn{18}{c}{III-V semiconductors}  \\
    \hline
    BN & 1 & 2.50 & 2.50 & 6.61 & 2.50 & 2.50 & 6.66 & 3.15 & 4.20 & 4.32 & 5.98\footnotemark[1] & 5.4\footnotemark[7] & 5.95 & \cite{blase1995quasiparticle} & \cite{cassabois2016hexagonal} & \cite{heyd2005energy} & \cite{tran2009accurate} \\
	BP & 1 & 3.19 & 3.19 & 3.19 & 3.20 & 3.20 & 3.20 & 1.21 & 1.93 & 1.95 & 2.16\footnotemark[1] & - & 2.1 & \cite{madelung2012semiconductors} & \cite{madelung2012semiconductors} & \cite{heyd2005energy} &  \\
	GaP & 2 & 3.82 & 3.82 & 3.82 & 3.85 & 3.85 & 3.85 & 1.56 & 2.25 & 2.22 & 2.47\footnotemark[1] & 2.48 & 2.35 & \cite{madelung2012semiconductors} & \cite{madelung2012semiconductors} & \cite{heyd2005energy} & \cite{lee2016prediction} \\
	BAs & 1 & 3.36 & 3.36 & 5.57 & 3.37 & 3.37 & 3.37 & 1.12 & 1.77 & 1.73 & 1.92\footnotemark[1] & 1.93 & - & \cite{merrill1977behavior} &  & \cite{heyd2005energy} & \cite{lee2016prediction} \\
	BSb & 3 & 3.71 & 3.71 & 3.71 & 3.62 & 3.62 & 3.62 & 0.64 & 1.16 & 1.04 & 1.37\footnotemark[1] & 1.28 & - & \cite{madelung2012semiconductors} &  & \cite{heyd2005energy} & \cite{lee2016prediction} \\
	AlN & 1 & 3.10 & 3.10 & 4.98 & 3.11 & 3.11 & 4.97 & 4.29 & 5.69 & 6.03 & 6.45\footnotemark[1] & 5.83 & 6.19 & \cite{madelung2012semiconductors} & \cite{madelung2012semiconductors} & \cite{heyd2005energy} & \cite{van2006adequacy} \\
	AlAs & 1 & 3.99 & 3.99 & 3.99 & 3.96 & 3.96 & 3.96 & 1.42 & 2.08 & 2.09 & 2.24\footnotemark[1] & 2.59 & 2.23 & \cite{madelung2012semiconductors} & \cite{madelung2012semiconductors} & \cite{heyd2005energy} & \cite{lee2016prediction} \\
	GaAs & 2 & 3.98 & 3.98 & 3.98 & 4.00 & 4.00 & 4.00 & 0.63 & 1.52 & 1.74 & 1.21\footnotemark[1] & 1.30 & 1.51 & \cite{madelung2012semiconductors} & \cite{madelung2012semiconductors} & \cite{heyd2005energy} & \cite{van2006adequacy} \\
	GaN & 4 & 3.19 & 3.19 & 3.19 & 3.20 & 3.20 & 3.20 & 1.71 & 2.98 & 2.98 & 3.03\footnotemark[1] & 2.80 & 3.17 & \cite{powell1993heteroepitaxial} & \cite{madelung2012semiconductors} & \cite{heyd2005energy} & \cite{van2006adequacy} \\
	YN & 1 & 3.43 & 3.43 & 3.43 & 3.44 & 3.44 & 3.44 & 0.16 & 1.04 & 0.75 & - & 0.97 & - & \cite{saha2011electronic} &  &  & \cite{saha2011electronic} \\
    \hline
    \multicolumn{18}{c}{II-VI semiconductors}  \\
    \hline
    ZnS & 1 & 3.79 & 3.79 & 3.79 & 3.82 & 3.82 & 3.82 & 2.22 & 3.48 & 3.51 & 3.42\footnotemark[1] & 3.29 & 3.54 & \cite{madelung2012semiconductors} & \cite{madelung2012semiconductors} & \cite{heyd2005energy} & \cite{van2006adequacy} \\
	BeS & 1 & 3.42 & 3.42 & 3.42 & 3.44 & 3.44 & 3.44 & 3.1 & 4.05 & 4.47 & 4.14 & 4.92 & 5.5 & \cite{madelung2012semiconductors} & \cite{madelung2012semiconductors} & \cite{laref2013comparative} & \cite{lee2016prediction} \\
	BeSe & 1 & 3.64 & 3.64 & 3.64 & 3.64 & 3.64 & 3.64 & 2.62 & 3.49 & 3.81 & 3.54 & 4.19 & 4.00 & \cite{madelung2012semiconductors} & \cite{madelung2012semiconductors} & \cite{laref2013comparative} & \cite{lee2016prediction} \\
	BeTe & 3 & 3.98 & 3.98 & 3.98 & 3.98 & 3.98 & 3.98 & 1.67 & 2.34 & 2.65 & 2.68 & 3.17 & 2.80 & \cite{madelung2012semiconductors} & \cite{madelung2012semiconductors} & \cite{laref2013comparative} & \cite{lee2016prediction} \\
	MgS & 1 & 3.65 & 3.65 & 3.65 & 3.67 & 3.67 & 3.67 & 2.84 & 3.84 & 4.46 & 4.78\footnotemark[1] & 4.044 & 4.50 &  & \cite{madelung2012semiconductors} & \cite{heyd2005energy} & \cite{nejatipour2015excitonic} \\
	BaSe & 4 & 4.49 & 4.49 & 4.49 & 4.66 & 4.66 & 4.66 & 1.48 & 2.21 & 3.09\footnotemark[4]\footnotemark[5] & 2.87\footnotemark[1] & 2.99 & 3.60 & \cite{grzybowski1983high} & \cite{madelung2012semiconductors} & \cite{heyd2005energy} & \cite{nejatipour2015excitonic} \\
	BaTe & 4 & 4.90 & 4.90 & 4.90 & 4.95 & 4.95 & 4.95 & 1.18 & 1.82 & 2.56\footnotemark[4]\footnotemark[5] & 2.50\footnotemark[1] & 2.33 & 3.40 & \cite{grzybowski1984band} & \cite{madelung2012semiconductors} & \cite{heyd2005energy} & \cite{nejatipour2015excitonic} \\
	CaSe & 2 & 4.13 & 4.13 & 4.13 & 4.18 & 4.18 & 4.18 & 1.93 & 2.64 & 3.5 & 3.02\footnotemark[1] & 3.94 & - & \cite{luo1994structural} &  & \cite{heyd2005energy} &  \\
	Na$_2$S & 2 & 4.56 & 4.56 & 4.56 & 4.62 & 4.62 & 4.62 & 2.63 & 3.71 & 4.67 & - & 4.77 & 5.00 & \cite{zintl1934gitterstruktur} &  &  &  \\
	MgSe & 1 & 4.20 & 4.20 & 6.80 & 4.15 & 4.15 & 6.72 & 2.71 & 3.71 & 4.26 & 2.62\footnotemark[1] & 4.58 & 4.05 & \cite{mittendorf1965rontgenographische} & \cite{madelung2012semiconductors} & \cite{heyd2005energy} &  \\
	MgTe & 3 & 4.56 & 4.56 & 7.41 & 4.53 & 4.53 & 7.40 & 2.26 & 3.00 & 3.54 & 3.74\footnotemark[1] & 4.19 & 3.49 & \cite{madelung2012semiconductors} & \cite{madelung2012semiconductors} & \cite{heyd2005energy} & \cite{lee2016prediction} \\
    \hline
    \multicolumn{18}{c}{Dichalcogenides}  \\
    \hline
    MoS$_2$ & 1 & 3.17 & 3.17 & 12.37 & 3.16 & 1.16 & 12.29 & 0.95 & 1.45 & 1.37 & 1.06 & 1.28 & 1.29 & \cite{bronsema1986structure} & & \cite{wickramaratne2014electronic} & \cite{cheiwchanchamnangij2012quasiparticle} \\
	HfSe$_2$ & 3 & 3.71 & 3.71 & 6.04 & 3.67 & 3.67 & 6.00 & 0.07 & 0.63 & 0.77 & 1.07 & 1.08 & 1.10 & \cite{hodul1984anomalies} &  & \cite{heyd2005energy} & \cite{abdulsalam2016optical} \\
	TiS$_2$ & 2 & 3.36 & 3.36 & 6.62 & 3.41 & 3.41 & 5.69 & 0.00 & 0.38 & 0.09\footnotemark[4]\footnotemark[6] & 0.40 & - & 0.30 & \cite{wiegers1992structures} &  & \cite{suga2015momentum} &  \\
	CrS$_2$ & 5 & 3.04 & 3.04 & 6.77 &  & - &  & 0.00 & 0.00 & 0.34\footnotemark[8]\footnotemark[3] & - & - & - &  &  &  &  \\
	MnS$_2$ & 5 & 3.28 & 3.28 & 6.57 &  & - &  & 0.00 & 0.00 & 0.00\footnotemark[9] & - & - & 0.00 &  & \cite{ennaoui1993iron} &  &  \\
\end{tabular}
\end{ruledtabular}
\settasks{
  label-width = 1em ,
  before-skip = -\parskip , 
  after-skip = -\parskip , 
  after-item-skip = -\parskip 
}
\scriptsize{
    \begin{tasks}(5)
        \task HSE03
        \task Mixing is tuned
        \task KPPRA 400
        \task Number of bands 1,000
        \task KPPRA 200
        \task KPPRA 650
        \task GW calculation
        \task Number of bands 864
        \task KPPRA 500
    \end{tasks}
}
\vspace{-3mm}
\caption{
    Data for materials studied in this work. ``Diff." - difficulty level. ``Calc." and ``Expt." have lattice constants of the relaxed structure and the experimental values, respectively. Lattice constants of the primitive unit cell are given, unless otherwise noted. A linear relationship between the two materials is assumed to determine the experimental lattice constant for alloys. The HSE and G$_0$W$_0$ values are compared with references when available: "HSE$^\prime$" and "G$_0$W$_0^\prime$". Reduced precision calculations are indicated in footnotes. "HSE03" - HSE03 approach\cite{heyd2003hybrid}. "GW" indicates the full GW approximation, "sc--GW" - self-consistent GW calculations. Materials for which the DFT$+$U approach was used are also noted.
}
\label{table:materials-data}
\end{table*}

\begin{table*}[th!]
\centering
\begin{ruledtabular}
\begin{tabular}{@{}cccccccccccccccccc@{}}
    Formula & Diff. & \multicolumn{3}{c}{Calculated} & \multicolumn{3}{c}{Experiment}   & \multicolumn{6}{c}{Band gaps (eV)}\Bstrut& \multicolumn{4}{c}{References}  \\ \cline{3-5} \cline{6-8} \cline{9-14} \cline{15-18}
    & & a & b & c & a & b & c & GGA & HSE & G$_0$W$_0$ & HSE$^\prime$ & G$_0$W$_0^\prime$ & Expt. & Lat. & Gap & HSE$^\prime$ & G$_0$W$_0^\prime$ \\
     \hline
    \multicolumn{18}{c}{Binary oxides}  \\
    \hline
    Li$_2$O & 2 & 3.15 & 3.15 & 3.15 &  & - &  & 5.18 & 6.85 & 8.07 & - & 8.10 & 8.00 &  & \cite{ishii1999optical} &  & \cite{sommer2012quasiparticle} \\
	MgO & 1 & 2.95 & 2.95 & 2.95 & 2.98 & 2.98 & 2.98 & 5.05 & 6.89 & 8.01 & 6.50 & 7.25 & 7.67 & \cite{tsirelson1998x} & \cite{madelung2012semiconductors} & \cite{heyd2005energy} &  \\
	BeO & 1 & 2.68 & 2.68 & 2.68 & 2.70 & 2.70 & 2.70 & 6.92 & 8.71 & 9.62 & 10.09 & 10.29 & 10.59 & \cite{madelung2012semiconductors} & \cite{madelung2012semiconductors} & \cite{shi2014strain} &  \\
	B$_2$O$_3$ & 1 & 4.12 & 4.49 & 4.49 & 4.13 & 4.61 & 4.61 & 9.97 & 11.00 & 12.04 & - & - & - & \cite{prewitt1968crystal} & \ &  &  \\
	SnO$_2$ & 4 & 3.23 & 4.74 & 4.75 & 3.19 & 4.74 & 4.74\footnotemark[1] & 1.00 & 2.84 & 2.74\footnotemark[2]$^,$\footnotemark[12] & 3.50\footnotemark[3] & 2.88 & 3.60 & \cite{mccarthy1989x} & \cite{madelung2012semiconductors} & \cite{janotti2011lda+} & \cite{berger2010ab} \\
	Al$_2$O$_3$ & 1 & 4.75 & 4.75 & 5.11 & 4.76 & 4.76 & 4.76 & 6.28 & 8.22 & 9.29 & 8.82 & - & 8.80\footnotemark[3] & \cite{finger1978crystal} & \cite{robertson2000band} & \cite{janotti2011lda+} &  \\
	$\alpha$-SiO$_2$ & 1 & 4.84 & 4.84 & 4.97 & 4.96 & 4.96 & 4.96 & 6.07 & 8.12 & 9.48 & 8.72 & 10.10\footnotemark[4] & 9.30 & \cite{pluth1985crystal} & \cite{weinberg1979transmission} & \cite{varley2012role} & \cite{kresse2012optical} \\
	BaO$_2$ & 4 & 3.72 & 3.72 & 4.15 & 3.78 & 3.78 & 4.30 & 2.02 & 3.60 & 3.73\footnotemark[2]$^,$\footnotemark[8] & - & - & 4.29 & \cite{bernal1935structure} & \cite{rao1979logarithmic} &  &  \\
	NaO$_3$ & 2 & 3.17 & 4.21 & 4.21 & 2.94 & 4.12 & 4.12 & 0.00 & 0.00 & 0.00 & - & - & - & \cite{klein1998synthesis} &  &  &  \\
	VO$_2$ & 5 & 2.80 & 4.52 & 4.52 & 2.85 & 4.55 & 4.55\footnotemark[1] & 0.00 & 0.18 & 0.00\footnotemark[2]$^,$\footnotemark[6] & - & - & - & \cite{kucharczyk1979accurate} &  &  &  \\
	TiO$_2$ & 2 & 12.19 & 3.72 & 6.55 & 12.16 & 3.74 & 6.51\footnotemark[5] & 2.78 & 4.28 & 4.66 & 3.13\footnotemark[3] & 3.10 & - & \cite{banfield1991identification} &  & \cite{janotti2011lda+} & \cite{berger2012computational} \\
	NiO & 5 & 2.93 & 2.93 & 2.93 & 2.95 & 2.95 & 2.95\footnotemark[5] & 0.00 & 2.98 & 4.07 & 4.10 & 3.60 & 4.30\footnotemark[7] & \cite{sasaki1979x} &  & \cite{gillen2013accurate} & \cite{toroker2011first} \\
	CaO & 2 & 3.94 & 3.94 & 4.73 &  & - &  & 3.30 & 4.72 & 4.97 & 4.23 & 4.40 & 6.93 &  & \cite{madelung2012semiconductors} & \cite{riefer2011interplay} & \cite{riefer2011interplay} \\
	ZnO-cub & 1 & 3.04 & 3.04 & 3.04 & 3.02 & 3.02 & 3.02 & 0.83 & 2.42 & 2.42 & 2.49 & 2.12 & 3.44 & \cite{bates1962new} & \cite{madelung2012semiconductors} & \cite{heyd2005energy} &  \\
	SrO & 2 & 3.60 & 3.60 & 3.60 & 3.64 & 3.64 & 3.64 & 3.25 & 4.75 & 5.10 & 4.70 & 5.57 & 5.22 & \cite{primak1948x} & \cite{madelung2012semiconductors} & \cite{bajdich2015surface} &  \\
    \hline
    \multicolumn{18}{c}{Ternary oxides}  \\
    \hline
    NdClO & 3 & 3.85 & 3.85 & 6.77 & 4.03 & 4.03 & 6.76\footnotemark[5] & 0.00 & 0.00 & 0.00\footnotemark[2]\footnotemark[8] & - & - & - & \cite{zachariasen1949crystal} &  &  &  \\
	SmNiO$_3$ & 7 & 5.31 & 5.33 & 7.41 & 5.33 & 5.44 & 7.57\footnotemark[5] & 0.00 & 0.00 & 0.00 & - & - & - & \cite{lacorre1991synthesis} &  &  &  \\
	NaOsO$_3$ & 4 & 5.25 & 5.25 & 5.25 &  & - &  & 0.00 & 1.34 & 0.12\footnotemark[2]$^,$\footnotemark[14] & - & - & - &  &  &  &  \\
	GdTiO$_3$ & 4 & 5.38 & 5.46 & 7.60 & 5.41 & 5.67 & 7.69 & 0.00 & 0.00 & 0.00\footnotemark[2]\footnotemark[12] & - & - & - & \cite{mccarthy1969preparation} &  &  &  \\
	SrTiO$_3$ & 2 & 5.50 & 5.50 & 5.51 & 5.51 & 5.51 & 5.51 & 1.94 & 3.24 & 2.90 & 3.20 & 3.57 & 3.25 & \cite{jauch1999anomalous, long2013lattice} & \cite{van2001bulk} & &  \\
	LaCoO$_3$ & 7 & 5.30 & 5.30 & 5.30 & 5.34 & 5.34 & 5.34 & 0.00 & 2.52 & 0.24\footnotemark[2]\footnotemark[12] & 2.52 & - & 0.60 & \cite{thornton1986neutron} & \cite{chainani1992electron} & \cite{zhang2014density} &  \\
	LaNiO$_3$ & 7 & 5.32 & 5.32 & 5.32 & 5.45 & 5.45 & 5.45 & 0.00 & 0.00 & 0.00\footnotemark[2]\footnotemark[12] & - & - & 0.0001 & \cite{garcia1992neutron} & \cite{ruegg2012electronic} &  &  \\
	LaMnO$_3$ & 7 & 3.84 & 3.84 & 3.84 & 3.88 & 3.88 & 3.88 & 0.00 & 0.00 & 0.00\footnotemark[2]\footnotemark[12] & - & - & - & \cite{v1943strukturen} &  &  &  \\
	GdMn$_2$O$_5$ & 7 & 5.63 & 7.23 & 8.5 & 5.68 & 7.35 & 8.54\footnotemark[1] & 0.62 & 2.29 & 2.27 & - & - & - & \cite{kagomiya2002structure} &  &  &  \\
	GdCoO$_3$ & 7 & 3.72 & 3.72 & 3.74 & 3.80 & 3.80 & 3.80 & 0.00 & 0.00 & 0.00\footnotemark[2]$^,$\footnotemark[11] & - & - & - & \cite{casalot1971evolution} &  &  &  \\
    \hline
    \multicolumn{18}{c}{Semiconductor Alloys}  \\
    \hline
    SiGe & 2 & 3.84 & 3.84 & 6.33 & 3.85 & 3.85 & 3.85 & 0.36 & 0.78 & 0.94 & - & - & 0.91 & \cite{ferrari2008strain} &  &  &  \\
	SiSn & 6 & 4.15 & 4.15 & 4.15 & 4.21 & 4.21 & 4.21 & 0.49 & 0.95 & 1.00 & - & - & 1.11 & \cite{soref1992electro} & \cite{hussain2015sisn} &  &  \\
	AlGaAs & 2 & 5.59 & 5.59 & 5.59 & 5.66 & 5.66 & 5.66 & 1.21 & 1.95 & 2.20 & - & - & 2.02 & \cite{agostini2011characterization} & \cite{agostini2011characterization} &  &  \\
	InGaAs & 6 & 5.84 & 5.84 & 5.85 & 5.84 & 5.84 & 5.84 & 0.00 & 0.82 & 1.19\footnotemark[2]$^,$\footnotemark[10] & - & - & 0.77 & \cite{agostini2011characterization} & \cite{agostini2011characterization} &  &  \\
	InGaP & 6 & 5.60 & 5.60 & 5.65 & 5.66 & 5.66 & 5.66\footnotemark[13] & 1.10 & 1.95 & 1.89\footnotemark[2]\footnotemark[10] & - & - & 1.90 &  &  &  &  \\
	AlInAs & 6 & 5.79 & 5.79 & 5.80 & 5.86 & 5.86 & 5.86\footnotemark[13] & 0.97 & 1.86 & 2.16\footnotemark[2]\footnotemark[10] & - & - & 1.50 &  &  &  &  \\
	AlInSb & 6 & 6.19 & 6.19 & 6.21 & 6.30 & 6.30 & 6.30\footnotemark[13] & 0.78 & 1.31 & 1.35\footnotemark[2]\footnotemark[10] & - & - & 1.13 &  &  &  &  \\
	GaAsP & 2 & 5.51 & 5.51 & 5.51 & 5.55 & 5.55 & 5.55\footnotemark[13] & 1.20 & 1.77 & 1.96\footnotemark[2]\footnotemark[6] & - & - & 2.03 &  &  &  &  \\
	GaAsSb & 4 & 5.84 & 5.84 & 5.83 & 5.87 & 5.87 & 5.87\footnotemark[13] & 0.08 & 0.53 & 0.54\footnotemark[9]$^,$\footnotemark[12] & - & - & 0.72 &  &  &  &  \\
	AlGaN & 2 & 4.46 & 4.46 & 4.46 & 4.45 & 4.45 & 4.45\footnotemark[13] & 2.36 & 3.66 & 3.83 & - & - & 4.60 &  &  &  &  \\
\end{tabular}
\end{ruledtabular}
\settasks{
  label-width = 1em ,
  before-skip = -\parskip , 
  after-skip = -\parskip , 
  after-item-skip = -\parskip 
}
\scriptsize{
    \begin{tasks}(2)
        \task Lattice parameters are shuffled.
        \task Number of bands 1,000
        \task The mixing parameter is tuned to get experimental value
        \task Self-consistent GW
        \task Conventional unit cell
        \task KPPRA 1,000
        \task DFT+U
        \task KPPRA 500
        \task Number of bands 500
        \task KPPRA 1,500
        \task KPPRA 650
        \task KPPRA 200
        \task KPPRA 250
    \end{tasks}
}
\vspace{0.5cm}
\caption{Table \ref{table:materials-data} continued.}
\vspace{2cm}
\end{table*}

    \section{Results}
\label{sec:results}
    
\begin{figure}[ht!]
        \includegraphics[width = 0.48\textwidth]{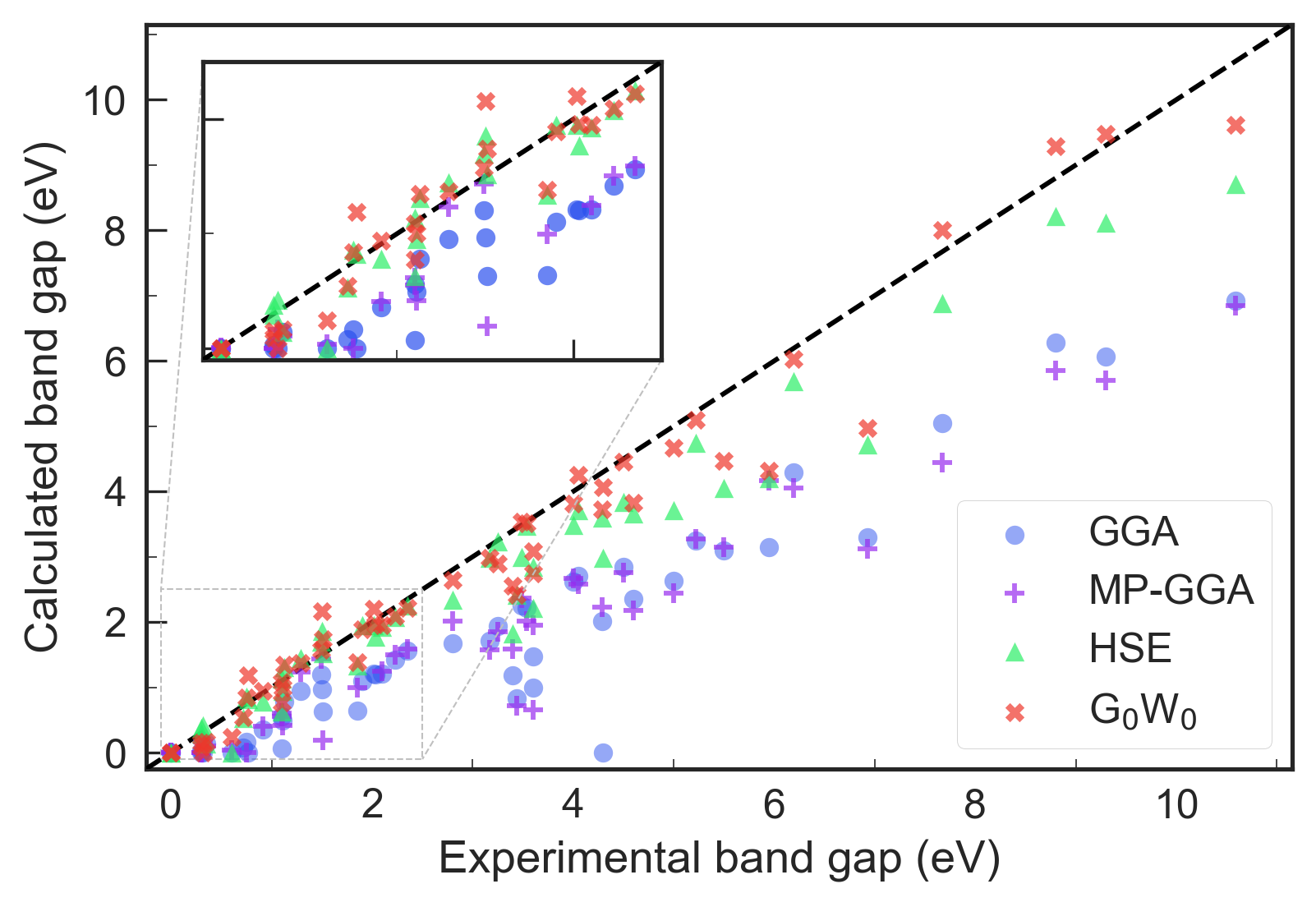}
        \caption{
            Comparative plot of the calculated and experimentally available values for all the electronic band gaps obtained in the current work. Legend: GGA, HSE, and $G_0W_0$ denote the results of this work for the corresponding level of theory. MP-GGA denote the results of Materials Project\cite{jain2013materialsproject} available at the moment of this writing.
        }
        \label{fig-all}
    \end{figure}

\begin{figure}[ht!]
        \includegraphics[width = 0.48\textwidth]{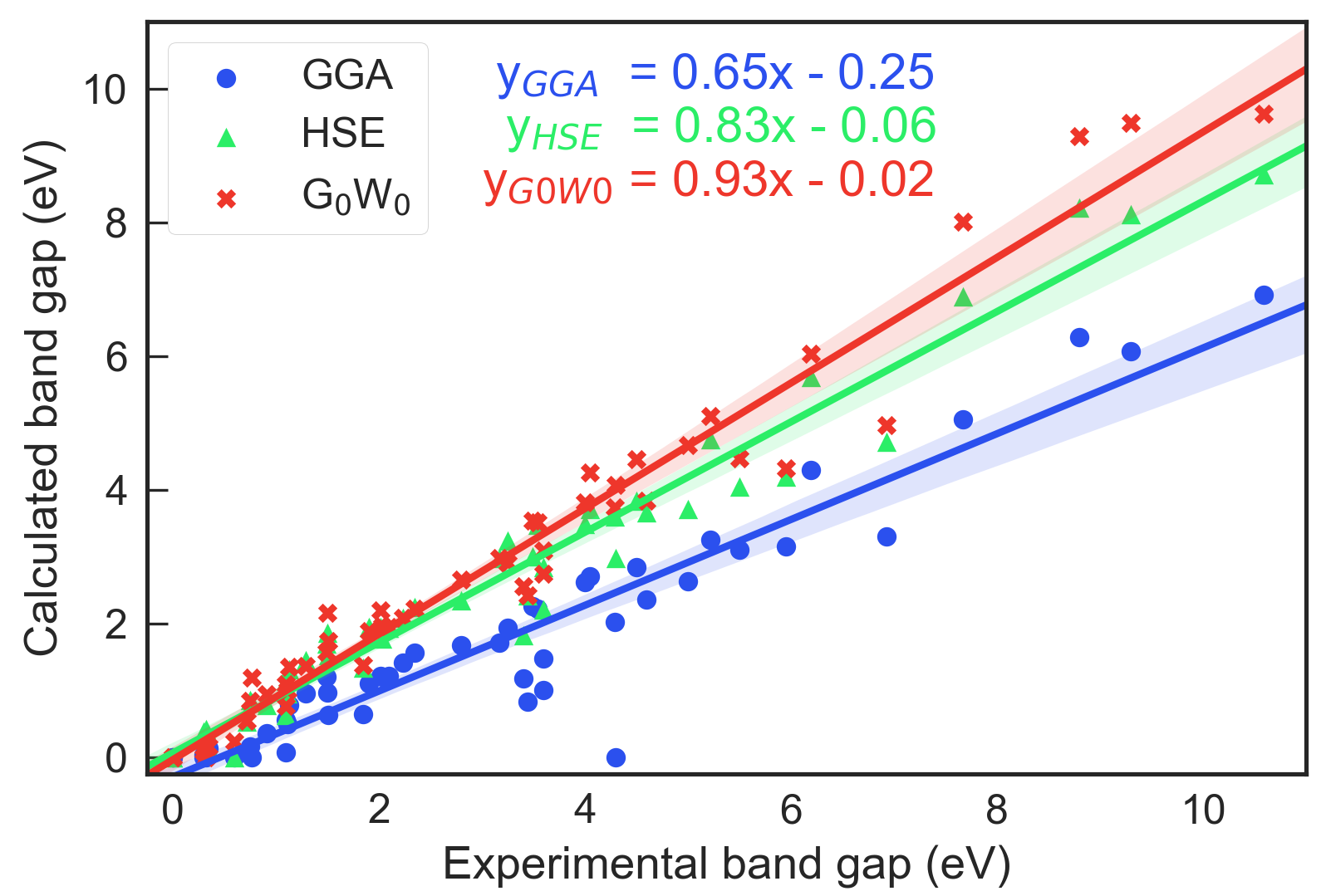}
        \caption{
            Comparative plot for all the band gaps calculated in the current work, including the linear $y = kx + b$ fits to data per each model. The legend is same as in Fig.\ref{fig-all}. The equations for each of the linear fits are shown in the figure.
        }
        \label{fig-reg}
    \end{figure}

    Fig. \ref{fig-all} shows a comparison of the calculated band gaps within GGA, HSE and G$_0$W$_0$ with their experimental values for all the materials where the experimental data is available. We also include the results of Materials Project\cite{jain2013materialsproject} (further referred to as MP) calculated within GGA (or GGA+U approach when specifically noted) for reference. As expected, in can be seen that the GGA underestimates the band gaps, and HSE and G$_0$W$_0$ both significantly improve the results. A linear regression model fit to the three different levels of theory is shown in Fig.\ref{fig-reg}. From the figure it can be seen that when a simple linear fit $y = kx + b$ to the data is used, the resulting values for the model-wise errors based on the coefficient of proportionality $k$ are as follows: GGA - 35\%, HSE - 17\%, G$_0$W$_0$ - 7\%.
    
    
    \begin{figure}[ht!]
        \includegraphics[width = 0.48\textwidth]{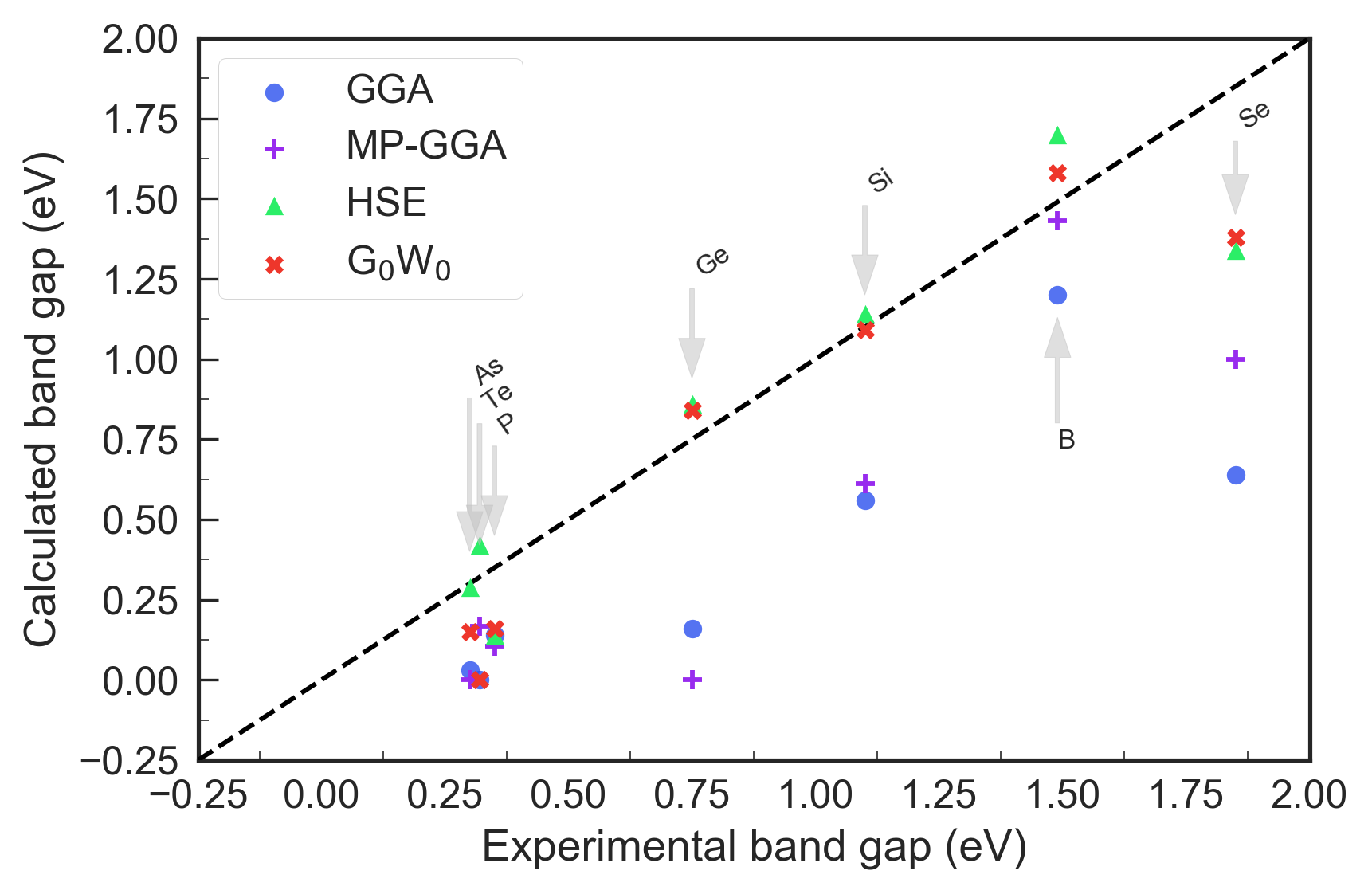}
        \caption{
            Comparative plot of the calculated and experimentally available values for all the electronic band gaps in the elemental (EL) category.
        }
        \label{fig-el}
    \end{figure}
    
     \begin{figure}[ht!]
        \includegraphics[width = 0.48\textwidth]{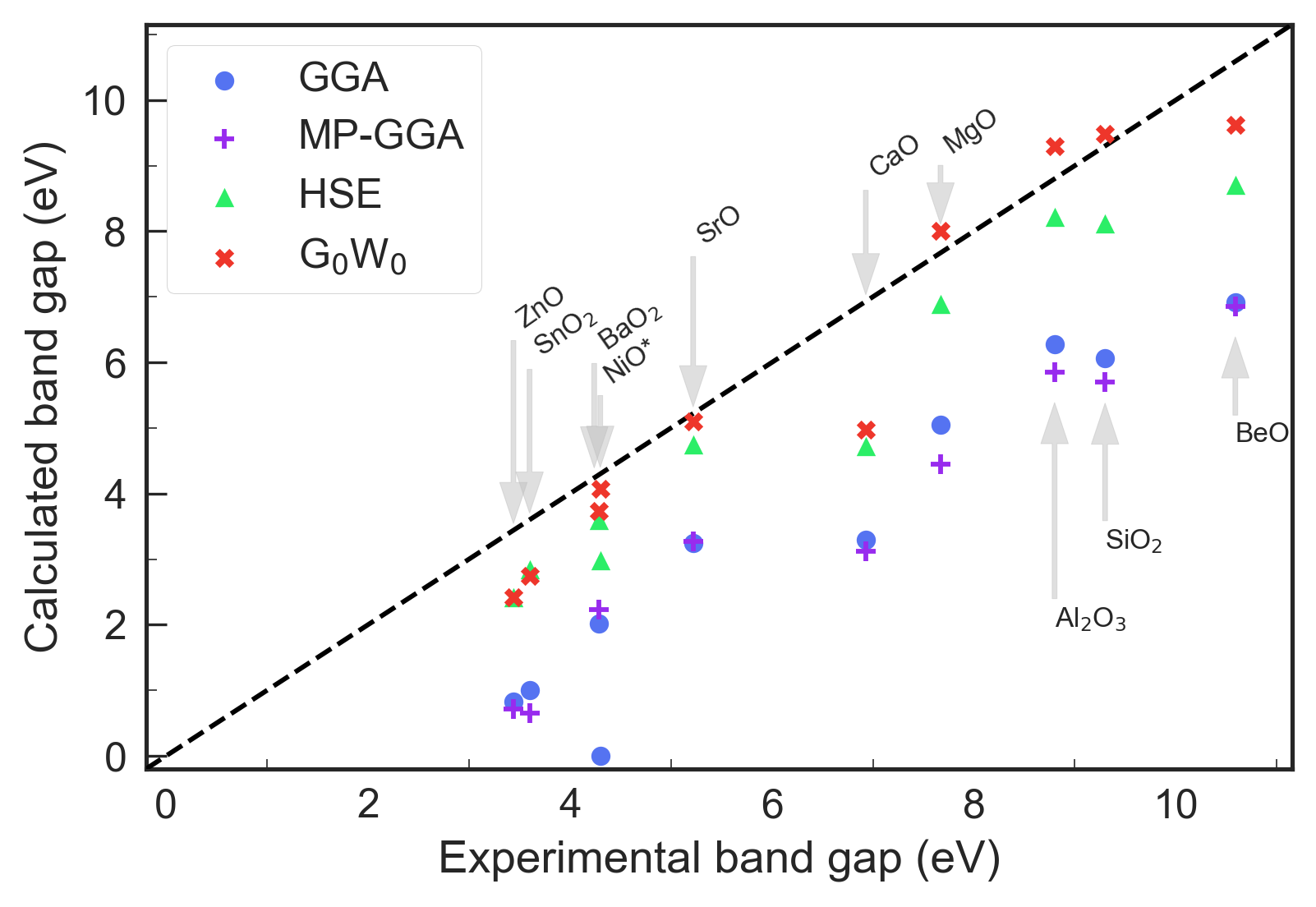}
        \caption{
            Same as Fig. \ref{fig-el} for the binary oxides (BO).
        }
        \label{fig-bo}
    \end{figure}   
    
     \begin{figure}[ht!]
        \includegraphics[width = 0.48\textwidth]{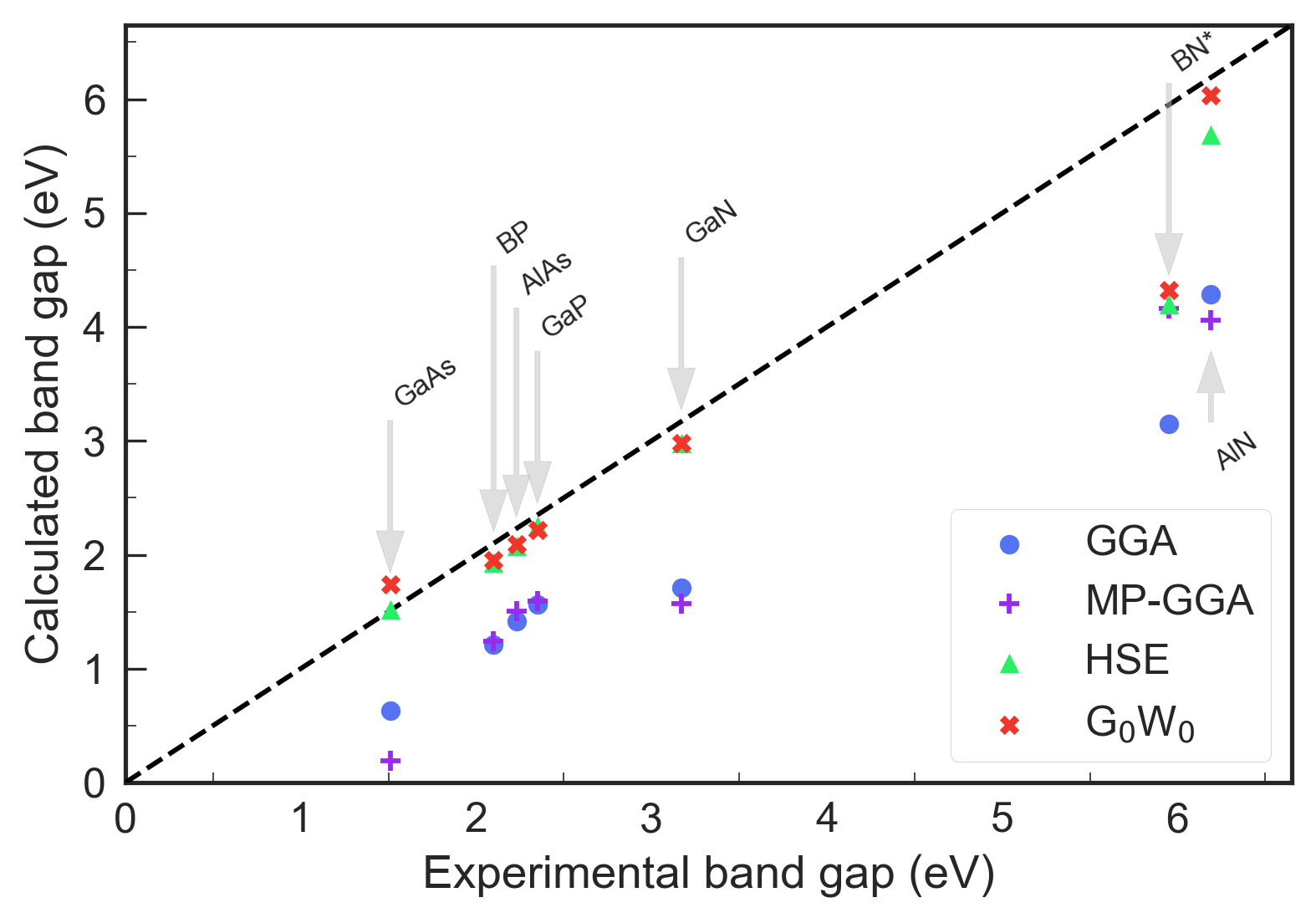}
        \caption{
            Same as Fig. \ref{fig-el} for the III-V compounds (35).
        }
        \label{fig-35}
    \end{figure}   
    
     \begin{figure}[ht!]
        \includegraphics[width = 0.48\textwidth]{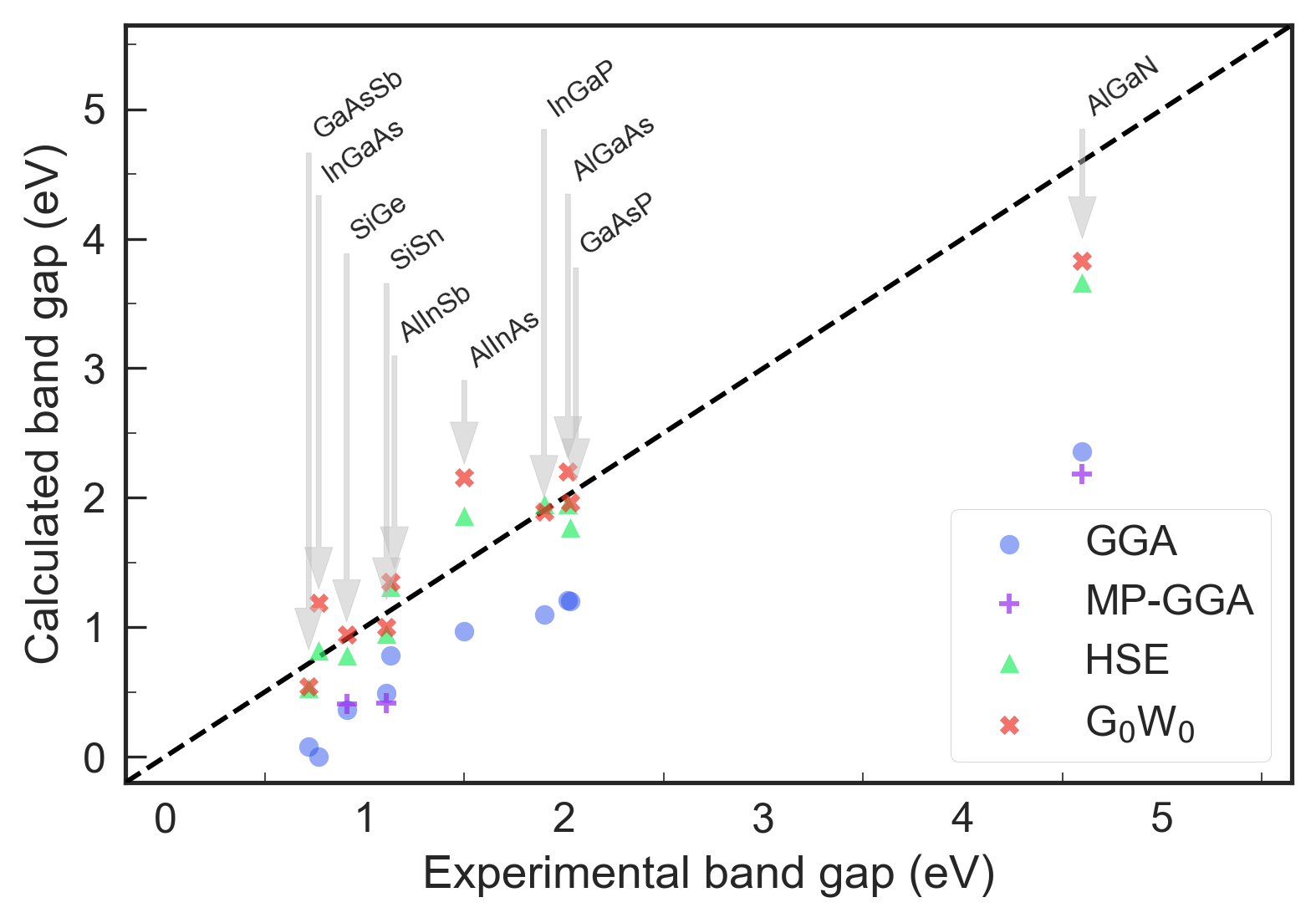}
        \caption{
            Same as Fig. \ref{fig-el} for the semiconductor alloys (AL).
        }
        \label{fig-al}
    \end{figure}   
    
     \begin{figure}[ht!]
        \includegraphics[width = 0.48\textwidth]{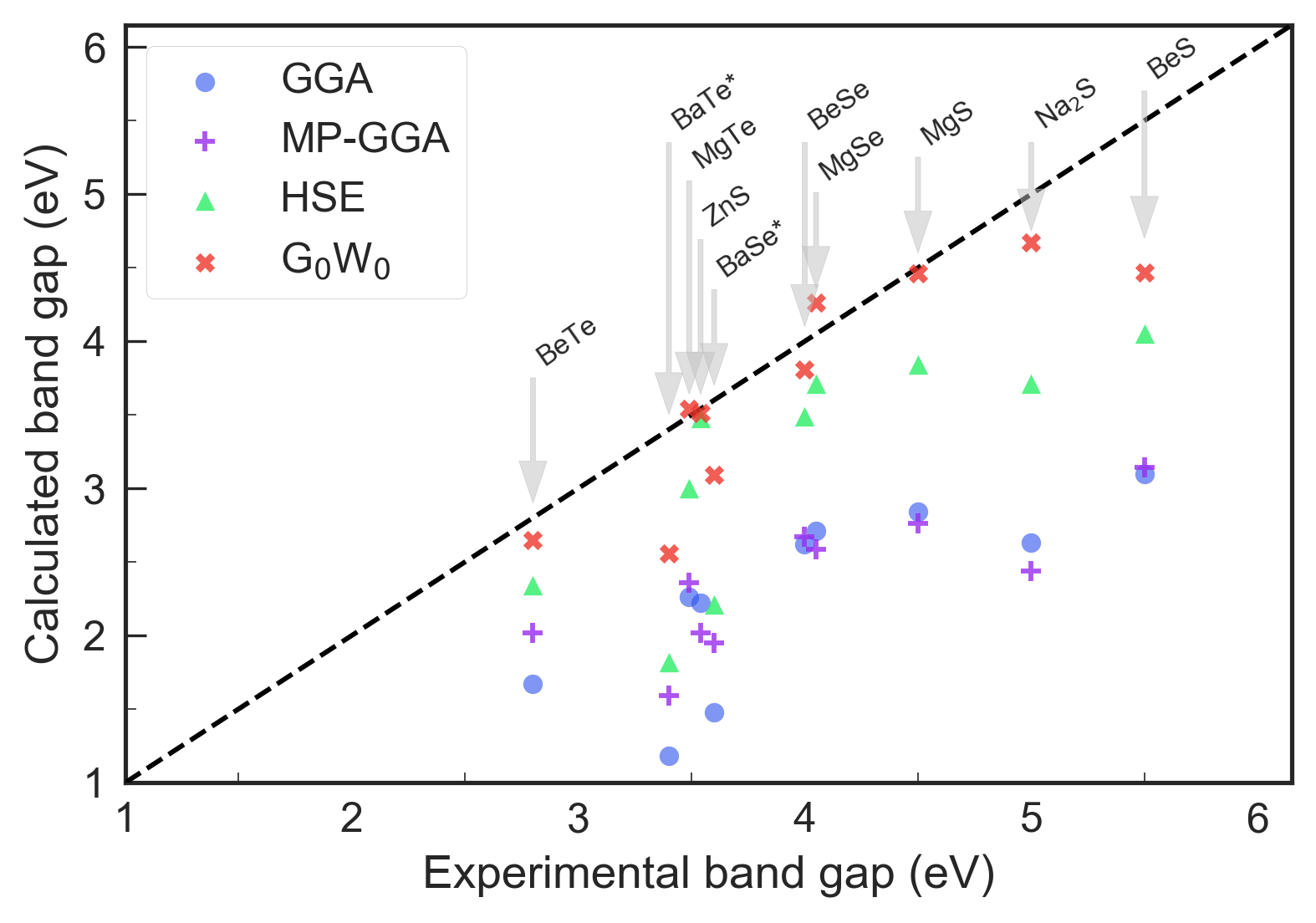}
        \caption{
            Same as Fig. \ref{fig-el} for the II-VI compounds (26).
        }
        \label{fig-26}
    \end{figure}   
    
     \begin{figure}[ht!]
        \includegraphics[width = 0.48\textwidth]{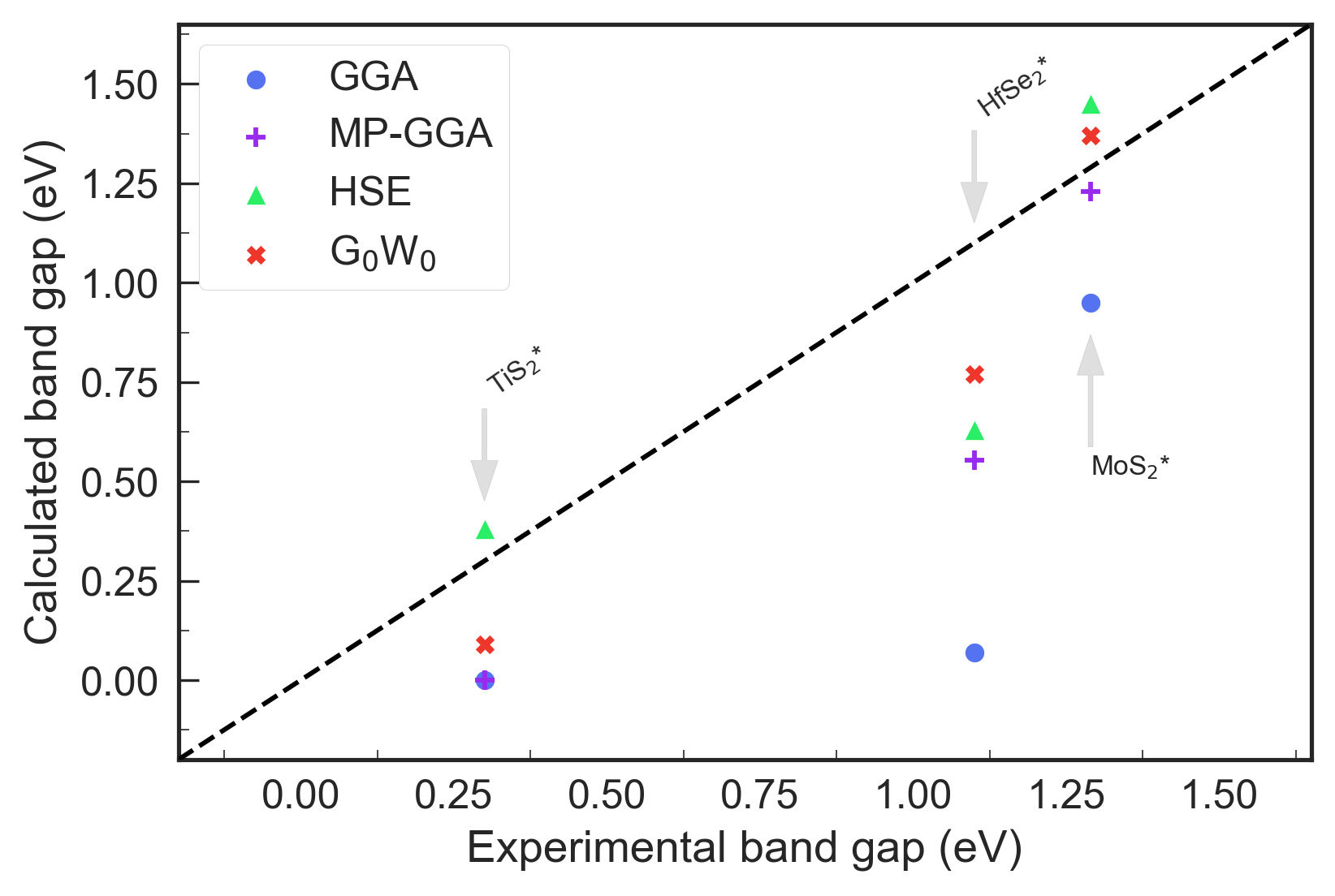}
        \caption{
            Same as Fig. \ref{fig-el} for the dichalcogeniges (DC).
        }
        \label{fig-dc}
    \end{figure}   
    
     \begin{figure}[ht!]
        \includegraphics[width = 0.48\textwidth]{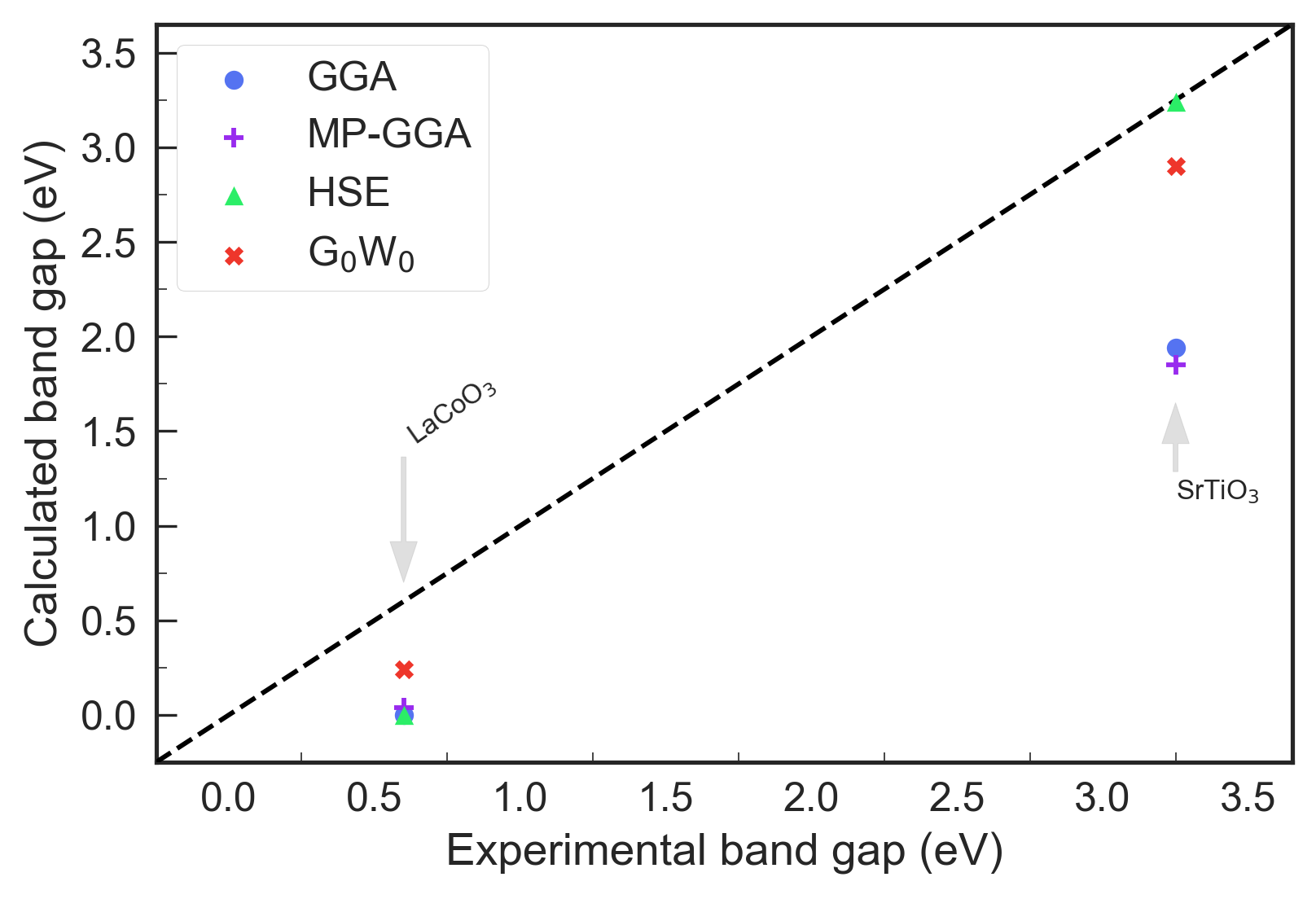}
        \caption{
            Same as Fig. \ref{fig-el} for the ternary oxides (TO)..
        }
        \label{fig-to}
    \end{figure}  
    
     \begin{figure}[ht!]
        \includegraphics[width = 0.48\textwidth]{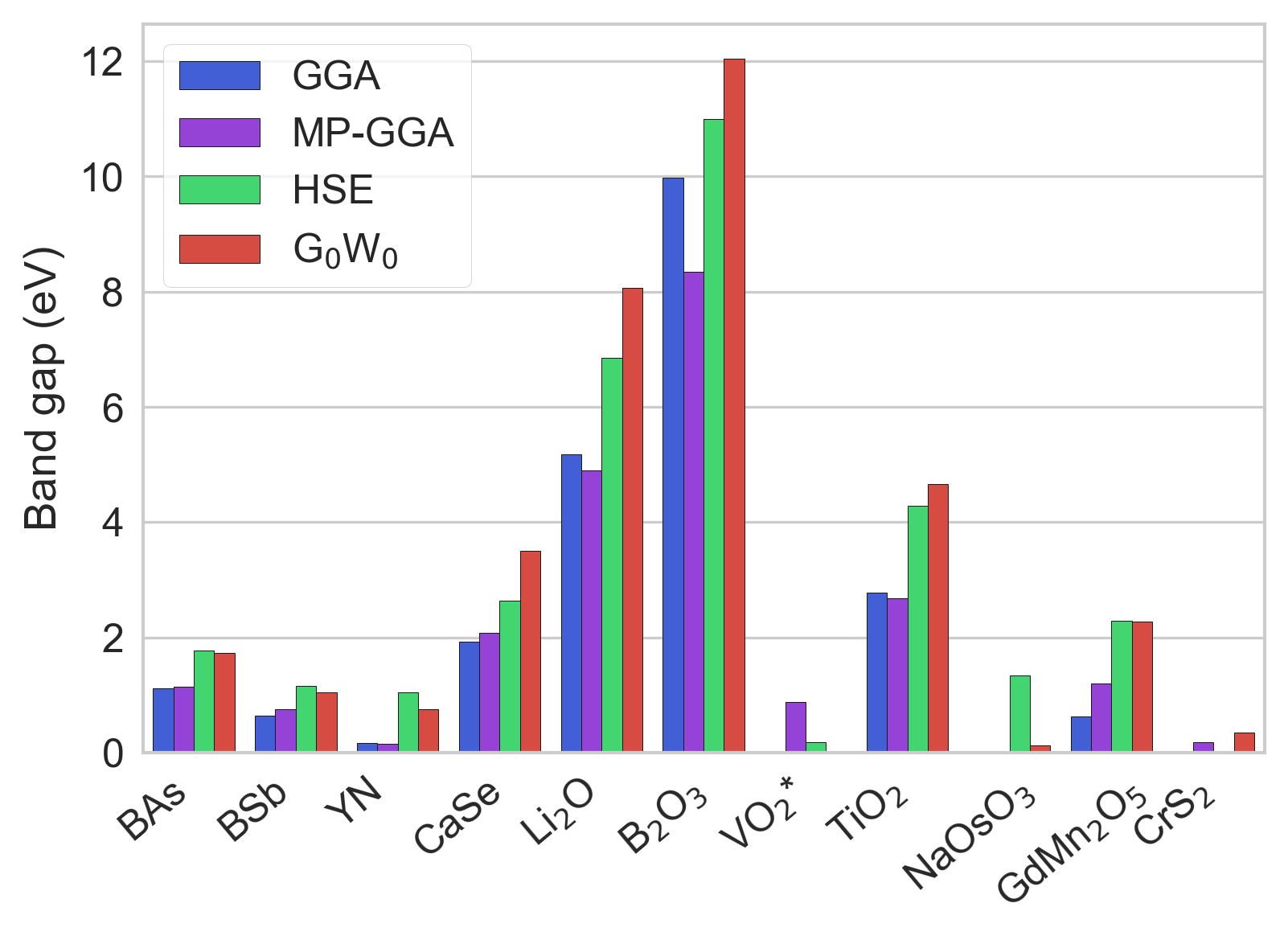}
        \caption{
            Calculated band gap values for different levels of theory for the materials without experimental data. The legend is same as in Fig.\ref{fig-all} (color-wise). For materials with an asterisk sign, the MP band gaps are calculated using DFT$+$U.
        }
        \label{fig-other}
    \end{figure}   

    \subsection{Elements}
    \label{subsec:el}
    
        Fig. \ref{fig-el} shows band gaps of elemental materials compared with experimental values. The gap for Te was measured to be 0.3 eV in \cite{Te_exp_gap}, however our GGA band structure does not have a gap. The lattice parameter $a$ of our relaxed structure is underestimated by 2.23\%. MP-GGA value is 0.186 eV, however in their case the lattice parameter is overestimated by 1.37\% \cite{Se_Te_abc}. The reduction in lattice parameter in our case can be attributed to the vdW correction. In another study the meta-GGA-SCAN functional is seen to predict the lattice parameters well \cite{te_vdw}, however the calculated HSE band gap is larger than the experimental value by 40\%. An exact-exchange mixing parameter of $0.125$ was needed to reproduce the correct experimental gap for Te in \cite{te_vdw}. Our G$_0$W$_0$ calculation predicts Te to be metallic.
        
        Ge has the experimental band gap of 0.75 eV\cite{madelung2012semiconductors}. We find the lattice parameter of the relaxed configuration within 0.5\% of the experimental value\cite{madelung2012semiconductors}. Both HSE and G$_0$W$_0$ overestimate the band gap by 16.21\% and 13.51\% correspondingly. While MP predicts Ge to be metallic, our calculation predicts a band gap of 0.16 eV within the GGA. We attribute this to the inclusion of semi-core states in our case. 
        
        For Si, the GGA band gap is underestimated by 48.7\%. Both HSE and G$_0$W$_0$ predict band gap of Si very well: within 3.6\% and 0.9\% correspondingly.
        
        For boron, our GGA band gap is 19.5\% less compared to the experimental value\cite{madelung2012semiconductors}. The MP GGA band gap for boron is underestimated by 4\%. We attribute this difference to the inclusion of vdW correction in our case. Our lattice parameter $a$ is about 4\% smaller compared to that of MP which results in smaller band gap. Similar trend is observed for Se where our lattice parameter is underestimated by 7\%.
        
        For other materials in this category - P, As and Se we find G$_0$W$_0$ to underestimate the gaps by 54\%, 50\% and 25.5\%, respectively. Incorporating partially self-consistent GW$_0$ corrects the errors to 11.1\%, 21\% and 10.23\%, respectively. Due to the small value of the gaps for P and As, the Gaussian smearing used during the calculation (50meV) can contribute to the error significantly.
        
        \begin{table}[h!]
    	\centering
    	\begin{tabular}{l |  c |  c | c | c}
    		\hline
    		\hline
    		\multirow{2}{*}{Material}& G$_0$W$_0$ & \multicolumn{3}{c}{sc-GW}  \\ \cline{2-5}
    		& Gap (eV) & Gap (eV) & Iterations & Energy cut-off (eV) \\
    	    \hline
    	    P   & 0.16  &   0.39  & 4 &   330 \\
    	    As  & 0.15  &   0.24  & 4 &   250 \\
    	    Se  & 1.38  &   2.04  & 3 &   300 \\
    	    BN  & 4.32  &   5.02  & 4 &   520 \\
    	    NiO & 4.07  &   5.30  & 4 &   415 \\
    		\hline
    		\hline
    	\end{tabular}
    	\caption{
    	    Band gaps calculated with partially self-consistent GW calculations. For all the calculations the self-energy of the non-diagonal components were included. The iteration of the quasi-particle (QP) energy shifts and the energy cut-off used are also tabulated.
    	    }
    	\label{table:sc-gw}
    \end{table}

    \subsection{III-V semiconductors}
    \label{subsec:35}
    
        Fig. \ref{fig-35} has the band gaps for category 35. For BP, AlAs, GaP, GaN and AlN, HSE and G$_0$W$_0$ predicts band gap within 8.1\% and 7\% accuracy, respectively. For GaAs, HSE predicts the band gap within 0.4\%. The G$_0$W$_0$ overestimates the band gap by 15.4\%, which might be due to the fact that the band gap estimate is performed on a grid as explained in sections \ref{sec:methodology} and \ref{sec:discussion}. Among the materials considered, BN has smaller GGA band gap compared to MP. Inclusion of vdW in our case reduces the lattice parameter by 2.4\% which reduces the GGA band gap.
        
        Our GGA band gap of BN is 26.3\% smaller compared to MP value. Also the MP GGA band gap predicts the band gap within 1.7\% of HSE band gap of BN. This is due to the absence of vdW interaction within the BN layers in MP calculations. While the lattice parameter $a$ of MP GGA matches within 0.8\% \cite{solozhenko1995isothermal}, lattice parameter $c$ is overestimated by 23\% which in turn causes the larger gap value that is close to experiment. We also find that within G$_0$W$_0$, the band gap has an error of 27.5\%. We employed self-consistent GW calculation to improve the band gap, and find that 4 iterations reduce the band gap error to 11.8\%.
    
    
    \subsection{II-VI}
    \label{subsec:26}
    
        Fig. \ref{fig-26} shows the band gaps for category 26. The bonding nature of the materials in this category is ionic and covalent. As a result, the inclusion of vdW interaction does not affect the lattice parameters much, except for BaSe and BaTe. Due to the inclusion of vdW interaction, our lattice parameters are smaller compared to MP results and hence the gaps are smaller as well. Our HSE band gap of MgS is underestimated by 19.67\% compared to \cite{heyd2005energy}. We attribute this difference to the smaller lattice constant due to vdW correction in our case.

    
    \subsection{Dichalcogenides}
    \label{subsec:dc}
    
        Fig. \ref{fig-dc} has the results for category DC. All the materials within this category are layered two-dimensional structures. For these materials, the inclusion of vdW interaction is critical to get the lattice parameter $c$ correctly. For HfSe$_2$ and MoS$_2$, our calculated $c$ value is within 1.65\%\cite{tis2_hfse2_abc} and 0.8\%\cite{mos2_mose2_abc} of the experimental value. The MP relaxed structures overestimate the values by 21\% and 12.7\%, respectively. Within GGA, TiS$_2$ is predicted not to have a band gap, while the HSE and G$_0$W$_0$ calculations open it. HSE overestimates the gap by 27.7\%.
      
    
    \subsection{Binary Oxides}
    \label{subsec:bo}
    
        Fig. \ref{fig-dc} has results for binary oxides. Within this category MP GGA band gaps match closely with our values, except in the case of NiO. While the MP has a gap of 2.498 eV, our calculation predicts material to be metallic. This discrepancy is due to the use Hubbard correction by MP, since a $U$ value of 6.2 eV was employed there. Our HSE and G$_0$W$_0$ calculations predict the band gaps within 30\% and 5.4\% of the experimental value, respectively. Further improvement can be obtained using the self-consistent GW approach as demonstrated in Table. \ref{table:sc-gw}.
        
    
    \subsection{Ternary oxides}
    \label{subsec:to}
    
        The calculated band gaps for materials in category TO is compared with experimental band gap in Fig. \ref{fig-to}. Within this category the experimental band gaps are found only for LaCoO$_3$ and SrTiO$_3$, as shown in Table.\ref{table:materials-data}. For LaCoO$_3$, both GGA and HSE predict the material to be metallic. In G$_0$W$_0$ we get a gap of 0.24 eV for LaCoO3, however, since we do not calculate the full band structure as explained in section \ref{sec:methodology}, a further study might be required to confirm the result. For SrTiO3 HSE predicts the gap very well with a 0.4\% error while G$_0$W$_0$ has an error of 10.7\%.
    
    
    \subsection{Semiconductor alloys}
    \label{subsec:al}
    
        Fig. \ref{fig-al} has results for AL category. InGaAs has no gap within GGA, while the HSE predicts the gap within 6.1\% of the experimental band gap. The G$_0$W$_0$ calculation overestimates the band gap by 54.7\%, we attribute this to the lack of the full band structure calculation and indirect nature of the gap. In the case of AlInAs G$_0$W$_0$ overestimates the band gap by 43.9\% for the same reason, while HSE has a 23.7\% larger value than experimental.

    \subsection{Other materials}
    \label{subsec:other}
    
        Materials for which experimental data is not found are plotted in Fig. \ref{fig-other}. Our GGA band gap matches well with MP band gap. HSE and G$_0$W$_0$ improves the result. The band gap difference of B$_2$O$_3$ within GGA between our calculation and MP can be attributed to the vdW interaction included in our calculations. Due to the localization (DFT+U) effect included in MP, VO$_2$ is semiconducting while our calculation shows VO$_2$ as metallic. 
        
        Within GGA, NaOsO$_3$ is predicted to be metallic both by us and MP. HSE opens a band gap of 1.34 eV. G$_0$W$_0$ predicts a band gap of 0.12 eV. We believe that self-consistent GW may increase the band gap closer to the HSE value. All levels of theory predict GdMn$_2$O$_5$ to be semiconducting.
     
    \section{Discussion}
\label{sec:discussion}
    
    We meant this study as a practical "end-to-end" benchmark of the ability of the current generation of pseudopotential density functional theory (DFT) to predict the electronic properties of materials. We also focused our attention on how it can be applied in an accessible way with minimal additional computational setup (i.e. no specialized hardware or compilation routines). As it was recently demonstrated in a comprehensive overview of the DFT simulation engines in \cite{DFTReproducibility2016lejaeghere}, most of them are inter-changeable with respect to the results delivered within the same model approximation. We selected VASP\cite{kresse1996software} as one of the most used tools in the space. Unlike the previous benchmarks, however, that considered computing aspects exclusively \cite{exabyte2018hp3c, 2010-jackson-cholia-lbl-cloud-con}, we went further and calculated the properties for a diverse set of material compounds.
    
    \subsection{Accessibility}
    \label{subsec:accessibility}
    
    The problem of accurate calculations of the electronic band gaps has been around for nearly as long as the computing itself. Although much effort was put into producing a way to obtain reliable high fidelity results, an accessible and repeatable option is still largely missing\cite{pizzi2016aiida, nomad}. Our work is an attempt to demonstrate how a standardized approach to the creation and execution of the first-principles modeling workflows developed by Exabyte Inc. can resolve the above. We present an accessible, repeatable and cost-effective way to deploy first-principles modeling workflows. Furthermore, we make the data freely available on the web, and provide an intuitive way to reproduce our work. Recently there has been much attention to high-throughput first-principles calculations of materials properties, which lead to the proliferation of the online databases and the development of the associated software tools\cite{jain2013materialsproject, curtarolo2012aflowlib, pizzi2016aiida, nomad, saal2013openQMD}. Our approach has similar capabilities, as demonstrated by this work, and is accessible to a larger community, in particular, to those without first-hand knowledge of DFT. 
    
    Another important recent advancement came from the data-centric approaches where large repositories of data can be used together with machine learning techniques in order to build predictive models\cite{isayev2017ml-descriptors, ward2016ml-framework}. Such models are able to deliver the predictions much faster, as they do not require the solution of physical equations. Our data-centric platform can power the construction of such models, with the potential to achieve improved accuracy of predictions by basing them on more accurate DFT results. Others approached the problem from a more traditional perspective attempting to construct DFT functionals capable of delivering high accuracy for the calculations of the electronic band structures\cite{goddard2016JPCL}. We will refrain here from discussing transferability from one material class to another for any such functional. Instead, we would like to point out that in practical applications the bottleneck that prevents the adoption of any of the aforementioned techniques is the human time required to get a prediction with a certain expected level of precision. Surely, reliably delivering high-fidelity results quickly is the best, however, an approach that takes a long time to compute but little to set up and oversee can work just as well.

    
    \begin{figure}[ht!]
        \includegraphics[width = 0.48\textwidth]{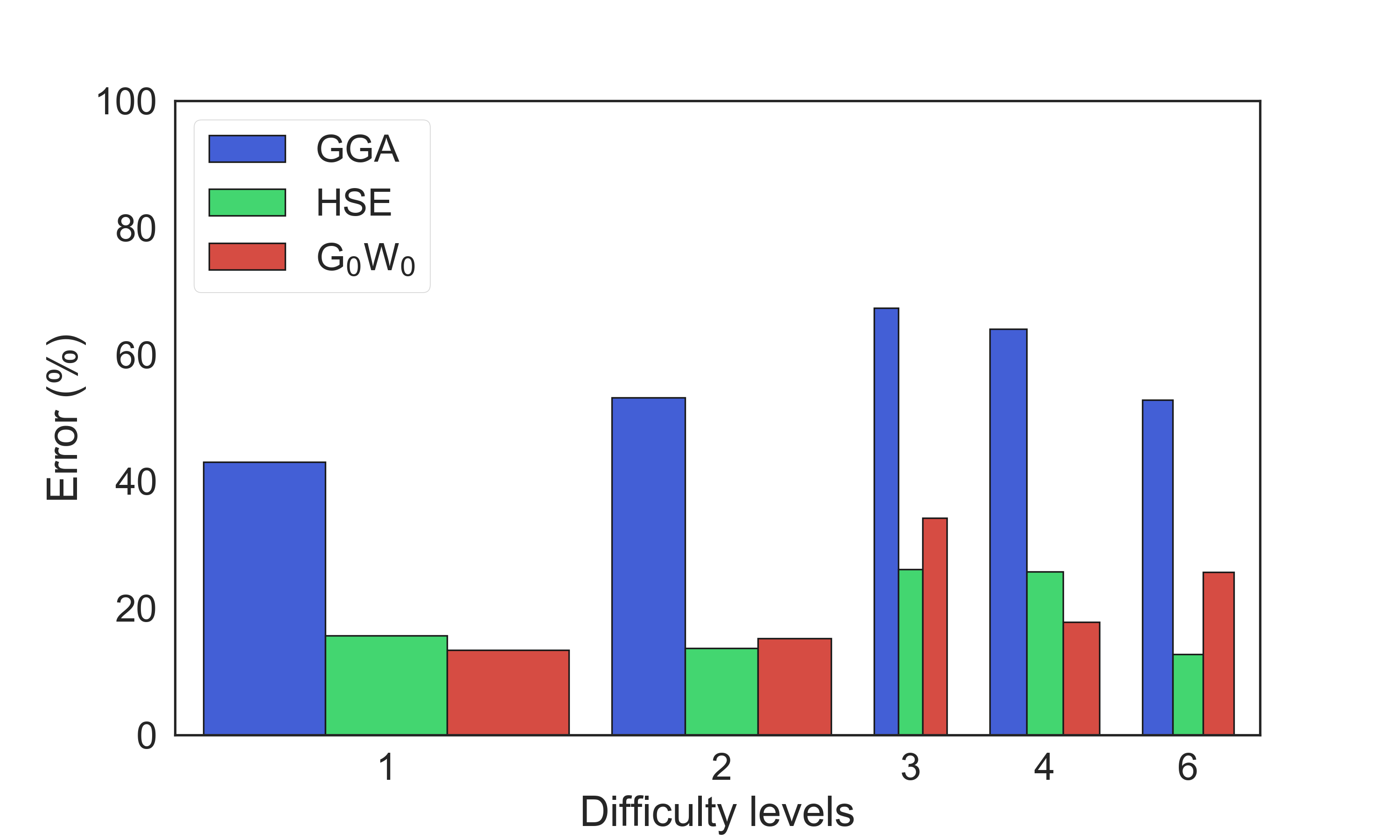}
        \caption{
            Difficulty-wise average errors. The width of the bars are proportional to the number of materials in category. Difficulty 5 and 7 is excluded due to low count ($<$3).
        }
        \label{figure:difficulty-wise-error}
    \end{figure}    
    
    \begin{figure}[ht!]
        \includegraphics[width = 0.48\textwidth]{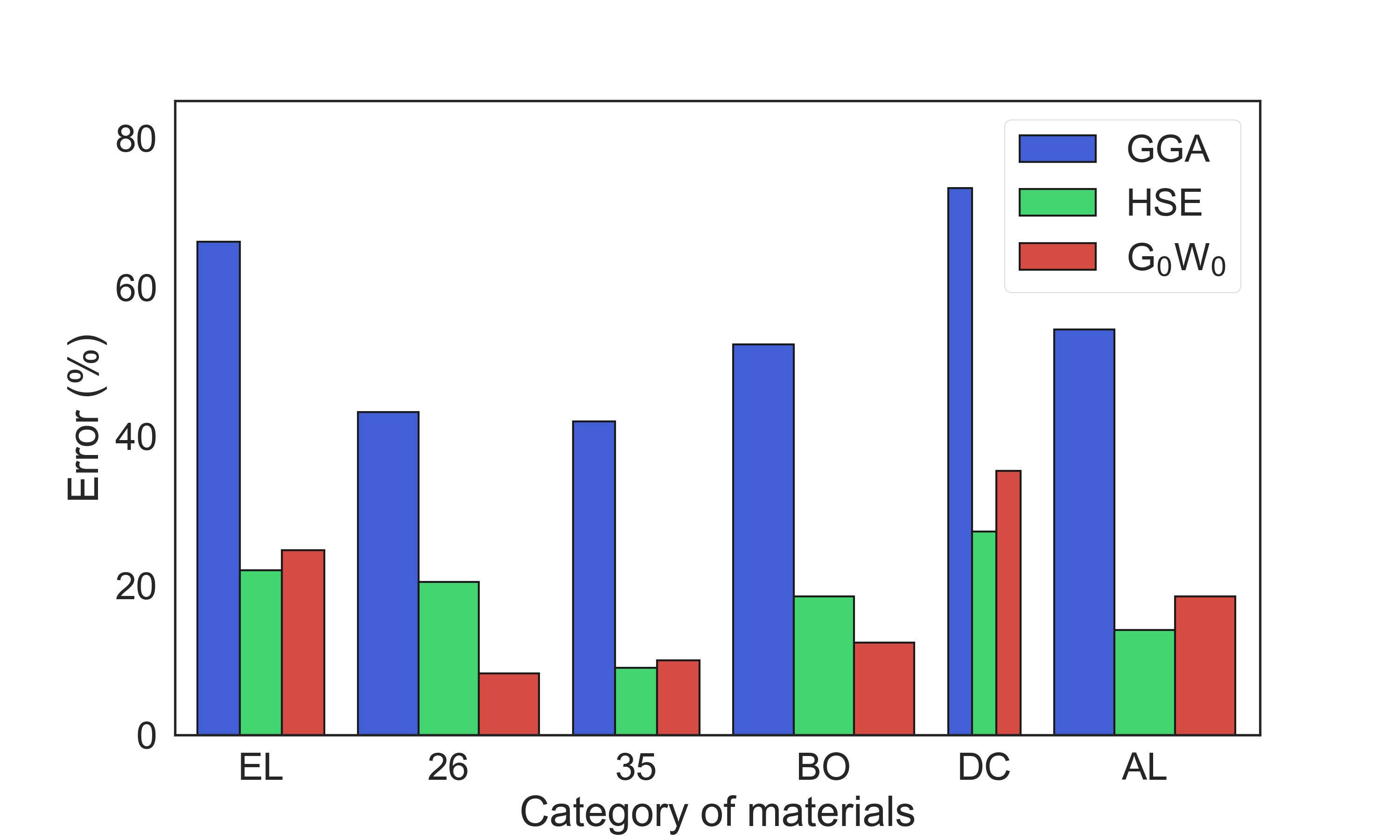}
        \caption{
            The average errors per each stoichiometric category. The width of the bars are proportional to the number of materials in category. Ternary oxides excluded due to low count ($<$3).
        }
        \label{figure:category-wise-error}
    \end{figure}    
    
    \subsection{Fidelity and error analysis}
    \label{subsec:fidelity}
    
    When comparing with the available experimental data we point out some important conditions used within our approach that are known to affect the calculation results. Firstly, we conduct the structural relaxation within the GGA and subsequently use the resulting structure for HSE and GW calculations. GGA is largely believed to work well for the ground-state properties of materials, and thus little change is expected when the structures are relaxed with HSE, for example\cite{heyd2005energy}. Secondly, in order to improve the treatment of van-der-Waals (vdW) interaction within our models we introduce a correction as implemented in VASP\cite{kresse2010vdWCorrection, grimme2006vdWcorrection}. This improves the results for layered materials especially, where the layered materials are considered. This is due to the fact that layered materials are self--passivated and the inter--layer interaction is dominated by the vdW interaction. Lastly, due to the computational complexity of the current implementation for $G_0W_0$ within VASP, we calculate the band gaps using the electronic eigenvalues on a grid of points inside the Brillouin zone, instead of using the standard path\cite{curtarolo2012aflowlib} as we did for GGA and HSE. The latter fact affects the fidelity of results for indirect gap semiconductors especially where the band extrema are located far from high symmetry points sampled by the grid. Due to constraints to the availability of memory, we reduced precision for few materials as indicated in Table \ref{table:materials-data}. The reduced number of k-points within the irreducible Brillouin zone may have contributed to the error as well.
    
    Figures \ref{figure:difficulty-wise-error} and \ref{figure:category-wise-error} have the data about the average errors per material category. Width of each column are proportional to the number or materials in each difficulty or category. We have omitted categories and difficulties that have experimental band gap available for less than 3 materials. We find that the GGA calculations have the largest error. HSE and G$_0$W$_0$ improve the band gap by similar margin. Category D3 has the largest error, although, notably, the sampling per this category is substantially less than for D1 and D2, for example. The dichalcogenides (DC) produced the largest error by material type, although it also has to be noted that the sampling in this category is lowest. We attribute this to the applied vdW correction in the layered materials. 
    
    The lattice constants $c$ of layered materials are sensitive to the type of vdW correction applied. Case specific vdW correction to the material can improve error in lattice constants and reduce the error. Even though HSE and G$_0$W$_0$ calculations improve the band gap for this category, we suspect correct inter--layer spacing can improve the band gaps further. We also find that for category EL, the G$_0$W$_0$ band gaps give larger error compared to HSE calculations. The self--consistent GW calculations improve the band gap for EL materials substantially (discussed in Sec. \ref{subsec:improvements} in detail). Category 26 and 35 compounds had most accurate predictions within HSE and G$_0$W$_0$. Category TO materials are excluded from Fig. \ref{figure:category-wise-error} due to low sample count.

    \subsection{Further improvements to accuracy}
    \label{subsec:improvements}

    There exist multiple ways to further improve the accuracy of the results obtained in this work. For the GW calculations, the self-consistent GW approach can improve the results. Table \ref{table:sc-gw} summarizes the band gaps calculated with self-consistent GW. We find the approach where the non-diagonal components of the self-energy are included to provide the best accuracy\cite{shishkin2007accurate} within a manageable time frame. We believe that executing the self-consistent GW calculations in a high-throughput manner is already possible for the materials studied in this work. In practice, it would presently require using compute nodes with extra large memory. Although such nodes are readily available from public cloud providers, the current computational implementation in VASP is not optimized for this regime and thus we would expect the resulting calculations to be more expensive and less reliable.
    
    Another way to improve the accuracy of the results would be to use a dynamically adjustable value for the HSE mixing parameter similar to how it is done in \cite{scHSE2014galli}. This approach would be more computationally intensive as it requires the convergence of the static dielectric constant with respect to the mixing parameter to be achieved during the calculation. Alternatively, a "hybrid" scheme could be considered where initially the improved value for the mixing parameter is calculated "on-the-fly" based on a statistical model, and then a "single-shot" HSE calculation is executed. 
    
    Lastly, the precision of the resulting calculations can be improved by addressing the concerns stated in the previous sub-section related to the sampling in the Brillouin zone. We used an approach based on the KPPRA, whereas introducing the logic for explicit convergence into the resulting workflows could be beneficial. For $G_0W_0$ calculations in particular a more thorough approach to treating the convergence of the end results with respect to the size of the pseudopotential basis set might be beneficial. We assumed the default recommended value of the cutoff and used all available planewave states (bands) for the summation whenever possible. When memory concerns arised we reduced the fidelity in a controlled way as explained in the section \ref{sec:methodology}.

    \begin{table}[bh!]
        \label{table:exact-accuracy}
    	\centering
    	\begin{tabular}{l |  c   l  c | l}
    		\hline
    		\hline
    		Model          & Avg. err, ($\%$) & Avg. runtime & Cost ($\$$) & Note \\
    	    \hline
    	    \hline
    	    *Exact*        & 0        & 30 days   & 5,000   & extrapolated   \\
      	    HSE            & 20       & 43 hrs    & 250     & factual   \\
      	    GGA            & 54       & 18 min    & 5       & factual   \\
      	    *Zero*         & 100      & 0.1 sec   & 0       & extrapolated   \\
    		\hline
    		\hline
    	\end{tabular}
    	\caption{
    	    Average errors and the associated average calculation time for the HSE and GGA cases studied in this work. *Exact* and *Zero* values are constructed through a simple logarithmic fit of the HSE/GGA data for the (hypothetic) models that would produce exact and zero-fidelity results correspondingly.
    	    }
    \end{table}    
    
    \subsection{Computational time and cost}
    
    In order to provide insights about the feasibility of further improved approaches and the ability to obtain the ultimate exact accuracy, we construct a simple logarithmic regression using the data obtained for the GGA and the HSE results. We exclude G0W0 because the results for it did not include the full band structure calculations, thus its set of computed properties is different. We assume that the average simulation lifetime increases exponentially as the average error is dropping. This is, of course, an overly simplified treatment and is only meant to produce qualitative results. We base our logic on the fact that the calculation of exchange interaction as employed within the HSE formalism includes the integral sums over the electronic states, and thus increasing the number of individual computations to the square of the number of wavefunctions. As can be seen from Table \ref{table:exact-accuracy}, within this logic one would need to run a simulation for about 30 days on average in order to produce an exact result. On the opposite side, a simulation with a runtime of less than 0.1 sec would fail to produce a meaningful result.
    
    Our motivation for the above is to provide a metric of the extent to which the physics-based first-principles modeling can augment the trial-and-error experimental approach when compared with respect to the capital and time investments required. We suggest that for the equivalent of one month of calculation time (human time) on a commodity compute server readily available from a cloud provider it is possible to obtain results that are accurate well within 20$\%$ and potentially within 1-5$\%$ range for the properties that we study in the current work. There are, admittedly, many factors that can adversely affect the result and many ways to optimize and improve upon the setup we used. Nevertheless, it is clear that the high fidelity results are not prohibitively expensive already today, and with the advancements in computing technology will become more and more prevalent. Furthermore, when compared with the capital spends required to manufacture and prototype the materials in experiment, even the "Exact" scenario we considered above appears attractive. Moreover, when data-centric community efforts without repetition are taken into account, the costs are further amortized. We believe that the correct approach to materials development from nanoscale is to use both high-fidelity simulations and experiments in a collaborative "funnel"-like scenario similar to how the computer-aided design and engineering is applied at present.

    \subsection{Future outlook}

    We believe that the landscape of computational materials design is rapidly evolving toward a data-driven science where the modeling results are aggregated and classified by their precision/accuracy. We believe, however, that the major improvements in way computational materials science is used would be significantly delayed if possible at all when only performed by means of the selected few. As the volume and variety of data available to community is growing at an accelerated speed, the veracity of this data also becomes increasingly important. The approach described in this work can solve both aforementioned concerns. The improved creative ideation with contributions from people with multiple backgrounds is enabled by modeling workflows accessible in a standardized and repeatable way and the shift away from the "medieval artisan-like" model\cite{pizzi2016aiida} still prevalent nowadays. On the other hand, this work provides first proof that high precision is also achievable, perhaps only for electronic materials at this moment, using existing first-principles modeling techniques. We believe that a hybrid data-driven approach with roots in high-fidelity modeling is most powerful.

    \\ \\
    \section{Conclusions}
\label{sec:conclusions}
    
    We report on the application of a novel approach to materials modeling from nanoscale implemented within the Exabyte platform\cite{exabytePlatform} to a diverse representative set of \NMaterials semiconducting materials (ESC-\NMaterialsNoSpace). The approach makes high-fidelity techniques such as pseudopotential Density Functional Theory with Hybrid Screened Exchange (HSE) and G$_0$W$_0$ approximation available in an accessible, repeatable and data-centric manner. We introduce a categorization for the materials according to the level of approximation used and explain the implementation of the corresponding modeling workflows. We present the results for the electronic band gaps obtained within the Generalized Gradient Approximation (GGA), HSE and G$_0$W$_0$, analyze the level of fidelity for the prediction delivered by each of the models used, and discuss the corresponding computational costs. 
    
    We compare the results with experimental data and prior similar calculation attempts, when available. We find the average relative error to be within 20$\%$ for HSE and GW results and within 55$\%$ for GGA. We further find the average calculation time on a current up-to-date compute server centrally available from a public cloud provider to fit within 30 min and 48 hours respectively for GGA and HSE. For the first time ever we present not only the results and the associated data, but also an easy-to-access way to reproduce and extend the results by means of Exabyte platform.\cite{exabytePlatformBGPhaseIURL} Our work provides an accessible, repeatable, and extensible practical recipe for performing high-fidelity first-principles calculations in a high-throughput manner. 
    \section{Acknowledgement}
\label{sec:data}
    
    This study was conceived, executed and sponsored in full by Exabyte Inc. All computations for this work were performed using Microsoft Azure cloud computing platform. We are grateful for the support from the Microsoft BizSpark team in particular. We thank Steven G. Louie, Marvin L. Cohen, Roger K. Lake, and Georg Kresse for their advice and fruitful discussions.

    \bibliography{references}

\begin{thebibliography}{133}%
\makeatletter
\providecommand \@ifxundefined [1]{%
 \@ifx{#1\undefined}
}%
\providecommand \@ifnum [1]{%
 \ifnum #1\expandafter \@firstoftwo
 \else \expandafter \@secondoftwo
 \fi
}%
\providecommand \@ifx [1]{%
 \ifx #1\expandafter \@firstoftwo
 \else \expandafter \@secondoftwo
 \fi
}%
\providecommand \natexlab [1]{#1}%
\providecommand \enquote  [1]{``#1''}%
\providecommand \bibnamefont  [1]{#1}%
\providecommand \bibfnamefont [1]{#1}%
\providecommand \citenamefont [1]{#1}%
\providecommand \href@noop [0]{\@secondoftwo}%
\providecommand \href [0]{\begingroup \@sanitize@url \@href}%
\providecommand \@href[1]{\@@startlink{#1}\@@href}%
\providecommand \@@href[1]{\endgroup#1\@@endlink}%
\providecommand \@sanitize@url [0]{\catcode `\\12\catcode `\$12\catcode
  `\&12\catcode `\#12\catcode `\^12\catcode `\_12\catcode `\%12\relax}%
\providecommand \@@startlink[1]{}%
\providecommand \@@endlink[0]{}%
\providecommand \url  [0]{\begingroup\@sanitize@url \@url }%
\providecommand \@url [1]{\endgroup\@href {#1}{\urlprefix }}%
\providecommand \urlprefix  [0]{URL }%
\providecommand \Eprint [0]{\href }%
\providecommand \doibase [0]{http://dx.doi.org/}%
\providecommand \selectlanguage [0]{\@gobble}%
\providecommand \bibinfo  [0]{\@secondoftwo}%
\providecommand \bibfield  [0]{\@secondoftwo}%
\providecommand \translation [1]{[#1]}%
\providecommand \BibitemOpen [0]{}%
\providecommand \bibitemStop [0]{}%
\providecommand \bibitemNoStop [0]{.\EOS\space}%
\providecommand \EOS [0]{\spacefactor3000\relax}%
\providecommand \BibitemShut  [1]{\csname bibitem#1\endcsname}%
\let\auto@bib@innerbib\@empty
\bibitem [{\citenamefont {Jain}\ \emph {et~al.}(2013)\citenamefont {Jain},
  \citenamefont {Ong}, \citenamefont {Hautier}, \citenamefont {Chen},
  \citenamefont {Richards}, \citenamefont {Dacek}, \citenamefont {Cholia},
  \citenamefont {Gunter}, \citenamefont {Skinner}, \citenamefont {Ceder} \emph
  {et~al.}}]{jain2013materialsproject}%
  \BibitemOpen
  \bibfield  {author} {\bibinfo {author} {\bibfnamefont {A.}~\bibnamefont
  {Jain}}, \bibinfo {author} {\bibfnamefont {S.~P.}\ \bibnamefont {Ong}},
  \bibinfo {author} {\bibfnamefont {G.}~\bibnamefont {Hautier}}, \bibinfo
  {author} {\bibfnamefont {W.}~\bibnamefont {Chen}}, \bibinfo {author}
  {\bibfnamefont {W.~D.}\ \bibnamefont {Richards}}, \bibinfo {author}
  {\bibfnamefont {S.}~\bibnamefont {Dacek}}, \bibinfo {author} {\bibfnamefont
  {S.}~\bibnamefont {Cholia}}, \bibinfo {author} {\bibfnamefont
  {D.}~\bibnamefont {Gunter}}, \bibinfo {author} {\bibfnamefont
  {D.}~\bibnamefont {Skinner}}, \bibinfo {author} {\bibfnamefont
  {G.}~\bibnamefont {Ceder}},  \emph {et~al.},\ }\href@noop {} {\bibfield
  {journal} {\bibinfo  {journal} {Apl Materials}\ }\textbf {\bibinfo {volume}
  {1}},\ \bibinfo {pages} {011002} (\bibinfo {year} {2013})}\BibitemShut
  {NoStop}%
\bibitem [{\citenamefont {Curtarolo}\ \emph {et~al.}(2012)\citenamefont
  {Curtarolo}, \citenamefont {Setyawan}, \citenamefont {Wang}, \citenamefont
  {Xue}, \citenamefont {Yang}, \citenamefont {Taylor}, \citenamefont {Nelson},
  \citenamefont {Hart}, \citenamefont {Sanvito}, \citenamefont
  {Buongiorno-Nardelli} \emph {et~al.}}]{curtarolo2012aflowlib}%
  \BibitemOpen
  \bibfield  {author} {\bibinfo {author} {\bibfnamefont {S.}~\bibnamefont
  {Curtarolo}}, \bibinfo {author} {\bibfnamefont {W.}~\bibnamefont {Setyawan}},
  \bibinfo {author} {\bibfnamefont {S.}~\bibnamefont {Wang}}, \bibinfo {author}
  {\bibfnamefont {J.}~\bibnamefont {Xue}}, \bibinfo {author} {\bibfnamefont
  {K.}~\bibnamefont {Yang}}, \bibinfo {author} {\bibfnamefont {R.~H.}\
  \bibnamefont {Taylor}}, \bibinfo {author} {\bibfnamefont {L.~J.}\
  \bibnamefont {Nelson}}, \bibinfo {author} {\bibfnamefont {G.~L.}\
  \bibnamefont {Hart}}, \bibinfo {author} {\bibfnamefont {S.}~\bibnamefont
  {Sanvito}}, \bibinfo {author} {\bibfnamefont {M.}~\bibnamefont
  {Buongiorno-Nardelli}},  \emph {et~al.},\ }\href@noop {} {\bibfield
  {journal} {\bibinfo  {journal} {Computational Materials Science}\ }\textbf
  {\bibinfo {volume} {58}},\ \bibinfo {pages} {227} (\bibinfo {year}
  {2012})}\BibitemShut {NoStop}%
\bibitem [{\citenamefont {Saal}\ \emph {et~al.}(2013)\citenamefont {Saal},
  \citenamefont {Kirklin}, \citenamefont {Aykol}, \citenamefont {Meredig},\
  and\ \citenamefont {Wolverton}}]{saal2013openQMD}%
  \BibitemOpen
  \bibfield  {author} {\bibinfo {author} {\bibfnamefont {J.~E.}\ \bibnamefont
  {Saal}}, \bibinfo {author} {\bibfnamefont {S.}~\bibnamefont {Kirklin}},
  \bibinfo {author} {\bibfnamefont {M.}~\bibnamefont {Aykol}}, \bibinfo
  {author} {\bibfnamefont {B.}~\bibnamefont {Meredig}}, \ and\ \bibinfo
  {author} {\bibfnamefont {C.}~\bibnamefont {Wolverton}},\ }\href@noop {}
  {\bibfield  {journal} {\bibinfo  {journal} {Jom}\ }\textbf {\bibinfo {volume}
  {65}},\ \bibinfo {pages} {1501} (\bibinfo {year} {2013})}\BibitemShut
  {NoStop}%
\bibitem [{\citenamefont {Pizzi}\ \emph {et~al.}(2016)\citenamefont {Pizzi},
  \citenamefont {Cepellotti}, \citenamefont {Sabatini}, \citenamefont
  {Marzari},\ and\ \citenamefont {Kozinsky}}]{pizzi2016aiida}%
  \BibitemOpen
  \bibfield  {author} {\bibinfo {author} {\bibfnamefont {G.}~\bibnamefont
  {Pizzi}}, \bibinfo {author} {\bibfnamefont {A.}~\bibnamefont {Cepellotti}},
  \bibinfo {author} {\bibfnamefont {R.}~\bibnamefont {Sabatini}}, \bibinfo
  {author} {\bibfnamefont {N.}~\bibnamefont {Marzari}}, \ and\ \bibinfo
  {author} {\bibfnamefont {B.}~\bibnamefont {Kozinsky}},\ }\href@noop {}
  {\bibfield  {journal} {\bibinfo  {journal} {Computational Materials Science}\
  }\textbf {\bibinfo {volume} {111}},\ \bibinfo {pages} {218} (\bibinfo {year}
  {2016})}\BibitemShut {NoStop}%
\bibitem [{nom()}]{nomad}%
  \BibitemOpen
  \href {https://www.nomad-coe.eu/} {\emph {\bibinfo {title} {The NOMAD
  laboratory: A European Centre of Excellence}}}\BibitemShut {NoStop}%
\bibitem [{\citenamefont {Haastrup}\ \emph {et~al.}(2018)\citenamefont
  {Haastrup}, \citenamefont {Strange}, \citenamefont {Pandey}, \citenamefont
  {Deilmann}, \citenamefont {Schmidt}, \citenamefont {Hinsche}, \citenamefont
  {Gjerding}, \citenamefont {Torelli}, \citenamefont {Larsen}, \citenamefont
  {Riis-Jensen} \emph {et~al.}}]{haastrup2018computational}%
  \BibitemOpen
  \bibfield  {author} {\bibinfo {author} {\bibfnamefont {S.}~\bibnamefont
  {Haastrup}}, \bibinfo {author} {\bibfnamefont {M.}~\bibnamefont {Strange}},
  \bibinfo {author} {\bibfnamefont {M.}~\bibnamefont {Pandey}}, \bibinfo
  {author} {\bibfnamefont {T.}~\bibnamefont {Deilmann}}, \bibinfo {author}
  {\bibfnamefont {P.~S.}\ \bibnamefont {Schmidt}}, \bibinfo {author}
  {\bibfnamefont {N.~F.}\ \bibnamefont {Hinsche}}, \bibinfo {author}
  {\bibfnamefont {M.~N.}\ \bibnamefont {Gjerding}}, \bibinfo {author}
  {\bibfnamefont {D.}~\bibnamefont {Torelli}}, \bibinfo {author} {\bibfnamefont
  {P.~M.}\ \bibnamefont {Larsen}}, \bibinfo {author} {\bibfnamefont {A.~C.}\
  \bibnamefont {Riis-Jensen}},  \emph {et~al.},\ }\href@noop {} {\bibfield
  {journal} {\bibinfo  {journal} {arXiv preprint arXiv:1806.03173}\ } (\bibinfo
  {year} {2018})}\BibitemShut {NoStop}%
\bibitem [{\citenamefont {Rasmussen}\ and\ \citenamefont
  {Thygesen}(2015)}]{rasmussen2015computational}%
  \BibitemOpen
  \bibfield  {author} {\bibinfo {author} {\bibfnamefont {F.~A.}\ \bibnamefont
  {Rasmussen}}\ and\ \bibinfo {author} {\bibfnamefont {K.~S.}\ \bibnamefont
  {Thygesen}},\ }\href@noop {} {\bibfield  {journal} {\bibinfo  {journal} {The
  Journal of Physical Chemistry C}\ }\textbf {\bibinfo {volume} {119}},\
  \bibinfo {pages} {13169} (\bibinfo {year} {2015})}\BibitemShut {NoStop}%
\bibitem [{\citenamefont {Ong}\ \emph {et~al.}(2013)\citenamefont {Ong},
  \citenamefont {Richards}, \citenamefont {Jain}, \citenamefont {Hautier},
  \citenamefont {Kocher}, \citenamefont {Cholia}, \citenamefont {Gunter},
  \citenamefont {Chevrier}, \citenamefont {Persson},\ and\ \citenamefont
  {Ceder}}]{ong2013python}%
  \BibitemOpen
  \bibfield  {author} {\bibinfo {author} {\bibfnamefont {S.~P.}\ \bibnamefont
  {Ong}}, \bibinfo {author} {\bibfnamefont {W.~D.}\ \bibnamefont {Richards}},
  \bibinfo {author} {\bibfnamefont {A.}~\bibnamefont {Jain}}, \bibinfo {author}
  {\bibfnamefont {G.}~\bibnamefont {Hautier}}, \bibinfo {author} {\bibfnamefont
  {M.}~\bibnamefont {Kocher}}, \bibinfo {author} {\bibfnamefont
  {S.}~\bibnamefont {Cholia}}, \bibinfo {author} {\bibfnamefont
  {D.}~\bibnamefont {Gunter}}, \bibinfo {author} {\bibfnamefont {V.~L.}\
  \bibnamefont {Chevrier}}, \bibinfo {author} {\bibfnamefont {K.~A.}\
  \bibnamefont {Persson}}, \ and\ \bibinfo {author} {\bibfnamefont
  {G.}~\bibnamefont {Ceder}},\ }\href@noop {} {\bibfield  {journal} {\bibinfo
  {journal} {Computational Materials Science}\ }\textbf {\bibinfo {volume}
  {68}},\ \bibinfo {pages} {314} (\bibinfo {year} {2013})}\BibitemShut
  {NoStop}%
\bibitem [{\citenamefont {Larsen}\ \emph {et~al.}(2017)\citenamefont {Larsen},
  \citenamefont {Mortensen}, \citenamefont {Blomqvist}, \citenamefont
  {Castelli}, \citenamefont {Christensen}, \citenamefont {Du{\l}ak},
  \citenamefont {Friis}, \citenamefont {Groves}, \citenamefont {Hammer},
  \citenamefont {Hargus} \emph {et~al.}}]{larsen2017atomic}%
  \BibitemOpen
  \bibfield  {author} {\bibinfo {author} {\bibfnamefont {A.~H.}\ \bibnamefont
  {Larsen}}, \bibinfo {author} {\bibfnamefont {J.~J.}\ \bibnamefont
  {Mortensen}}, \bibinfo {author} {\bibfnamefont {J.}~\bibnamefont
  {Blomqvist}}, \bibinfo {author} {\bibfnamefont {I.~E.}\ \bibnamefont
  {Castelli}}, \bibinfo {author} {\bibfnamefont {R.}~\bibnamefont
  {Christensen}}, \bibinfo {author} {\bibfnamefont {M.}~\bibnamefont
  {Du{\l}ak}}, \bibinfo {author} {\bibfnamefont {J.}~\bibnamefont {Friis}},
  \bibinfo {author} {\bibfnamefont {M.~N.}\ \bibnamefont {Groves}}, \bibinfo
  {author} {\bibfnamefont {B.}~\bibnamefont {Hammer}}, \bibinfo {author}
  {\bibfnamefont {C.}~\bibnamefont {Hargus}},  \emph {et~al.},\ }\href@noop {}
  {\bibfield  {journal} {\bibinfo  {journal} {Journal of Physics: Condensed
  Matter}\ }\textbf {\bibinfo {volume} {29}},\ \bibinfo {pages} {273002}
  (\bibinfo {year} {2017})}\BibitemShut {NoStop}%
\bibitem [{cit()}]{citrine}%
  \BibitemOpen
  \href {https://citrination.com/} {\emph {\bibinfo {title} {Citrine
  Informatics: Materials Data Platform}}}\BibitemShut {NoStop}%
\bibitem [{til()}]{tilde}%
  \BibitemOpen
  \href {https://tilde.pro/} {\emph {\bibinfo {title} {Tilde Materials
  Informatics}}}\BibitemShut {NoStop}%
\bibitem [{\citenamefont {Villars}\ \emph {et~al.}(2004)\citenamefont
  {Villars}, \citenamefont {Berndt}, \citenamefont {Brandenburg}, \citenamefont
  {Cenzual}, \citenamefont {Daams}, \citenamefont {Hulliger}, \citenamefont
  {Massalski}, \citenamefont {Okamoto}, \citenamefont {Osaki}, \citenamefont
  {Prince} \emph {et~al.}}]{villars2004pauling}%
  \BibitemOpen
  \bibfield  {author} {\bibinfo {author} {\bibfnamefont {P.}~\bibnamefont
  {Villars}}, \bibinfo {author} {\bibfnamefont {M.}~\bibnamefont {Berndt}},
  \bibinfo {author} {\bibfnamefont {K.}~\bibnamefont {Brandenburg}}, \bibinfo
  {author} {\bibfnamefont {K.}~\bibnamefont {Cenzual}}, \bibinfo {author}
  {\bibfnamefont {J.}~\bibnamefont {Daams}}, \bibinfo {author} {\bibfnamefont
  {F.}~\bibnamefont {Hulliger}}, \bibinfo {author} {\bibfnamefont
  {T.}~\bibnamefont {Massalski}}, \bibinfo {author} {\bibfnamefont
  {H.}~\bibnamefont {Okamoto}}, \bibinfo {author} {\bibfnamefont
  {K.}~\bibnamefont {Osaki}}, \bibinfo {author} {\bibfnamefont
  {A.}~\bibnamefont {Prince}},  \emph {et~al.},\ }\href@noop {} {\bibfield
  {journal} {\bibinfo  {journal} {Journal of Alloys and Compounds}\ }\textbf
  {\bibinfo {volume} {367}},\ \bibinfo {pages} {293} (\bibinfo {year}
  {2004})}\BibitemShut {NoStop}%
\bibitem [{\citenamefont {Isayev}\ \emph {et~al.}(2017)\citenamefont {Isayev},
  \citenamefont {Oses}, \citenamefont {Toher}, \citenamefont {Gossett},
  \citenamefont {Curtarolo},\ and\ \citenamefont
  {Tropsha}}]{isayev2017ml-descriptors}%
  \BibitemOpen
  \bibfield  {author} {\bibinfo {author} {\bibfnamefont {O.}~\bibnamefont
  {Isayev}}, \bibinfo {author} {\bibfnamefont {C.}~\bibnamefont {Oses}},
  \bibinfo {author} {\bibfnamefont {C.}~\bibnamefont {Toher}}, \bibinfo
  {author} {\bibfnamefont {E.}~\bibnamefont {Gossett}}, \bibinfo {author}
  {\bibfnamefont {S.}~\bibnamefont {Curtarolo}}, \ and\ \bibinfo {author}
  {\bibfnamefont {A.}~\bibnamefont {Tropsha}},\ }\href
  {http://dx.doi.org/10.1038/ncomms15679} {\bibfield  {journal} {\bibinfo
  {journal} {Nature Communications}\ }\textbf {\bibinfo {volume} {8}},\
  \bibinfo {pages} {15679 EP } (\bibinfo {year} {2017})},\ \bibinfo {note}
  {article}\BibitemShut {NoStop}%
\bibitem [{\citenamefont {Ward}\ \emph {et~al.}(2016)\citenamefont {Ward},
  \citenamefont {Agrawal}, \citenamefont {Choudhary},\ and\ \citenamefont
  {Wolverton}}]{ward2016ml-framework}%
  \BibitemOpen
  \bibfield  {author} {\bibinfo {author} {\bibfnamefont {L.}~\bibnamefont
  {Ward}}, \bibinfo {author} {\bibfnamefont {A.}~\bibnamefont {Agrawal}},
  \bibinfo {author} {\bibfnamefont {A.}~\bibnamefont {Choudhary}}, \ and\
  \bibinfo {author} {\bibfnamefont {C.}~\bibnamefont {Wolverton}},\ }\href
  {http://dx.doi.org/10.1038/npjcompumats.2016.28} {\bibfield  {journal}
  {\bibinfo  {journal} {Npj Computational Materials}\ }\textbf {\bibinfo
  {volume} {2}},\ \bibinfo {pages} {16028 EP } (\bibinfo {year} {2016})},\
  \bibinfo {note} {article}\BibitemShut {NoStop}%
\bibitem [{\citenamefont {Yang}\ \emph {et~al.}(2018)\citenamefont {Yang},
  \citenamefont {Wang}, \citenamefont {Zhao}, \citenamefont {Song},
  \citenamefont {Zhang},\ and\ \citenamefont {Liu}}]{yang2018matcloud}%
  \BibitemOpen
  \bibfield  {author} {\bibinfo {author} {\bibfnamefont {X.}~\bibnamefont
  {Yang}}, \bibinfo {author} {\bibfnamefont {Z.}~\bibnamefont {Wang}}, \bibinfo
  {author} {\bibfnamefont {X.}~\bibnamefont {Zhao}}, \bibinfo {author}
  {\bibfnamefont {J.}~\bibnamefont {Song}}, \bibinfo {author} {\bibfnamefont
  {M.}~\bibnamefont {Zhang}}, \ and\ \bibinfo {author} {\bibfnamefont
  {H.}~\bibnamefont {Liu}},\ }\href@noop {} {\bibfield  {journal} {\bibinfo
  {journal} {Computational Materials Science}\ }\textbf {\bibinfo {volume}
  {146}},\ \bibinfo {pages} {319} (\bibinfo {year} {2018})}\BibitemShut
  {NoStop}%
\bibitem [{\citenamefont {Bazhirov}\ \emph {et~al.}(2017)\citenamefont
  {Bazhirov}, \citenamefont {Mohammadi}, \citenamefont {Ding},\ and\
  \citenamefont {Barabash}}]{2016-exabyte-aps-abstract}%
  \BibitemOpen
  \bibfield  {author} {\bibinfo {author} {\bibfnamefont {T.}~\bibnamefont
  {Bazhirov}}, \bibinfo {author} {\bibfnamefont {M.}~\bibnamefont {Mohammadi}},
  \bibinfo {author} {\bibfnamefont {K.}~\bibnamefont {Ding}}, \ and\ \bibinfo
  {author} {\bibfnamefont {S.}~\bibnamefont {Barabash}},\ }\href
  {http://meetings.aps.org/Meeting/MAR17/Session/C1.7} {\bibfield  {journal}
  {\bibinfo  {journal} {Proceedings of the American Physical Society March
  Meeting 2017}\ } (\bibinfo {year} {2017})}\BibitemShut {NoStop}%
\bibitem [{\citenamefont {Klimeck}\ \emph {et~al.}(2008)\citenamefont
  {Klimeck}, \citenamefont {McLennan}, \citenamefont {Brophy}, \citenamefont
  {Adams~III},\ and\ \citenamefont {Lundstrom}}]{klimeck2008nanohub}%
  \BibitemOpen
  \bibfield  {author} {\bibinfo {author} {\bibfnamefont {G.}~\bibnamefont
  {Klimeck}}, \bibinfo {author} {\bibfnamefont {M.}~\bibnamefont {McLennan}},
  \bibinfo {author} {\bibfnamefont {S.~P.}\ \bibnamefont {Brophy}}, \bibinfo
  {author} {\bibfnamefont {G.~B.}\ \bibnamefont {Adams~III}}, \ and\ \bibinfo
  {author} {\bibfnamefont {M.~S.}\ \bibnamefont {Lundstrom}},\ }\href@noop {}
  {\bibfield  {journal} {\bibinfo  {journal} {Computing in Science \&
  Engineering}\ }\textbf {\bibinfo {volume} {10}},\ \bibinfo {pages} {17}
  (\bibinfo {year} {2008})}\BibitemShut {NoStop}%
\bibitem [{\citenamefont {Mohammadi}\ and\ \citenamefont
  {Bazhirov}(2018)}]{exabyte2018hp3c}%
  \BibitemOpen
  \bibfield  {author} {\bibinfo {author} {\bibfnamefont {M.}~\bibnamefont
  {Mohammadi}}\ and\ \bibinfo {author} {\bibfnamefont {T.}~\bibnamefont
  {Bazhirov}},\ }in\ \href {\doibase 10.1145/3195612.3195613} {\emph {\bibinfo
  {booktitle} {Proceedings of the 2Nd International Conference on High
  Performance Compilation, Computing and Communications}}},\ \bibinfo {series
  and number} {HP3C}\ (\bibinfo  {publisher} {ACM},\ \bibinfo {address} {New
  York, NY, USA},\ \bibinfo {year} {2018})\ pp.\ \bibinfo {pages}
  {1--5}\BibitemShut {NoStop}%
\bibitem [{exa({\natexlab{a}})}]{exabytePlatform}%
  \BibitemOpen
  \href {https://exabyte.io/} {\emph {\bibinfo {title} {Exabyte.io: materials
  discovery cloud}}} ({\natexlab{a}})\BibitemShut {NoStop}%
\bibitem [{\citenamefont {Hohenberg}\ and\ \citenamefont
  {Kohn}(1964)}]{hohenberg-kohn1964DFT}%
  \BibitemOpen
  \bibfield  {author} {\bibinfo {author} {\bibfnamefont {P.}~\bibnamefont
  {Hohenberg}}\ and\ \bibinfo {author} {\bibfnamefont {W.}~\bibnamefont
  {Kohn}},\ }\href {\doibase 10.1103/PhysRev.136.B864} {\bibfield  {journal}
  {\bibinfo  {journal} {Phys. Rev.}\ }\textbf {\bibinfo {volume} {136}},\
  \bibinfo {pages} {B864} (\bibinfo {year} {1964})}\BibitemShut {NoStop}%
\bibitem [{\citenamefont {Ihm}\ \emph {et~al.}(1979)\citenamefont {Ihm},
  \citenamefont {Zunger},\ and\ \citenamefont
  {Cohen}}]{mlcohen1979pseudopotentialDFT}%
  \BibitemOpen
  \bibfield  {author} {\bibinfo {author} {\bibfnamefont {J.}~\bibnamefont
  {Ihm}}, \bibinfo {author} {\bibfnamefont {A.}~\bibnamefont {Zunger}}, \ and\
  \bibinfo {author} {\bibfnamefont {M.}~\bibnamefont {Cohen}},\ }\href@noop {}
  {\bibfield  {journal} {\bibinfo  {journal} {Journal of Physics C: Solid State
  Physics}\ }\textbf {\bibinfo {volume} {12}},\ \bibinfo {pages} {4409}
  (\bibinfo {year} {1979})}\BibitemShut {NoStop}%
\bibitem [{\citenamefont {Perdew}\ \emph {et~al.}(1996)\citenamefont {Perdew},
  \citenamefont {Burke},\ and\ \citenamefont
  {Ernzerhof}}]{perdew1996generalized}%
  \BibitemOpen
  \bibfield  {author} {\bibinfo {author} {\bibfnamefont {J.~P.}\ \bibnamefont
  {Perdew}}, \bibinfo {author} {\bibfnamefont {K.}~\bibnamefont {Burke}}, \
  and\ \bibinfo {author} {\bibfnamefont {M.}~\bibnamefont {Ernzerhof}},\
  }\href@noop {} {\bibfield  {journal} {\bibinfo  {journal} {Physical review
  letters}\ }\textbf {\bibinfo {volume} {77}},\ \bibinfo {pages} {3865}
  (\bibinfo {year} {1996})}\BibitemShut {NoStop}%
\bibitem [{\citenamefont {Heyd}\ \emph {et~al.}(2003)\citenamefont {Heyd},
  \citenamefont {Scuseria},\ and\ \citenamefont {Ernzerhof}}]{heyd2003hybrid}%
  \BibitemOpen
  \bibfield  {author} {\bibinfo {author} {\bibfnamefont {J.}~\bibnamefont
  {Heyd}}, \bibinfo {author} {\bibfnamefont {G.~E.}\ \bibnamefont {Scuseria}},
  \ and\ \bibinfo {author} {\bibfnamefont {M.}~\bibnamefont {Ernzerhof}},\
  }\href@noop {} {\bibfield  {journal} {\bibinfo  {journal} {The Journal of
  chemical physics}\ }\textbf {\bibinfo {volume} {118}},\ \bibinfo {pages}
  {8207} (\bibinfo {year} {2003})}\BibitemShut {NoStop}%
\bibitem [{\citenamefont {Hybertsen}\ and\ \citenamefont
  {Louie}(1985)}]{hybertsen-louie1985GW}%
  \BibitemOpen
  \bibfield  {author} {\bibinfo {author} {\bibfnamefont {M.~S.}\ \bibnamefont
  {Hybertsen}}\ and\ \bibinfo {author} {\bibfnamefont {S.~G.}\ \bibnamefont
  {Louie}},\ }\href {\doibase 10.1103/PhysRevLett.55.1418} {\bibfield
  {journal} {\bibinfo  {journal} {Phys. Rev. Lett.}\ }\textbf {\bibinfo
  {volume} {55}},\ \bibinfo {pages} {1418} (\bibinfo {year}
  {1985})}\BibitemShut {NoStop}%
\bibitem [{exa({\natexlab{b}})}]{exabytePlatformBGPhaseIURL}%
  \BibitemOpen
  \href {https://exabyte.io/} {\emph {\bibinfo {title} {Exabyte platform:
  project URL with data about simulations}}} ({\natexlab{b}})\BibitemShut
  {NoStop}%
\bibitem [{\citenamefont {Heyd}\ \emph {et~al.}(2005)\citenamefont {Heyd},
  \citenamefont {Peralta}, \citenamefont {Scuseria},\ and\ \citenamefont
  {Martin}}]{heyd2005energy}%
  \BibitemOpen
  \bibfield  {author} {\bibinfo {author} {\bibfnamefont {J.}~\bibnamefont
  {Heyd}}, \bibinfo {author} {\bibfnamefont {J.~E.}\ \bibnamefont {Peralta}},
  \bibinfo {author} {\bibfnamefont {G.~E.}\ \bibnamefont {Scuseria}}, \ and\
  \bibinfo {author} {\bibfnamefont {R.~L.}\ \bibnamefont {Martin}},\
  }\href@noop {} {\bibfield  {journal} {\bibinfo  {journal} {The Journal of
  chemical physics}\ }\textbf {\bibinfo {volume} {123}},\ \bibinfo {pages}
  {174101} (\bibinfo {year} {2005})}\BibitemShut {NoStop}%
\bibitem [{\citenamefont {Kresse}(1996)}]{kresse1996software}%
  \BibitemOpen
  \bibfield  {author} {\bibinfo {author} {\bibfnamefont {G.}~\bibnamefont
  {Kresse}},\ }\href@noop {} {\bibfield  {journal} {\bibinfo  {journal} {Phys.
  Rev. B}\ }\textbf {\bibinfo {volume} {54}},\ \bibinfo {pages} {169} (\bibinfo
  {year} {1996})}\BibitemShut {NoStop}%
\bibitem [{\citenamefont {Surh}\ \emph {et~al.}(1991)\citenamefont {Surh},
  \citenamefont {Li},\ and\ \citenamefont {Louie}}]{louie1991gaas-soc}%
  \BibitemOpen
  \bibfield  {author} {\bibinfo {author} {\bibfnamefont {M.~P.}\ \bibnamefont
  {Surh}}, \bibinfo {author} {\bibfnamefont {M.-F.}\ \bibnamefont {Li}}, \ and\
  \bibinfo {author} {\bibfnamefont {S.~G.}\ \bibnamefont {Louie}},\ }\href
  {\doibase 10.1103/PhysRevB.43.4286} {\bibfield  {journal} {\bibinfo
  {journal} {Phys. Rev. B}\ }\textbf {\bibinfo {volume} {43}},\ \bibinfo
  {pages} {4286} (\bibinfo {year} {1991})}\BibitemShut {NoStop}%
\bibitem [{\citenamefont {Kresse}\ and\ \citenamefont
  {Furthm{\"u}ller}(1996)}]{kresse1996efficient}%
  \BibitemOpen
  \bibfield  {author} {\bibinfo {author} {\bibfnamefont {G.}~\bibnamefont
  {Kresse}}\ and\ \bibinfo {author} {\bibfnamefont {J.}~\bibnamefont
  {Furthm{\"u}ller}},\ }\href@noop {} {\bibfield  {journal} {\bibinfo
  {journal} {Physical review B}\ }\textbf {\bibinfo {volume} {54}},\ \bibinfo
  {pages} {11169} (\bibinfo {year} {1996})}\BibitemShut {NoStop}%
\bibitem [{\citenamefont {Kohn}\ and\ \citenamefont
  {Sham}(1965)}]{kohn1965self}%
  \BibitemOpen
  \bibfield  {author} {\bibinfo {author} {\bibfnamefont {W.}~\bibnamefont
  {Kohn}}\ and\ \bibinfo {author} {\bibfnamefont {L.~J.}\ \bibnamefont
  {Sham}},\ }\href@noop {} {\bibfield  {journal} {\bibinfo  {journal} {Physical
  review}\ }\textbf {\bibinfo {volume} {140}},\ \bibinfo {pages} {A1133}
  (\bibinfo {year} {1965})}\BibitemShut {NoStop}%
\bibitem [{\citenamefont {Bl{\"o}chl}(1994)}]{blochl1994projector}%
  \BibitemOpen
  \bibfield  {author} {\bibinfo {author} {\bibfnamefont {P.~E.}\ \bibnamefont
  {Bl{\"o}chl}},\ }\href@noop {} {\bibfield  {journal} {\bibinfo  {journal}
  {Physical review B}\ }\textbf {\bibinfo {volume} {50}},\ \bibinfo {pages}
  {17953} (\bibinfo {year} {1994})}\BibitemShut {NoStop}%
\bibitem [{\citenamefont {Hacene}\ \emph {et~al.}(2012)\citenamefont {Hacene},
  \citenamefont {Anciaux-Sedrakian}, \citenamefont {Rozanska}, \citenamefont
  {Klahr}, \citenamefont {Guignon},\ and\ \citenamefont
  {Fleurat-Lessard}}]{hacene2012accelerating}%
  \BibitemOpen
  \bibfield  {author} {\bibinfo {author} {\bibfnamefont {M.}~\bibnamefont
  {Hacene}}, \bibinfo {author} {\bibfnamefont {A.}~\bibnamefont
  {Anciaux-Sedrakian}}, \bibinfo {author} {\bibfnamefont {X.}~\bibnamefont
  {Rozanska}}, \bibinfo {author} {\bibfnamefont {D.}~\bibnamefont {Klahr}},
  \bibinfo {author} {\bibfnamefont {T.}~\bibnamefont {Guignon}}, \ and\
  \bibinfo {author} {\bibfnamefont {P.}~\bibnamefont {Fleurat-Lessard}},\
  }\href@noop {} {\bibfield  {journal} {\bibinfo  {journal} {Journal of
  computational chemistry}\ }\textbf {\bibinfo {volume} {33}},\ \bibinfo
  {pages} {2581} (\bibinfo {year} {2012})}\BibitemShut {NoStop}%
\bibitem [{\citenamefont {Johnson}\ and\ \citenamefont
  {Joannopoulos}(2001)}]{johnson2001block}%
  \BibitemOpen
  \bibfield  {author} {\bibinfo {author} {\bibfnamefont {S.~G.}\ \bibnamefont
  {Johnson}}\ and\ \bibinfo {author} {\bibfnamefont {J.~D.}\ \bibnamefont
  {Joannopoulos}},\ }\href@noop {} {\bibfield  {journal} {\bibinfo  {journal}
  {Optics express}\ }\textbf {\bibinfo {volume} {8}},\ \bibinfo {pages} {173}
  (\bibinfo {year} {2001})}\BibitemShut {NoStop}%
\bibitem [{\citenamefont {Grimme}(2006)}]{grimme2006vdWcorrection}%
  \BibitemOpen
  \bibfield  {author} {\bibinfo {author} {\bibfnamefont {S.}~\bibnamefont
  {Grimme}},\ }\href {\doibase 10.1002/jcc.20495} {\bibfield  {journal}
  {\bibinfo  {journal} {Journal of computational chemistry}\ }\textbf {\bibinfo
  {volume} {27}},\ \bibinfo {pages} {1787} (\bibinfo {year}
  {2006})}\BibitemShut {NoStop}%
\bibitem [{azu()}]{azure-instance-types}%
  \BibitemOpen
  \href
  {https://docs.microsoft.com/en-us/azure/virtual-machines/virtual-machines-linux-sizes}
  {\emph {\bibinfo {title} {Microsoft Azure Cloud Computing platform: web
  page.}}}\BibitemShut {Stop}%
\bibitem [{exa({\natexlab{c}})}]{exabyteRESTAPIClient}%
  \BibitemOpen
  \href {https://github.com/exabyte-io/exabyte-api-client/} {\emph {\bibinfo
  {title} {Exabyte RESTful API client: online URL}}}
  ({\natexlab{c}})\BibitemShut {NoStop}%
\bibitem [{\citenamefont {Jette}\ and\ \citenamefont
  {Foote}(1935)}]{jette1935precision}%
  \BibitemOpen
  \bibfield  {author} {\bibinfo {author} {\bibfnamefont {E.~R.}\ \bibnamefont
  {Jette}}\ and\ \bibinfo {author} {\bibfnamefont {F.}~\bibnamefont {Foote}},\
  }\href@noop {} {\bibfield  {journal} {\bibinfo  {journal} {The Journal of
  Chemical Physics}\ }\textbf {\bibinfo {volume} {3}},\ \bibinfo {pages} {605}
  (\bibinfo {year} {1935})}\BibitemShut {NoStop}%
\bibitem [{\citenamefont {Kittel}\ \emph {et~al.}(1996)\citenamefont {Kittel},
  \citenamefont {McEuen},\ and\ \citenamefont
  {McEuen}}]{kittel1996introduction}%
  \BibitemOpen
  \bibfield  {author} {\bibinfo {author} {\bibfnamefont {C.}~\bibnamefont
  {Kittel}}, \bibinfo {author} {\bibfnamefont {P.}~\bibnamefont {McEuen}}, \
  and\ \bibinfo {author} {\bibfnamefont {P.}~\bibnamefont {McEuen}},\
  }\href@noop {} {\emph {\bibinfo {title} {Introduction to solid state
  physics}}},\ Vol.~\bibinfo {volume} {8}\ (\bibinfo  {publisher} {Wiley New
  York},\ \bibinfo {year} {1996})\BibitemShut {NoStop}%
\bibitem [{\citenamefont {Shishkin}\ and\ \citenamefont
  {Kresse}(2007)}]{shishkin2007self}%
  \BibitemOpen
  \bibfield  {author} {\bibinfo {author} {\bibfnamefont {M.}~\bibnamefont
  {Shishkin}}\ and\ \bibinfo {author} {\bibfnamefont {G.}~\bibnamefont
  {Kresse}},\ }\href@noop {} {\bibfield  {journal} {\bibinfo  {journal}
  {Physical Review B}\ }\textbf {\bibinfo {volume} {75}},\ \bibinfo {pages}
  {235102} (\bibinfo {year} {2007})}\BibitemShut {NoStop}%
\bibitem [{\citenamefont {Madelung}(2012)}]{madelung2012semiconductors}%
  \BibitemOpen
  \bibfield  {author} {\bibinfo {author} {\bibfnamefont {O.}~\bibnamefont
  {Madelung}},\ }\href@noop {} {\emph {\bibinfo {title} {Semiconductors: data
  handbook}}}\ (\bibinfo  {publisher} {Springer Science \& Business Media},\
  \bibinfo {year} {2012})\BibitemShut {NoStop}%
\bibitem [{\citenamefont {van Schilfgaarde}\ \emph {et~al.}(2006)\citenamefont
  {van Schilfgaarde}, \citenamefont {Kotani},\ and\ \citenamefont
  {Faleev}}]{van2006adequacy}%
  \BibitemOpen
  \bibfield  {author} {\bibinfo {author} {\bibfnamefont {M.}~\bibnamefont {van
  Schilfgaarde}}, \bibinfo {author} {\bibfnamefont {T.}~\bibnamefont {Kotani}},
  \ and\ \bibinfo {author} {\bibfnamefont {S.~V.}\ \bibnamefont {Faleev}},\
  }\href@noop {} {\bibfield  {journal} {\bibinfo  {journal} {Physical Review
  B}\ }\textbf {\bibinfo {volume} {74}},\ \bibinfo {pages} {245125} (\bibinfo
  {year} {2006})}\BibitemShut {NoStop}%
\bibitem [{\citenamefont {Keller}\ \emph {et~al.}(1977)\citenamefont {Keller},
  \citenamefont {Holzapfel},\ and\ \citenamefont {Schulz}}]{Se_Te_abc}%
  \BibitemOpen
  \bibfield  {author} {\bibinfo {author} {\bibfnamefont {R.}~\bibnamefont
  {Keller}}, \bibinfo {author} {\bibfnamefont {W.}~\bibnamefont {Holzapfel}}, \
  and\ \bibinfo {author} {\bibfnamefont {H.}~\bibnamefont {Schulz}},\
  }\href@noop {} {\bibfield  {journal} {\bibinfo  {journal} {Physical Review
  B}\ }\textbf {\bibinfo {volume} {16}},\ \bibinfo {pages} {4404} (\bibinfo
  {year} {1977})}\BibitemShut {NoStop}%
\bibitem [{\citenamefont {Anzin}\ \emph
  {et~al.}(1977{\natexlab{a}})\citenamefont {Anzin}, \citenamefont {Eremets},
  \citenamefont {Kosichkin}, \citenamefont {Nadezhdinskii},\ and\ \citenamefont
  {Shirokov}}]{anzin1977measurement}%
  \BibitemOpen
  \bibfield  {author} {\bibinfo {author} {\bibfnamefont {V.}~\bibnamefont
  {Anzin}}, \bibinfo {author} {\bibfnamefont {M.}~\bibnamefont {Eremets}},
  \bibinfo {author} {\bibfnamefont {Y.~V.}\ \bibnamefont {Kosichkin}}, \bibinfo
  {author} {\bibfnamefont {A.}~\bibnamefont {Nadezhdinskii}}, \ and\ \bibinfo
  {author} {\bibfnamefont {A.}~\bibnamefont {Shirokov}},\ }\href@noop {}
  {\bibfield  {journal} {\bibinfo  {journal} {physica status solidi (a)}\
  }\textbf {\bibinfo {volume} {42}},\ \bibinfo {pages} {385} (\bibinfo {year}
  {1977}{\natexlab{a}})}\BibitemShut {NoStop}%
\bibitem [{\citenamefont {Yi}\ \emph {et~al.}(2018{\natexlab{a}})\citenamefont
  {Yi}, \citenamefont {Zhu}, \citenamefont {Cai}, \citenamefont {Jia},\ and\
  \citenamefont {Cho}}]{yi2018nature}%
  \BibitemOpen
  \bibfield  {author} {\bibinfo {author} {\bibfnamefont {S.}~\bibnamefont
  {Yi}}, \bibinfo {author} {\bibfnamefont {Z.}~\bibnamefont {Zhu}}, \bibinfo
  {author} {\bibfnamefont {X.}~\bibnamefont {Cai}}, \bibinfo {author}
  {\bibfnamefont {Y.}~\bibnamefont {Jia}}, \ and\ \bibinfo {author}
  {\bibfnamefont {J.-H.}\ \bibnamefont {Cho}},\ }\href@noop {} {\bibfield
  {journal} {\bibinfo  {journal} {Inorganic chemistry}\ }\textbf {\bibinfo
  {volume} {57}},\ \bibinfo {pages} {5083} (\bibinfo {year}
  {2018}{\natexlab{a}})}\BibitemShut {NoStop}%
\bibitem [{\citenamefont {Decker}\ and\ \citenamefont
  {Kasper}(1959)}]{decker1959crystal}%
  \BibitemOpen
  \bibfield  {author} {\bibinfo {author} {\bibfnamefont {B.}~\bibnamefont
  {Decker}}\ and\ \bibinfo {author} {\bibfnamefont {J.}~\bibnamefont
  {Kasper}},\ }\href@noop {} {\bibfield  {journal} {\bibinfo  {journal} {Acta
  Crystallographica}\ }\textbf {\bibinfo {volume} {12}},\ \bibinfo {pages}
  {503} (\bibinfo {year} {1959})}\BibitemShut {NoStop}%
\bibitem [{\citenamefont {Cucka}\ and\ \citenamefont
  {Barrett}(1962)}]{cucka1962crystal}%
  \BibitemOpen
  \bibfield  {author} {\bibinfo {author} {\bibfnamefont {P.}~\bibnamefont
  {Cucka}}\ and\ \bibinfo {author} {\bibfnamefont {C.}~\bibnamefont
  {Barrett}},\ }\href@noop {} {\bibfield  {journal} {\bibinfo  {journal} {Acta
  Crystallographica}\ }\textbf {\bibinfo {volume} {15}},\ \bibinfo {pages}
  {865} (\bibinfo {year} {1962})}\BibitemShut {NoStop}%
\bibitem [{\citenamefont {Brown}\ and\ \citenamefont
  {Rundqvist}(1965)}]{brown1965refinement}%
  \BibitemOpen
  \bibfield  {author} {\bibinfo {author} {\bibfnamefont {A.}~\bibnamefont
  {Brown}}\ and\ \bibinfo {author} {\bibfnamefont {S.}~\bibnamefont
  {Rundqvist}},\ }\href@noop {} {\bibfield  {journal} {\bibinfo  {journal}
  {Acta Crystallographica}\ }\textbf {\bibinfo {volume} {19}},\ \bibinfo
  {pages} {684} (\bibinfo {year} {1965})}\BibitemShut {NoStop}%
\bibitem [{\citenamefont {Asahina}\ and\ \citenamefont
  {Morita}(1984)}]{asahina1984band}%
  \BibitemOpen
  \bibfield  {author} {\bibinfo {author} {\bibfnamefont {H.}~\bibnamefont
  {Asahina}}\ and\ \bibinfo {author} {\bibfnamefont {A.}~\bibnamefont
  {Morita}},\ }\href@noop {} {\bibfield  {journal} {\bibinfo  {journal}
  {Journal of Physics C: Solid State Physics}\ }\textbf {\bibinfo {volume}
  {17}},\ \bibinfo {pages} {1839} (\bibinfo {year} {1984})}\BibitemShut
  {NoStop}%
\bibitem [{\citenamefont {Gomes}\ and\ \citenamefont
  {Carvalho}(2015)}]{gomes2015phosphorene}%
  \BibitemOpen
  \bibfield  {author} {\bibinfo {author} {\bibfnamefont {L.~C.}\ \bibnamefont
  {Gomes}}\ and\ \bibinfo {author} {\bibfnamefont {A.}~\bibnamefont
  {Carvalho}},\ }\href@noop {} {\bibfield  {journal} {\bibinfo  {journal}
  {Physical Review B}\ }\textbf {\bibinfo {volume} {92}},\ \bibinfo {pages}
  {085406} (\bibinfo {year} {2015})}\BibitemShut {NoStop}%
\bibitem [{\citenamefont {Tran}\ \emph {et~al.}(2014)\citenamefont {Tran},
  \citenamefont {Soklaski}, \citenamefont {Liang},\ and\ \citenamefont
  {Yang}}]{tran2014layer}%
  \BibitemOpen
  \bibfield  {author} {\bibinfo {author} {\bibfnamefont {V.}~\bibnamefont
  {Tran}}, \bibinfo {author} {\bibfnamefont {R.}~\bibnamefont {Soklaski}},
  \bibinfo {author} {\bibfnamefont {Y.}~\bibnamefont {Liang}}, \ and\ \bibinfo
  {author} {\bibfnamefont {L.}~\bibnamefont {Yang}},\ }\href@noop {} {\bibfield
   {journal} {\bibinfo  {journal} {Physical Review B}\ }\textbf {\bibinfo
  {volume} {89}},\ \bibinfo {pages} {235319} (\bibinfo {year}
  {2014})}\BibitemShut {NoStop}%
\bibitem [{\citenamefont {Smith}\ \emph {et~al.}(1975)\citenamefont {Smith},
  \citenamefont {Leadbetter},\ and\ \citenamefont
  {Apling}}]{smith1975structures}%
  \BibitemOpen
  \bibfield  {author} {\bibinfo {author} {\bibfnamefont {P.}~\bibnamefont
  {Smith}}, \bibinfo {author} {\bibfnamefont {A.}~\bibnamefont {Leadbetter}}, \
  and\ \bibinfo {author} {\bibfnamefont {A.}~\bibnamefont {Apling}},\
  }\href@noop {} {\bibfield  {journal} {\bibinfo  {journal} {Philosophical
  Magazine}\ }\textbf {\bibinfo {volume} {31}},\ \bibinfo {pages} {57}
  (\bibinfo {year} {1975})}\BibitemShut {NoStop}%
\bibitem [{\citenamefont {Kecik}\ \emph {et~al.}(2016)\citenamefont {Kecik},
  \citenamefont {Durgun},\ and\ \citenamefont {Ciraci}}]{kecik2016stability}%
  \BibitemOpen
  \bibfield  {author} {\bibinfo {author} {\bibfnamefont {D.}~\bibnamefont
  {Kecik}}, \bibinfo {author} {\bibfnamefont {E.}~\bibnamefont {Durgun}}, \
  and\ \bibinfo {author} {\bibfnamefont {S.}~\bibnamefont {Ciraci}},\
  }\href@noop {} {\bibfield  {journal} {\bibinfo  {journal} {Physical Review
  B}\ }\textbf {\bibinfo {volume} {94}},\ \bibinfo {pages} {205409} (\bibinfo
  {year} {2016})}\BibitemShut {NoStop}%
\bibitem [{\citenamefont {Barrett}\ \emph {et~al.}(1963)\citenamefont
  {Barrett}, \citenamefont {Cucka},\ and\ \citenamefont
  {Haefner}}]{barrett1963crystal}%
  \BibitemOpen
  \bibfield  {author} {\bibinfo {author} {\bibfnamefont {C.}~\bibnamefont
  {Barrett}}, \bibinfo {author} {\bibfnamefont {P.}~\bibnamefont {Cucka}}, \
  and\ \bibinfo {author} {\bibfnamefont {K.}~\bibnamefont {Haefner}},\
  }\href@noop {} {\bibfield  {journal} {\bibinfo  {journal} {Acta
  Crystallographica}\ }\textbf {\bibinfo {volume} {16}},\ \bibinfo {pages}
  {451} (\bibinfo {year} {1963})}\BibitemShut {NoStop}%
\bibitem [{\citenamefont {Brownlee}(1950)}]{brownlee1950lattice}%
  \BibitemOpen
  \bibfield  {author} {\bibinfo {author} {\bibfnamefont {L.}~\bibnamefont
  {Brownlee}},\ }\href@noop {} {\bibfield  {journal} {\bibinfo  {journal}
  {Nature}\ }\textbf {\bibinfo {volume} {166}},\ \bibinfo {pages} {482}
  (\bibinfo {year} {1950})}\BibitemShut {NoStop}%
\bibitem [{\citenamefont {Hummer}\ \emph {et~al.}(2009)\citenamefont {Hummer},
  \citenamefont {Harl},\ and\ \citenamefont {Kresse}}]{hummer2009heyd}%
  \BibitemOpen
  \bibfield  {author} {\bibinfo {author} {\bibfnamefont {K.}~\bibnamefont
  {Hummer}}, \bibinfo {author} {\bibfnamefont {J.}~\bibnamefont {Harl}}, \ and\
  \bibinfo {author} {\bibfnamefont {G.}~\bibnamefont {Kresse}},\ }\href@noop {}
  {\bibfield  {journal} {\bibinfo  {journal} {Physical Review B}\ }\textbf
  {\bibinfo {volume} {80}},\ \bibinfo {pages} {115205} (\bibinfo {year}
  {2009})}\BibitemShut {NoStop}%
\bibitem [{\citenamefont {Blase}\ \emph {et~al.}(1995)\citenamefont {Blase},
  \citenamefont {Rubio}, \citenamefont {Louie},\ and\ \citenamefont
  {Cohen}}]{blase1995quasiparticle}%
  \BibitemOpen
  \bibfield  {author} {\bibinfo {author} {\bibfnamefont {X.}~\bibnamefont
  {Blase}}, \bibinfo {author} {\bibfnamefont {A.}~\bibnamefont {Rubio}},
  \bibinfo {author} {\bibfnamefont {S.~G.}\ \bibnamefont {Louie}}, \ and\
  \bibinfo {author} {\bibfnamefont {M.~L.}\ \bibnamefont {Cohen}},\ }\href@noop
  {} {\bibfield  {journal} {\bibinfo  {journal} {Physical review B}\ }\textbf
  {\bibinfo {volume} {51}},\ \bibinfo {pages} {6868} (\bibinfo {year}
  {1995})}\BibitemShut {NoStop}%
\bibitem [{\citenamefont {Cassabois}\ \emph {et~al.}(2016)\citenamefont
  {Cassabois}, \citenamefont {Valvin},\ and\ \citenamefont
  {Gil}}]{cassabois2016hexagonal}%
  \BibitemOpen
  \bibfield  {author} {\bibinfo {author} {\bibfnamefont {G.}~\bibnamefont
  {Cassabois}}, \bibinfo {author} {\bibfnamefont {P.}~\bibnamefont {Valvin}}, \
  and\ \bibinfo {author} {\bibfnamefont {B.}~\bibnamefont {Gil}},\ }\href@noop
  {} {\bibfield  {journal} {\bibinfo  {journal} {Nature Photonics}\ }\textbf
  {\bibinfo {volume} {10}},\ \bibinfo {pages} {nphoton} (\bibinfo {year}
  {2016})}\BibitemShut {NoStop}%
\bibitem [{\citenamefont {Tran}\ and\ \citenamefont
  {Blaha}(2009)}]{tran2009accurate}%
  \BibitemOpen
  \bibfield  {author} {\bibinfo {author} {\bibfnamefont {F.}~\bibnamefont
  {Tran}}\ and\ \bibinfo {author} {\bibfnamefont {P.}~\bibnamefont {Blaha}},\
  }\href@noop {} {\bibfield  {journal} {\bibinfo  {journal} {Physical review
  letters}\ }\textbf {\bibinfo {volume} {102}},\ \bibinfo {pages} {226401}
  (\bibinfo {year} {2009})}\BibitemShut {NoStop}%
\bibitem [{\citenamefont {Lee}\ \emph {et~al.}(2016)\citenamefont {Lee},
  \citenamefont {Seko}, \citenamefont {Shitara}, \citenamefont {Nakayama},\
  and\ \citenamefont {Tanaka}}]{lee2016prediction}%
  \BibitemOpen
  \bibfield  {author} {\bibinfo {author} {\bibfnamefont {J.}~\bibnamefont
  {Lee}}, \bibinfo {author} {\bibfnamefont {A.}~\bibnamefont {Seko}}, \bibinfo
  {author} {\bibfnamefont {K.}~\bibnamefont {Shitara}}, \bibinfo {author}
  {\bibfnamefont {K.}~\bibnamefont {Nakayama}}, \ and\ \bibinfo {author}
  {\bibfnamefont {I.}~\bibnamefont {Tanaka}},\ }\href@noop {} {\bibfield
  {journal} {\bibinfo  {journal} {Physical Review B}\ }\textbf {\bibinfo
  {volume} {93}},\ \bibinfo {pages} {115104} (\bibinfo {year}
  {2016})}\BibitemShut {NoStop}%
\bibitem [{\citenamefont {Merrill}(1977)}]{merrill1977behavior}%
  \BibitemOpen
  \bibfield  {author} {\bibinfo {author} {\bibfnamefont {L.}~\bibnamefont
  {Merrill}},\ }\href@noop {} {\bibfield  {journal} {\bibinfo  {journal}
  {Journal of Physical and Chemical Reference Data}\ }\textbf {\bibinfo
  {volume} {6}},\ \bibinfo {pages} {1205} (\bibinfo {year} {1977})}\BibitemShut
  {NoStop}%
\bibitem [{\citenamefont {Powell}\ \emph {et~al.}(1993)\citenamefont {Powell},
  \citenamefont {Lee}, \citenamefont {Kim},\ and\ \citenamefont
  {Greene}}]{powell1993heteroepitaxial}%
  \BibitemOpen
  \bibfield  {author} {\bibinfo {author} {\bibfnamefont {R.}~\bibnamefont
  {Powell}}, \bibinfo {author} {\bibfnamefont {N.-E.}\ \bibnamefont {Lee}},
  \bibinfo {author} {\bibfnamefont {Y.-W.}\ \bibnamefont {Kim}}, \ and\
  \bibinfo {author} {\bibfnamefont {J.}~\bibnamefont {Greene}},\ }\href@noop {}
  {\bibfield  {journal} {\bibinfo  {journal} {Journal of applied physics}\
  }\textbf {\bibinfo {volume} {73}},\ \bibinfo {pages} {189} (\bibinfo {year}
  {1993})}\BibitemShut {NoStop}%
\bibitem [{\citenamefont {Saha}\ \emph {et~al.}(2011)\citenamefont {Saha},
  \citenamefont {Sands},\ and\ \citenamefont {Waghmare}}]{saha2011electronic}%
  \BibitemOpen
  \bibfield  {author} {\bibinfo {author} {\bibfnamefont {B.}~\bibnamefont
  {Saha}}, \bibinfo {author} {\bibfnamefont {T.~D.}\ \bibnamefont {Sands}}, \
  and\ \bibinfo {author} {\bibfnamefont {U.~V.}\ \bibnamefont {Waghmare}},\
  }\href@noop {} {\bibfield  {journal} {\bibinfo  {journal} {Journal of Applied
  Physics}\ }\textbf {\bibinfo {volume} {109}},\ \bibinfo {pages} {073720}
  (\bibinfo {year} {2011})}\BibitemShut {NoStop}%
\bibitem [{\citenamefont {Laref}\ and\ \citenamefont
  {Laref}(2013)}]{laref2013comparative}%
  \BibitemOpen
  \bibfield  {author} {\bibinfo {author} {\bibfnamefont {S.}~\bibnamefont
  {Laref}}\ and\ \bibinfo {author} {\bibfnamefont {A.}~\bibnamefont {Laref}},\
  }\href@noop {} {\bibfield  {journal} {\bibinfo  {journal} {Journal of
  Materials Science}\ }\textbf {\bibinfo {volume} {48}},\ \bibinfo {pages}
  {5499} (\bibinfo {year} {2013})}\BibitemShut {NoStop}%
\bibitem [{\citenamefont {Nejatipour}\ and\ \citenamefont
  {Dadsetani}(2015)}]{nejatipour2015excitonic}%
  \BibitemOpen
  \bibfield  {author} {\bibinfo {author} {\bibfnamefont {H.}~\bibnamefont
  {Nejatipour}}\ and\ \bibinfo {author} {\bibfnamefont {M.}~\bibnamefont
  {Dadsetani}},\ }\href@noop {} {\bibfield  {journal} {\bibinfo  {journal}
  {Physica Scripta}\ }\textbf {\bibinfo {volume} {90}},\ \bibinfo {pages}
  {085802} (\bibinfo {year} {2015})}\BibitemShut {NoStop}%
\bibitem [{\citenamefont {Grzybowski}\ and\ \citenamefont
  {Ruoff}(1983)}]{grzybowski1983high}%
  \BibitemOpen
  \bibfield  {author} {\bibinfo {author} {\bibfnamefont {T.}~\bibnamefont
  {Grzybowski}}\ and\ \bibinfo {author} {\bibfnamefont {A.}~\bibnamefont
  {Ruoff}},\ }\href@noop {} {\bibfield  {journal} {\bibinfo  {journal}
  {Physical Review B}\ }\textbf {\bibinfo {volume} {27}},\ \bibinfo {pages}
  {6502} (\bibinfo {year} {1983})}\BibitemShut {NoStop}%
\bibitem [{\citenamefont {Grzybowski}\ and\ \citenamefont
  {Ruoff}(1984)}]{grzybowski1984band}%
  \BibitemOpen
  \bibfield  {author} {\bibinfo {author} {\bibfnamefont {T.~A.}\ \bibnamefont
  {Grzybowski}}\ and\ \bibinfo {author} {\bibfnamefont {A.~L.}\ \bibnamefont
  {Ruoff}},\ }\href@noop {} {\bibfield  {journal} {\bibinfo  {journal}
  {Physical review letters}\ }\textbf {\bibinfo {volume} {53}},\ \bibinfo
  {pages} {489} (\bibinfo {year} {1984})}\BibitemShut {NoStop}%
\bibitem [{\citenamefont {Luo}\ \emph {et~al.}(1994)\citenamefont {Luo},
  \citenamefont {Greene}, \citenamefont {Ghandehari}, \citenamefont {Li},\ and\
  \citenamefont {Ruoff}}]{luo1994structural}%
  \BibitemOpen
  \bibfield  {author} {\bibinfo {author} {\bibfnamefont {H.}~\bibnamefont
  {Luo}}, \bibinfo {author} {\bibfnamefont {R.~G.}\ \bibnamefont {Greene}},
  \bibinfo {author} {\bibfnamefont {K.}~\bibnamefont {Ghandehari}}, \bibinfo
  {author} {\bibfnamefont {T.}~\bibnamefont {Li}}, \ and\ \bibinfo {author}
  {\bibfnamefont {A.~L.}\ \bibnamefont {Ruoff}},\ }\href@noop {} {\bibfield
  {journal} {\bibinfo  {journal} {Physical review B}\ }\textbf {\bibinfo
  {volume} {50}},\ \bibinfo {pages} {16232} (\bibinfo {year}
  {1994})}\BibitemShut {NoStop}%
\bibitem [{\citenamefont {Zintl}\ \emph {et~al.}(1934)\citenamefont {Zintl},
  \citenamefont {Harder},\ and\ \citenamefont
  {Dauth}}]{zintl1934gitterstruktur}%
  \BibitemOpen
  \bibfield  {author} {\bibinfo {author} {\bibfnamefont {E.}~\bibnamefont
  {Zintl}}, \bibinfo {author} {\bibfnamefont {A.}~\bibnamefont {Harder}}, \
  and\ \bibinfo {author} {\bibfnamefont {B.}~\bibnamefont {Dauth}},\
  }\href@noop {} {\bibfield  {journal} {\bibinfo  {journal} {Zeitschrift
  f{\"u}r Elektrochemie und angewandte physikalische Chemie}\ }\textbf
  {\bibinfo {volume} {40}},\ \bibinfo {pages} {588} (\bibinfo {year}
  {1934})}\BibitemShut {NoStop}%
\bibitem [{\citenamefont
  {Mittendorf}(1965)}]{mittendorf1965rontgenographische}%
  \BibitemOpen
  \bibfield  {author} {\bibinfo {author} {\bibfnamefont {H.}~\bibnamefont
  {Mittendorf}},\ }\href@noop {} {\bibfield  {journal} {\bibinfo  {journal}
  {Zeitschrift f{\"u}r Physik}\ }\textbf {\bibinfo {volume} {183}},\ \bibinfo
  {pages} {113} (\bibinfo {year} {1965})}\BibitemShut {NoStop}%
\bibitem [{\citenamefont {Bronsema}\ \emph
  {et~al.}(1986{\natexlab{a}})\citenamefont {Bronsema}, \citenamefont
  {De~Boer},\ and\ \citenamefont {Jellinek}}]{bronsema1986structure}%
  \BibitemOpen
  \bibfield  {author} {\bibinfo {author} {\bibfnamefont {K.}~\bibnamefont
  {Bronsema}}, \bibinfo {author} {\bibfnamefont {J.}~\bibnamefont {De~Boer}}, \
  and\ \bibinfo {author} {\bibfnamefont {F.}~\bibnamefont {Jellinek}},\
  }\href@noop {} {\bibfield  {journal} {\bibinfo  {journal} {Zeitschrift
  f{\"u}r anorganische und allgemeine Chemie}\ }\textbf {\bibinfo {volume}
  {540}},\ \bibinfo {pages} {15} (\bibinfo {year}
  {1986}{\natexlab{a}})}\BibitemShut {NoStop}%
\bibitem [{\citenamefont {Wickramaratne}\ \emph {et~al.}(2014)\citenamefont
  {Wickramaratne}, \citenamefont {Zahid},\ and\ \citenamefont
  {Lake}}]{wickramaratne2014electronic}%
  \BibitemOpen
  \bibfield  {author} {\bibinfo {author} {\bibfnamefont {D.}~\bibnamefont
  {Wickramaratne}}, \bibinfo {author} {\bibfnamefont {F.}~\bibnamefont
  {Zahid}}, \ and\ \bibinfo {author} {\bibfnamefont {R.~K.}\ \bibnamefont
  {Lake}},\ }\href@noop {} {\bibfield  {journal} {\bibinfo  {journal} {The
  Journal of chemical physics}\ }\textbf {\bibinfo {volume} {140}},\ \bibinfo
  {pages} {124710} (\bibinfo {year} {2014})}\BibitemShut {NoStop}%
\bibitem [{\citenamefont {Cheiwchanchamnangij}\ and\ \citenamefont
  {Lambrecht}(2012)}]{cheiwchanchamnangij2012quasiparticle}%
  \BibitemOpen
  \bibfield  {author} {\bibinfo {author} {\bibfnamefont {T.}~\bibnamefont
  {Cheiwchanchamnangij}}\ and\ \bibinfo {author} {\bibfnamefont {W.~R.}\
  \bibnamefont {Lambrecht}},\ }\href@noop {} {\bibfield  {journal} {\bibinfo
  {journal} {Physical Review B}\ }\textbf {\bibinfo {volume} {85}},\ \bibinfo
  {pages} {205302} (\bibinfo {year} {2012})}\BibitemShut {NoStop}%
\bibitem [{\citenamefont {Hodul}\ and\ \citenamefont
  {Stacy}(1984)}]{hodul1984anomalies}%
  \BibitemOpen
  \bibfield  {author} {\bibinfo {author} {\bibfnamefont {D.~T.}\ \bibnamefont
  {Hodul}}\ and\ \bibinfo {author} {\bibfnamefont {A.~M.}\ \bibnamefont
  {Stacy}},\ }\href@noop {} {\bibfield  {journal} {\bibinfo  {journal} {Journal
  of Solid State Chemistry}\ }\textbf {\bibinfo {volume} {54}},\ \bibinfo
  {pages} {438} (\bibinfo {year} {1984})}\BibitemShut {NoStop}%
\bibitem [{\citenamefont {Abdulsalam}\ and\ \citenamefont
  {Joubert}(2016)}]{abdulsalam2016optical}%
  \BibitemOpen
  \bibfield  {author} {\bibinfo {author} {\bibfnamefont {M.}~\bibnamefont
  {Abdulsalam}}\ and\ \bibinfo {author} {\bibfnamefont {D.~P.}\ \bibnamefont
  {Joubert}},\ }\href@noop {} {\bibfield  {journal} {\bibinfo  {journal}
  {physica status solidi (b)}\ }\textbf {\bibinfo {volume} {253}},\ \bibinfo
  {pages} {705} (\bibinfo {year} {2016})}\BibitemShut {NoStop}%
\bibitem [{\citenamefont {Wiegers}\ and\ \citenamefont
  {Meerschaut}(1992)}]{wiegers1992structures}%
  \BibitemOpen
  \bibfield  {author} {\bibinfo {author} {\bibfnamefont {G.}~\bibnamefont
  {Wiegers}}\ and\ \bibinfo {author} {\bibfnamefont {A.}~\bibnamefont
  {Meerschaut}},\ }\href@noop {} {\bibfield  {journal} {\bibinfo  {journal}
  {Journal of alloys and compounds}\ }\textbf {\bibinfo {volume} {178}},\
  \bibinfo {pages} {351} (\bibinfo {year} {1992})}\BibitemShut {NoStop}%
\bibitem [{\citenamefont {Suga}\ \emph {et~al.}(2015)\citenamefont {Suga},
  \citenamefont {Tusche}, \citenamefont {Matsushita}, \citenamefont {Ellguth},
  \citenamefont {Irizawa},\ and\ \citenamefont {Kirschner}}]{suga2015momentum}%
  \BibitemOpen
  \bibfield  {author} {\bibinfo {author} {\bibfnamefont {S.}~\bibnamefont
  {Suga}}, \bibinfo {author} {\bibfnamefont {C.}~\bibnamefont {Tusche}},
  \bibinfo {author} {\bibfnamefont {Y.-i.}\ \bibnamefont {Matsushita}},
  \bibinfo {author} {\bibfnamefont {M.}~\bibnamefont {Ellguth}}, \bibinfo
  {author} {\bibfnamefont {A.}~\bibnamefont {Irizawa}}, \ and\ \bibinfo
  {author} {\bibfnamefont {J.}~\bibnamefont {Kirschner}},\ }\href@noop {}
  {\bibfield  {journal} {\bibinfo  {journal} {New Journal of Physics}\ }\textbf
  {\bibinfo {volume} {17}},\ \bibinfo {pages} {083010} (\bibinfo {year}
  {2015})}\BibitemShut {NoStop}%
\bibitem [{\citenamefont {Ennaoui}\ \emph {et~al.}(1993)\citenamefont
  {Ennaoui}, \citenamefont {Fiechter}, \citenamefont {Pettenkofer},
  \citenamefont {Alonso-Vante}, \citenamefont {B{\"u}ker}, \citenamefont
  {Bronold}, \citenamefont {H{\"o}pfner},\ and\ \citenamefont
  {Tributsch}}]{ennaoui1993iron}%
  \BibitemOpen
  \bibfield  {author} {\bibinfo {author} {\bibfnamefont {A.}~\bibnamefont
  {Ennaoui}}, \bibinfo {author} {\bibfnamefont {S.}~\bibnamefont {Fiechter}},
  \bibinfo {author} {\bibfnamefont {C.}~\bibnamefont {Pettenkofer}}, \bibinfo
  {author} {\bibfnamefont {N.}~\bibnamefont {Alonso-Vante}}, \bibinfo {author}
  {\bibfnamefont {K.}~\bibnamefont {B{\"u}ker}}, \bibinfo {author}
  {\bibfnamefont {M.}~\bibnamefont {Bronold}}, \bibinfo {author} {\bibfnamefont
  {C.}~\bibnamefont {H{\"o}pfner}}, \ and\ \bibinfo {author} {\bibfnamefont
  {H.}~\bibnamefont {Tributsch}},\ }\href@noop {} {\bibfield  {journal}
  {\bibinfo  {journal} {Solar Energy Materials and Solar Cells}\ }\textbf
  {\bibinfo {volume} {29}},\ \bibinfo {pages} {289} (\bibinfo {year}
  {1993})}\BibitemShut {NoStop}%
\bibitem [{\citenamefont {Ishii}\ \emph {et~al.}(1999)\citenamefont {Ishii},
  \citenamefont {Murakami},\ and\ \citenamefont {Itoh}}]{ishii1999optical}%
  \BibitemOpen
  \bibfield  {author} {\bibinfo {author} {\bibfnamefont {Y.}~\bibnamefont
  {Ishii}}, \bibinfo {author} {\bibfnamefont {J.-i.}\ \bibnamefont {Murakami}},
  \ and\ \bibinfo {author} {\bibfnamefont {M.}~\bibnamefont {Itoh}},\
  }\href@noop {} {\bibfield  {journal} {\bibinfo  {journal} {Journal of the
  Physical Society of Japan}\ }\textbf {\bibinfo {volume} {68}},\ \bibinfo
  {pages} {696} (\bibinfo {year} {1999})}\BibitemShut {NoStop}%
\bibitem [{\citenamefont {Sommer}\ \emph {et~al.}(2012)\citenamefont {Sommer},
  \citenamefont {Kr{\"u}ger},\ and\ \citenamefont
  {Pollmann}}]{sommer2012quasiparticle}%
  \BibitemOpen
  \bibfield  {author} {\bibinfo {author} {\bibfnamefont {C.}~\bibnamefont
  {Sommer}}, \bibinfo {author} {\bibfnamefont {P.}~\bibnamefont {Kr{\"u}ger}},
  \ and\ \bibinfo {author} {\bibfnamefont {J.}~\bibnamefont {Pollmann}},\
  }\href@noop {} {\bibfield  {journal} {\bibinfo  {journal} {Physical Review
  B}\ }\textbf {\bibinfo {volume} {85}},\ \bibinfo {pages} {165119} (\bibinfo
  {year} {2012})}\BibitemShut {NoStop}%
\bibitem [{\citenamefont {Tsirelson}\ \emph {et~al.}(1998)\citenamefont
  {Tsirelson}, \citenamefont {Avilov}, \citenamefont {Abramov}, \citenamefont
  {Belokoneva}, \citenamefont {Kitaneh},\ and\ \citenamefont
  {Feil}}]{tsirelson1998x}%
  \BibitemOpen
  \bibfield  {author} {\bibinfo {author} {\bibfnamefont {V.}~\bibnamefont
  {Tsirelson}}, \bibinfo {author} {\bibfnamefont {A.}~\bibnamefont {Avilov}},
  \bibinfo {author} {\bibfnamefont {Y.~A.}\ \bibnamefont {Abramov}}, \bibinfo
  {author} {\bibfnamefont {E.}~\bibnamefont {Belokoneva}}, \bibinfo {author}
  {\bibfnamefont {R.}~\bibnamefont {Kitaneh}}, \ and\ \bibinfo {author}
  {\bibfnamefont {D.}~\bibnamefont {Feil}},\ }\href@noop {} {\bibfield
  {journal} {\bibinfo  {journal} {Acta Crystallographica Section B}\ }\textbf
  {\bibinfo {volume} {54}},\ \bibinfo {pages} {8} (\bibinfo {year}
  {1998})}\BibitemShut {NoStop}%
\bibitem [{\citenamefont {Shi}\ \emph {et~al.}(2014)\citenamefont {Shi},
  \citenamefont {Qin}, \citenamefont {Hu}, \citenamefont {Duan}, \citenamefont
  {Qu}, \citenamefont {Wu},\ and\ \citenamefont {Tang}}]{shi2014strain}%
  \BibitemOpen
  \bibfield  {author} {\bibinfo {author} {\bibfnamefont {L.}~\bibnamefont
  {Shi}}, \bibinfo {author} {\bibfnamefont {Y.}~\bibnamefont {Qin}}, \bibinfo
  {author} {\bibfnamefont {J.}~\bibnamefont {Hu}}, \bibinfo {author}
  {\bibfnamefont {Y.}~\bibnamefont {Duan}}, \bibinfo {author} {\bibfnamefont
  {L.}~\bibnamefont {Qu}}, \bibinfo {author} {\bibfnamefont {L.}~\bibnamefont
  {Wu}}, \ and\ \bibinfo {author} {\bibfnamefont {G.}~\bibnamefont {Tang}},\
  }\href@noop {} {\bibfield  {journal} {\bibinfo  {journal} {EPL (Europhysics
  Letters)}\ }\textbf {\bibinfo {volume} {106}},\ \bibinfo {pages} {57001}
  (\bibinfo {year} {2014})}\BibitemShut {NoStop}%
\bibitem [{\citenamefont {Prewitt}\ and\ \citenamefont
  {Shannon}(1968)}]{prewitt1968crystal}%
  \BibitemOpen
  \bibfield  {author} {\bibinfo {author} {\bibfnamefont {C.}~\bibnamefont
  {Prewitt}}\ and\ \bibinfo {author} {\bibfnamefont {R.}~\bibnamefont
  {Shannon}},\ }\href@noop {} {\bibfield  {journal} {\bibinfo  {journal} {Acta
  Crystallographica Section B: Structural Crystallography and Crystal
  Chemistry}\ }\textbf {\bibinfo {volume} {24}},\ \bibinfo {pages} {869}
  (\bibinfo {year} {1968})}\BibitemShut {NoStop}%
\bibitem [{\citenamefont {McCarthy}\ and\ \citenamefont
  {Welton}(1989)}]{mccarthy1989x}%
  \BibitemOpen
  \bibfield  {author} {\bibinfo {author} {\bibfnamefont {G.~J.}\ \bibnamefont
  {McCarthy}}\ and\ \bibinfo {author} {\bibfnamefont {J.~M.}\ \bibnamefont
  {Welton}},\ }\href@noop {} {\bibfield  {journal} {\bibinfo  {journal} {Powder
  Diffraction}\ }\textbf {\bibinfo {volume} {4}},\ \bibinfo {pages} {156}
  (\bibinfo {year} {1989})}\BibitemShut {NoStop}%
\bibitem [{\citenamefont {Janotti}\ and\ \citenamefont {Van~de
  Walle}(2011)}]{janotti2011lda+}%
  \BibitemOpen
  \bibfield  {author} {\bibinfo {author} {\bibfnamefont {A.}~\bibnamefont
  {Janotti}}\ and\ \bibinfo {author} {\bibfnamefont {C.~G.}\ \bibnamefont
  {Van~de Walle}},\ }\href@noop {} {\bibfield  {journal} {\bibinfo  {journal}
  {physica status solidi (b)}\ }\textbf {\bibinfo {volume} {248}},\ \bibinfo
  {pages} {799} (\bibinfo {year} {2011})}\BibitemShut {NoStop}%
\bibitem [{\citenamefont {Berger}\ \emph {et~al.}(2010)\citenamefont {Berger},
  \citenamefont {Reining},\ and\ \citenamefont {Sottile}}]{berger2010ab}%
  \BibitemOpen
  \bibfield  {author} {\bibinfo {author} {\bibfnamefont {J.}~\bibnamefont
  {Berger}}, \bibinfo {author} {\bibfnamefont {L.}~\bibnamefont {Reining}}, \
  and\ \bibinfo {author} {\bibfnamefont {F.}~\bibnamefont {Sottile}},\
  }\href@noop {} {\bibfield  {journal} {\bibinfo  {journal} {Physical Review
  B}\ }\textbf {\bibinfo {volume} {82}},\ \bibinfo {pages} {041103} (\bibinfo
  {year} {2010})}\BibitemShut {NoStop}%
\bibitem [{\citenamefont {Finger}\ and\ \citenamefont
  {Hazen}(1978)}]{finger1978crystal}%
  \BibitemOpen
  \bibfield  {author} {\bibinfo {author} {\bibfnamefont {L.~W.}\ \bibnamefont
  {Finger}}\ and\ \bibinfo {author} {\bibfnamefont {R.~M.}\ \bibnamefont
  {Hazen}},\ }\href@noop {} {\bibfield  {journal} {\bibinfo  {journal} {Journal
  of Applied Physics}\ }\textbf {\bibinfo {volume} {49}},\ \bibinfo {pages}
  {5823} (\bibinfo {year} {1978})}\BibitemShut {NoStop}%
\bibitem [{\citenamefont {Robertson}(2000)}]{robertson2000band}%
  \BibitemOpen
  \bibfield  {author} {\bibinfo {author} {\bibfnamefont {J.}~\bibnamefont
  {Robertson}},\ }\href@noop {} {\bibfield  {journal} {\bibinfo  {journal}
  {Journal of Vacuum Science \& Technology B: Microelectronics and Nanometer
  Structures Processing, Measurement, and Phenomena}\ }\textbf {\bibinfo
  {volume} {18}},\ \bibinfo {pages} {1785} (\bibinfo {year}
  {2000})}\BibitemShut {NoStop}%
\bibitem [{\citenamefont {Pluth}\ \emph {et~al.}(1985)\citenamefont {Pluth},
  \citenamefont {Smith},\ and\ \citenamefont {Faber~Jr}}]{pluth1985crystal}%
  \BibitemOpen
  \bibfield  {author} {\bibinfo {author} {\bibfnamefont {J.}~\bibnamefont
  {Pluth}}, \bibinfo {author} {\bibfnamefont {J.}~\bibnamefont {Smith}}, \ and\
  \bibinfo {author} {\bibfnamefont {J.}~\bibnamefont {Faber~Jr}},\ }\href@noop
  {} {\bibfield  {journal} {\bibinfo  {journal} {Journal of Applied Physics}\
  }\textbf {\bibinfo {volume} {57}},\ \bibinfo {pages} {1045} (\bibinfo {year}
  {1985})}\BibitemShut {NoStop}%
\bibitem [{\citenamefont {Weinberg}\ \emph {et~al.}(1979)\citenamefont
  {Weinberg}, \citenamefont {Rubloff},\ and\ \citenamefont
  {Bassous}}]{weinberg1979transmission}%
  \BibitemOpen
  \bibfield  {author} {\bibinfo {author} {\bibfnamefont {Z.}~\bibnamefont
  {Weinberg}}, \bibinfo {author} {\bibfnamefont {G.}~\bibnamefont {Rubloff}}, \
  and\ \bibinfo {author} {\bibfnamefont {E.}~\bibnamefont {Bassous}},\
  }\href@noop {} {\bibfield  {journal} {\bibinfo  {journal} {Physical Review
  B}\ }\textbf {\bibinfo {volume} {19}},\ \bibinfo {pages} {3107} (\bibinfo
  {year} {1979})}\BibitemShut {NoStop}%
\bibitem [{\citenamefont {Varley}\ \emph {et~al.}(2012)\citenamefont {Varley},
  \citenamefont {Janotti}, \citenamefont {Franchini},\ and\ \citenamefont
  {Van~de Walle}}]{varley2012role}%
  \BibitemOpen
  \bibfield  {author} {\bibinfo {author} {\bibfnamefont {J.}~\bibnamefont
  {Varley}}, \bibinfo {author} {\bibfnamefont {A.}~\bibnamefont {Janotti}},
  \bibinfo {author} {\bibfnamefont {C.}~\bibnamefont {Franchini}}, \ and\
  \bibinfo {author} {\bibfnamefont {C.~G.}\ \bibnamefont {Van~de Walle}},\
  }\href@noop {} {\bibfield  {journal} {\bibinfo  {journal} {Physical Review
  B}\ }\textbf {\bibinfo {volume} {85}},\ \bibinfo {pages} {081109} (\bibinfo
  {year} {2012})}\BibitemShut {NoStop}%
\bibitem [{\citenamefont {Kresse}\ \emph {et~al.}(2012)\citenamefont {Kresse},
  \citenamefont {Marsman}, \citenamefont {Hintzsche},\ and\ \citenamefont
  {Flage-Larsen}}]{kresse2012optical}%
  \BibitemOpen
  \bibfield  {author} {\bibinfo {author} {\bibfnamefont {G.}~\bibnamefont
  {Kresse}}, \bibinfo {author} {\bibfnamefont {M.}~\bibnamefont {Marsman}},
  \bibinfo {author} {\bibfnamefont {L.}~\bibnamefont {Hintzsche}}, \ and\
  \bibinfo {author} {\bibfnamefont {E.}~\bibnamefont {Flage-Larsen}},\
  }\href@noop {} {\bibfield  {journal} {\bibinfo  {journal} {Physical Review
  B}\ }\textbf {\bibinfo {volume} {85}},\ \bibinfo {pages} {045205} (\bibinfo
  {year} {2012})}\BibitemShut {NoStop}%
\bibitem [{\citenamefont {Bernal}\ \emph {et~al.}(1935)\citenamefont {Bernal},
  \citenamefont {Djatlowa}, \citenamefont {Kasarnowsky}, \citenamefont
  {Reichstein},\ and\ \citenamefont {Ward}}]{bernal1935structure}%
  \BibitemOpen
  \bibfield  {author} {\bibinfo {author} {\bibfnamefont {J.}~\bibnamefont
  {Bernal}}, \bibinfo {author} {\bibfnamefont {E.}~\bibnamefont {Djatlowa}},
  \bibinfo {author} {\bibfnamefont {I.}~\bibnamefont {Kasarnowsky}}, \bibinfo
  {author} {\bibfnamefont {S.}~\bibnamefont {Reichstein}}, \ and\ \bibinfo
  {author} {\bibfnamefont {A.}~\bibnamefont {Ward}},\ }\href@noop {} {\bibfield
   {journal} {\bibinfo  {journal} {Zeitschrift f{\"u}r
  Kristallographie-Crystalline Materials}\ }\textbf {\bibinfo {volume} {92}},\
  \bibinfo {pages} {344} (\bibinfo {year} {1935})}\BibitemShut {NoStop}%
\bibitem [{\citenamefont {Rao}\ and\ \citenamefont
  {Kearney}(1979)}]{rao1979logarithmic}%
  \BibitemOpen
  \bibfield  {author} {\bibinfo {author} {\bibfnamefont {A.}~\bibnamefont
  {Rao}}\ and\ \bibinfo {author} {\bibfnamefont {R.}~\bibnamefont {Kearney}},\
  }\href@noop {} {\bibfield  {journal} {\bibinfo  {journal} {physica status
  solidi (b)}\ }\textbf {\bibinfo {volume} {95}},\ \bibinfo {pages} {243}
  (\bibinfo {year} {1979})}\BibitemShut {NoStop}%
\bibitem [{\citenamefont {Klein}\ \emph {et~al.}(1998)\citenamefont {Klein},
  \citenamefont {Armbruster},\ and\ \citenamefont
  {Jansen}}]{klein1998synthesis}%
  \BibitemOpen
  \bibfield  {author} {\bibinfo {author} {\bibfnamefont {W.}~\bibnamefont
  {Klein}}, \bibinfo {author} {\bibfnamefont {K.}~\bibnamefont {Armbruster}}, \
  and\ \bibinfo {author} {\bibfnamefont {M.}~\bibnamefont {Jansen}},\
  }\href@noop {} {\bibfield  {journal} {\bibinfo  {journal} {Chemical
  Communications}\ ,\ \bibinfo {pages} {707}} (\bibinfo {year}
  {1998})}\BibitemShut {NoStop}%
\bibitem [{\citenamefont {Kucharczyk}\ and\ \citenamefont
  {Niklewski}(1979)}]{kucharczyk1979accurate}%
  \BibitemOpen
  \bibfield  {author} {\bibinfo {author} {\bibfnamefont {D.}~\bibnamefont
  {Kucharczyk}}\ and\ \bibinfo {author} {\bibfnamefont {T.}~\bibnamefont
  {Niklewski}},\ }\href@noop {} {\bibfield  {journal} {\bibinfo  {journal}
  {Journal of Applied Crystallography}\ }\textbf {\bibinfo {volume} {12}},\
  \bibinfo {pages} {370} (\bibinfo {year} {1979})}\BibitemShut {NoStop}%
\bibitem [{\citenamefont {Banfield}\ \emph {et~al.}(1991)\citenamefont
  {Banfield}, \citenamefont {Veblen},\ and\ \citenamefont
  {Smith}}]{banfield1991identification}%
  \BibitemOpen
  \bibfield  {author} {\bibinfo {author} {\bibfnamefont {J.~F.}\ \bibnamefont
  {Banfield}}, \bibinfo {author} {\bibfnamefont {D.~R.}\ \bibnamefont
  {Veblen}}, \ and\ \bibinfo {author} {\bibfnamefont {D.~J.}\ \bibnamefont
  {Smith}},\ }\href@noop {} {\bibfield  {journal} {\bibinfo  {journal}
  {American Mineralogist}\ }\textbf {\bibinfo {volume} {76}},\ \bibinfo {pages}
  {343} (\bibinfo {year} {1991})}\BibitemShut {NoStop}%
\bibitem [{\citenamefont {Berger}\ and\ \citenamefont
  {Neaton}(2012)}]{berger2012computational}%
  \BibitemOpen
  \bibfield  {author} {\bibinfo {author} {\bibfnamefont {R.~F.}\ \bibnamefont
  {Berger}}\ and\ \bibinfo {author} {\bibfnamefont {J.~B.}\ \bibnamefont
  {Neaton}},\ }\href@noop {} {\bibfield  {journal} {\bibinfo  {journal}
  {Physical Review B}\ }\textbf {\bibinfo {volume} {86}},\ \bibinfo {pages}
  {165211} (\bibinfo {year} {2012})}\BibitemShut {NoStop}%
\bibitem [{\citenamefont {SASAKI}\ \emph {et~al.}(1979)\citenamefont {SASAKI},
  \citenamefont {FUJINO},\ and\ \citenamefont {TAK{\'E}UCHI}}]{sasaki1979x}%
  \BibitemOpen
  \bibfield  {author} {\bibinfo {author} {\bibfnamefont {S.}~\bibnamefont
  {SASAKI}}, \bibinfo {author} {\bibfnamefont {K.}~\bibnamefont {FUJINO}}, \
  and\ \bibinfo {author} {\bibfnamefont {Y.}~\bibnamefont {TAK{\'E}UCHI}},\
  }\href@noop {} {\bibfield  {journal} {\bibinfo  {journal} {Proceedings of the
  Japan Academy, Series B}\ }\textbf {\bibinfo {volume} {55}},\ \bibinfo
  {pages} {43} (\bibinfo {year} {1979})}\BibitemShut {NoStop}%
\bibitem [{\citenamefont {Gillen}\ and\ \citenamefont
  {Robertson}(2013)}]{gillen2013accurate}%
  \BibitemOpen
  \bibfield  {author} {\bibinfo {author} {\bibfnamefont {R.}~\bibnamefont
  {Gillen}}\ and\ \bibinfo {author} {\bibfnamefont {J.}~\bibnamefont
  {Robertson}},\ }\href@noop {} {\bibfield  {journal} {\bibinfo  {journal}
  {Journal of Physics: Condensed Matter}\ }\textbf {\bibinfo {volume} {25}},\
  \bibinfo {pages} {165502} (\bibinfo {year} {2013})}\BibitemShut {NoStop}%
\bibitem [{\citenamefont {Toroker}\ \emph {et~al.}(2011)\citenamefont
  {Toroker}, \citenamefont {Kanan}, \citenamefont {Alidoust}, \citenamefont
  {Isseroff}, \citenamefont {Liao},\ and\ \citenamefont
  {Carter}}]{toroker2011first}%
  \BibitemOpen
  \bibfield  {author} {\bibinfo {author} {\bibfnamefont {M.~C.}\ \bibnamefont
  {Toroker}}, \bibinfo {author} {\bibfnamefont {D.~K.}\ \bibnamefont {Kanan}},
  \bibinfo {author} {\bibfnamefont {N.}~\bibnamefont {Alidoust}}, \bibinfo
  {author} {\bibfnamefont {L.~Y.}\ \bibnamefont {Isseroff}}, \bibinfo {author}
  {\bibfnamefont {P.}~\bibnamefont {Liao}}, \ and\ \bibinfo {author}
  {\bibfnamefont {E.~A.}\ \bibnamefont {Carter}},\ }\href@noop {} {\bibfield
  {journal} {\bibinfo  {journal} {Physical Chemistry Chemical Physics}\
  }\textbf {\bibinfo {volume} {13}},\ \bibinfo {pages} {16644} (\bibinfo {year}
  {2011})}\BibitemShut {NoStop}%
\bibitem [{\citenamefont {Riefer}\ \emph {et~al.}(2011)\citenamefont {Riefer},
  \citenamefont {Fuchs}, \citenamefont {R{\"o}dl}, \citenamefont {Schleife},
  \citenamefont {Bechstedt},\ and\ \citenamefont
  {Goldhahn}}]{riefer2011interplay}%
  \BibitemOpen
  \bibfield  {author} {\bibinfo {author} {\bibfnamefont {A.}~\bibnamefont
  {Riefer}}, \bibinfo {author} {\bibfnamefont {F.}~\bibnamefont {Fuchs}},
  \bibinfo {author} {\bibfnamefont {C.}~\bibnamefont {R{\"o}dl}}, \bibinfo
  {author} {\bibfnamefont {A.}~\bibnamefont {Schleife}}, \bibinfo {author}
  {\bibfnamefont {F.}~\bibnamefont {Bechstedt}}, \ and\ \bibinfo {author}
  {\bibfnamefont {R.}~\bibnamefont {Goldhahn}},\ }\href@noop {} {\bibfield
  {journal} {\bibinfo  {journal} {Physical Review B}\ }\textbf {\bibinfo
  {volume} {84}},\ \bibinfo {pages} {075218} (\bibinfo {year}
  {2011})}\BibitemShut {NoStop}%
\bibitem [{\citenamefont {Bates}\ \emph {et~al.}(1962)\citenamefont {Bates},
  \citenamefont {White},\ and\ \citenamefont {Roy}}]{bates1962new}%
  \BibitemOpen
  \bibfield  {author} {\bibinfo {author} {\bibfnamefont {C.~H.}\ \bibnamefont
  {Bates}}, \bibinfo {author} {\bibfnamefont {W.~B.}\ \bibnamefont {White}}, \
  and\ \bibinfo {author} {\bibfnamefont {R.}~\bibnamefont {Roy}},\ }\href@noop
  {} {\bibfield  {journal} {\bibinfo  {journal} {Science}\ }\textbf {\bibinfo
  {volume} {137}},\ \bibinfo {pages} {993} (\bibinfo {year}
  {1962})}\BibitemShut {NoStop}%
\bibitem [{\citenamefont {Primak}\ \emph {et~al.}(1948)\citenamefont {Primak},
  \citenamefont {Kaufman},\ and\ \citenamefont {Ward}}]{primak1948x}%
  \BibitemOpen
  \bibfield  {author} {\bibinfo {author} {\bibfnamefont {W.}~\bibnamefont
  {Primak}}, \bibinfo {author} {\bibfnamefont {H.}~\bibnamefont {Kaufman}}, \
  and\ \bibinfo {author} {\bibfnamefont {R.}~\bibnamefont {Ward}},\ }\href@noop
  {} {\bibfield  {journal} {\bibinfo  {journal} {Journal of the American
  Chemical Society}\ }\textbf {\bibinfo {volume} {70}},\ \bibinfo {pages}
  {2043} (\bibinfo {year} {1948})}\BibitemShut {NoStop}%
\bibitem [{\citenamefont {Bajdich}\ \emph {et~al.}(2015)\citenamefont
  {Bajdich}, \citenamefont {N{\o}rskov},\ and\ \citenamefont
  {Vojvodic}}]{bajdich2015surface}%
  \BibitemOpen
  \bibfield  {author} {\bibinfo {author} {\bibfnamefont {M.}~\bibnamefont
  {Bajdich}}, \bibinfo {author} {\bibfnamefont {J.~K.}\ \bibnamefont
  {N{\o}rskov}}, \ and\ \bibinfo {author} {\bibfnamefont {A.}~\bibnamefont
  {Vojvodic}},\ }\href@noop {} {\bibfield  {journal} {\bibinfo  {journal}
  {Physical Review B}\ }\textbf {\bibinfo {volume} {91}},\ \bibinfo {pages}
  {155401} (\bibinfo {year} {2015})}\BibitemShut {NoStop}%
\bibitem [{\citenamefont {Zachariasen}(1949)}]{zachariasen1949crystal}%
  \BibitemOpen
  \bibfield  {author} {\bibinfo {author} {\bibfnamefont {W.}~\bibnamefont
  {Zachariasen}},\ }\href@noop {} {\bibfield  {journal} {\bibinfo  {journal}
  {Acta Crystallographica}\ }\textbf {\bibinfo {volume} {2}},\ \bibinfo {pages}
  {388} (\bibinfo {year} {1949})}\BibitemShut {NoStop}%
\bibitem [{\citenamefont {Lacorre}\ \emph {et~al.}(1991)\citenamefont
  {Lacorre}, \citenamefont {Torrance}, \citenamefont {Pannetier}, \citenamefont
  {Nazzal}, \citenamefont {Wang},\ and\ \citenamefont
  {Huang}}]{lacorre1991synthesis}%
  \BibitemOpen
  \bibfield  {author} {\bibinfo {author} {\bibfnamefont {P.}~\bibnamefont
  {Lacorre}}, \bibinfo {author} {\bibfnamefont {J.}~\bibnamefont {Torrance}},
  \bibinfo {author} {\bibfnamefont {J.}~\bibnamefont {Pannetier}}, \bibinfo
  {author} {\bibfnamefont {A.}~\bibnamefont {Nazzal}}, \bibinfo {author}
  {\bibfnamefont {P.}~\bibnamefont {Wang}}, \ and\ \bibinfo {author}
  {\bibfnamefont {T.}~\bibnamefont {Huang}},\ }\href@noop {} {\bibfield
  {journal} {\bibinfo  {journal} {Journal of Solid State Chemistry}\ }\textbf
  {\bibinfo {volume} {91}},\ \bibinfo {pages} {225} (\bibinfo {year}
  {1991})}\BibitemShut {NoStop}%
\bibitem [{\citenamefont {McCarthy}\ \emph {et~al.}(1969)\citenamefont
  {McCarthy}, \citenamefont {White},\ and\ \citenamefont
  {Roy}}]{mccarthy1969preparation}%
  \BibitemOpen
  \bibfield  {author} {\bibinfo {author} {\bibfnamefont {G.~J.}\ \bibnamefont
  {McCarthy}}, \bibinfo {author} {\bibfnamefont {W.~B.}\ \bibnamefont {White}},
  \ and\ \bibinfo {author} {\bibfnamefont {R.}~\bibnamefont {Roy}},\
  }\href@noop {} {\bibfield  {journal} {\bibinfo  {journal} {Materials Research
  Bulletin}\ }\textbf {\bibinfo {volume} {4}},\ \bibinfo {pages} {251}
  (\bibinfo {year} {1969})}\BibitemShut {NoStop}%
\bibitem [{\citenamefont {Jauch}\ and\ \citenamefont
  {Palmer}(1999)}]{jauch1999anomalous}%
  \BibitemOpen
  \bibfield  {author} {\bibinfo {author} {\bibfnamefont {W.}~\bibnamefont
  {Jauch}}\ and\ \bibinfo {author} {\bibfnamefont {A.}~\bibnamefont {Palmer}},\
  }\href@noop {} {\bibfield  {journal} {\bibinfo  {journal} {Physical Review
  B}\ }\textbf {\bibinfo {volume} {60}},\ \bibinfo {pages} {2961} (\bibinfo
  {year} {1999})}\BibitemShut {NoStop}%
\bibitem [{\citenamefont {Long}\ \emph {et~al.}(2013)\citenamefont {Long},
  \citenamefont {Yang},\ and\ \citenamefont {Wei}}]{long2013lattice}%
  \BibitemOpen
  \bibfield  {author} {\bibinfo {author} {\bibfnamefont {J.}~\bibnamefont
  {Long}}, \bibinfo {author} {\bibfnamefont {L.}~\bibnamefont {Yang}}, \ and\
  \bibinfo {author} {\bibfnamefont {X.}~\bibnamefont {Wei}},\ }\href@noop {}
  {\bibfield  {journal} {\bibinfo  {journal} {Journal of Alloys and Compounds}\
  }\textbf {\bibinfo {volume} {549}},\ \bibinfo {pages} {336} (\bibinfo {year}
  {2013})}\BibitemShut {NoStop}%
\bibitem [{\citenamefont {Van~Benthem}\ \emph {et~al.}(2001)\citenamefont
  {Van~Benthem}, \citenamefont {Els{\"a}sser},\ and\ \citenamefont
  {French}}]{van2001bulk}%
  \BibitemOpen
  \bibfield  {author} {\bibinfo {author} {\bibfnamefont {K.}~\bibnamefont
  {Van~Benthem}}, \bibinfo {author} {\bibfnamefont {C.}~\bibnamefont
  {Els{\"a}sser}}, \ and\ \bibinfo {author} {\bibfnamefont {R.}~\bibnamefont
  {French}},\ }\href@noop {} {\bibfield  {journal} {\bibinfo  {journal}
  {Journal of applied physics}\ }\textbf {\bibinfo {volume} {90}},\ \bibinfo
  {pages} {6156} (\bibinfo {year} {2001})}\BibitemShut {NoStop}%
\bibitem [{\citenamefont {Thornton}\ \emph {et~al.}(1986)\citenamefont
  {Thornton}, \citenamefont {Tofield},\ and\ \citenamefont
  {Hewat}}]{thornton1986neutron}%
  \BibitemOpen
  \bibfield  {author} {\bibinfo {author} {\bibfnamefont {G.}~\bibnamefont
  {Thornton}}, \bibinfo {author} {\bibfnamefont {B.}~\bibnamefont {Tofield}}, \
  and\ \bibinfo {author} {\bibfnamefont {A.}~\bibnamefont {Hewat}},\
  }\href@noop {} {\bibfield  {journal} {\bibinfo  {journal} {Journal of Solid
  State Chemistry}\ }\textbf {\bibinfo {volume} {61}},\ \bibinfo {pages} {301}
  (\bibinfo {year} {1986})}\BibitemShut {NoStop}%
\bibitem [{\citenamefont {Chainani}\ \emph {et~al.}(1992)\citenamefont
  {Chainani}, \citenamefont {Mathew},\ and\ \citenamefont
  {Sarma}}]{chainani1992electron}%
  \BibitemOpen
  \bibfield  {author} {\bibinfo {author} {\bibfnamefont {A.}~\bibnamefont
  {Chainani}}, \bibinfo {author} {\bibfnamefont {M.}~\bibnamefont {Mathew}}, \
  and\ \bibinfo {author} {\bibfnamefont {D.}~\bibnamefont {Sarma}},\
  }\href@noop {} {\bibfield  {journal} {\bibinfo  {journal} {Physical Review
  B}\ }\textbf {\bibinfo {volume} {46}},\ \bibinfo {pages} {9976} (\bibinfo
  {year} {1992})}\BibitemShut {NoStop}%
\bibitem [{\citenamefont {Zhang}\ \emph {et~al.}(2014)\citenamefont {Zhang},
  \citenamefont {Gang},\ and\ \citenamefont {Wan}}]{zhang2014density}%
  \BibitemOpen
  \bibfield  {author} {\bibinfo {author} {\bibfnamefont {X.-b.}\ \bibnamefont
  {Zhang}}, \bibinfo {author} {\bibfnamefont {F.}~\bibnamefont {Gang}}, \ and\
  \bibinfo {author} {\bibfnamefont {H.-l.}\ \bibnamefont {Wan}},\ }\href@noop
  {} {\bibfield  {journal} {\bibinfo  {journal} {Chinese Journal of Chemical
  Physics}\ }\textbf {\bibinfo {volume} {27}},\ \bibinfo {pages} {274}
  (\bibinfo {year} {2014})}\BibitemShut {NoStop}%
\bibitem [{\citenamefont {Garcia-Munoz}\ \emph {et~al.}(1992)\citenamefont
  {Garcia-Munoz}, \citenamefont {Rodriguez-Carvajal}, \citenamefont {Lacorre},\
  and\ \citenamefont {Torrance}}]{garcia1992neutron}%
  \BibitemOpen
  \bibfield  {author} {\bibinfo {author} {\bibfnamefont {J.}~\bibnamefont
  {Garcia-Munoz}}, \bibinfo {author} {\bibfnamefont {J.}~\bibnamefont
  {Rodriguez-Carvajal}}, \bibinfo {author} {\bibfnamefont {P.}~\bibnamefont
  {Lacorre}}, \ and\ \bibinfo {author} {\bibfnamefont {J.}~\bibnamefont
  {Torrance}},\ }\href@noop {} {\bibfield  {journal} {\bibinfo  {journal}
  {Physical review B}\ }\textbf {\bibinfo {volume} {46}},\ \bibinfo {pages}
  {4414} (\bibinfo {year} {1992})}\BibitemShut {NoStop}%
\bibitem [{\citenamefont {R{\"u}egg}\ \emph {et~al.}(2012)\citenamefont
  {R{\"u}egg}, \citenamefont {Mitra}, \citenamefont {Demkov},\ and\
  \citenamefont {Fiete}}]{ruegg2012electronic}%
  \BibitemOpen
  \bibfield  {author} {\bibinfo {author} {\bibfnamefont {A.}~\bibnamefont
  {R{\"u}egg}}, \bibinfo {author} {\bibfnamefont {C.}~\bibnamefont {Mitra}},
  \bibinfo {author} {\bibfnamefont {A.~A.}\ \bibnamefont {Demkov}}, \ and\
  \bibinfo {author} {\bibfnamefont {G.~A.}\ \bibnamefont {Fiete}},\ }\href@noop
  {} {\bibfield  {journal} {\bibinfo  {journal} {Physical Review B}\ }\textbf
  {\bibinfo {volume} {85}},\ \bibinfo {pages} {245131} (\bibinfo {year}
  {2012})}\BibitemShut {NoStop}%
\bibitem [{\citenamefont {v.~N{\'a}ray-Szab{\'o}}(1943)}]{v1943strukturen}%
  \BibitemOpen
  \bibfield  {author} {\bibinfo {author} {\bibfnamefont {S.}~\bibnamefont
  {v.~N{\'a}ray-Szab{\'o}}},\ }\href@noop {} {\bibfield  {journal} {\bibinfo
  {journal} {Naturwissenschaften}\ }\textbf {\bibinfo {volume} {31}},\ \bibinfo
  {pages} {466} (\bibinfo {year} {1943})}\BibitemShut {NoStop}%
\bibitem [{\citenamefont {Kagomiya}\ \emph {et~al.}(2002)\citenamefont
  {Kagomiya}, \citenamefont {Kohn},\ and\ \citenamefont
  {Uchiyama}}]{kagomiya2002structure}%
  \BibitemOpen
  \bibfield  {author} {\bibinfo {author} {\bibfnamefont {I.}~\bibnamefont
  {Kagomiya}}, \bibinfo {author} {\bibfnamefont {K.}~\bibnamefont {Kohn}}, \
  and\ \bibinfo {author} {\bibfnamefont {T.}~\bibnamefont {Uchiyama}},\
  }\href@noop {} {\bibfield  {journal} {\bibinfo  {journal} {Ferroelectrics}\
  }\textbf {\bibinfo {volume} {280}},\ \bibinfo {pages} {131} (\bibinfo {year}
  {2002})}\BibitemShut {NoStop}%
\bibitem [{\citenamefont {Casalot}\ \emph {et~al.}(1971)\citenamefont
  {Casalot}, \citenamefont {Dougier},\ and\ \citenamefont
  {Hagenmuller}}]{casalot1971evolution}%
  \BibitemOpen
  \bibfield  {author} {\bibinfo {author} {\bibfnamefont {A.}~\bibnamefont
  {Casalot}}, \bibinfo {author} {\bibfnamefont {P.}~\bibnamefont {Dougier}}, \
  and\ \bibinfo {author} {\bibfnamefont {P.}~\bibnamefont {Hagenmuller}},\
  }\href@noop {} {\bibfield  {journal} {\bibinfo  {journal} {Journal of Physics
  and Chemistry of Solids}\ }\textbf {\bibinfo {volume} {32}},\ \bibinfo
  {pages} {407} (\bibinfo {year} {1971})}\BibitemShut {NoStop}%
\bibitem [{\citenamefont {Ferrari}\ and\ \citenamefont
  {Bocchi}(2008)}]{ferrari2008strain}%
  \BibitemOpen
  \bibfield  {author} {\bibinfo {author} {\bibfnamefont {C.}~\bibnamefont
  {Ferrari}}\ and\ \bibinfo {author} {\bibfnamefont {C.}~\bibnamefont
  {Bocchi}},\ }in\ \href@noop {} {\emph {\bibinfo {booktitle} {Characterization
  of Semiconductor Heterostructures and Nanostructures}}}\ (\bibinfo
  {publisher} {Elsevier},\ \bibinfo {year} {2008})\ pp.\ \bibinfo {pages}
  {93--132}\BibitemShut {NoStop}%
\bibitem [{\citenamefont {Soref}(1992)}]{soref1992electro}%
  \BibitemOpen
  \bibfield  {author} {\bibinfo {author} {\bibfnamefont {R.}~\bibnamefont
  {Soref}},\ }\href@noop {} {\bibfield  {journal} {\bibinfo  {journal} {Journal
  of applied physics}\ }\textbf {\bibinfo {volume} {72}},\ \bibinfo {pages}
  {626} (\bibinfo {year} {1992})}\BibitemShut {NoStop}%
\bibitem [{\citenamefont {Hussain}\ \emph {et~al.}(2015)\citenamefont
  {Hussain}, \citenamefont {Wehbe},\ and\ \citenamefont
  {Hussain}}]{hussain2015sisn}%
  \BibitemOpen
  \bibfield  {author} {\bibinfo {author} {\bibfnamefont {A.~M.}\ \bibnamefont
  {Hussain}}, \bibinfo {author} {\bibfnamefont {N.}~\bibnamefont {Wehbe}}, \
  and\ \bibinfo {author} {\bibfnamefont {M.~M.}\ \bibnamefont {Hussain}},\
  }\href@noop {} {\bibfield  {journal} {\bibinfo  {journal} {Applied Physics
  Letters}\ }\textbf {\bibinfo {volume} {107}},\ \bibinfo {pages} {082111}
  (\bibinfo {year} {2015})}\BibitemShut {NoStop}%
\bibitem [{\citenamefont {Agostini}\ and\ \citenamefont
  {Lamberti}(2011)}]{agostini2011characterization}%
  \BibitemOpen
  \bibfield  {author} {\bibinfo {author} {\bibfnamefont {G.}~\bibnamefont
  {Agostini}}\ and\ \bibinfo {author} {\bibfnamefont {C.}~\bibnamefont
  {Lamberti}},\ }\href@noop {} {\emph {\bibinfo {title} {Characterization of
  semiconductor heterostructures and nanostructures}}}\ (\bibinfo  {publisher}
  {Elsevier},\ \bibinfo {year} {2011})\BibitemShut {NoStop}%
\bibitem [{\citenamefont {Anzin}\ \emph
  {et~al.}(1977{\natexlab{b}})\citenamefont {Anzin}, \citenamefont {Eremets},
  \citenamefont {Kosichkin}, \citenamefont {Nadezhdinskii},\ and\ \citenamefont
  {Shirokov}}]{Te_exp_gap}%
  \BibitemOpen
  \bibfield  {author} {\bibinfo {author} {\bibfnamefont {V.}~\bibnamefont
  {Anzin}}, \bibinfo {author} {\bibfnamefont {M.}~\bibnamefont {Eremets}},
  \bibinfo {author} {\bibfnamefont {Y.~V.}\ \bibnamefont {Kosichkin}}, \bibinfo
  {author} {\bibfnamefont {A.}~\bibnamefont {Nadezhdinskii}}, \ and\ \bibinfo
  {author} {\bibfnamefont {A.}~\bibnamefont {Shirokov}},\ }\href@noop {}
  {\bibfield  {journal} {\bibinfo  {journal} {physica status solidi (a)}\
  }\textbf {\bibinfo {volume} {42}},\ \bibinfo {pages} {385} (\bibinfo {year}
  {1977}{\natexlab{b}})}\BibitemShut {NoStop}%
\bibitem [{\citenamefont {Yi}\ \emph {et~al.}(2018{\natexlab{b}})\citenamefont
  {Yi}, \citenamefont {Zhu}, \citenamefont {Cai}, \citenamefont {Jia},\ and\
  \citenamefont {Cho}}]{te_vdw}%
  \BibitemOpen
  \bibfield  {author} {\bibinfo {author} {\bibfnamefont {S.}~\bibnamefont
  {Yi}}, \bibinfo {author} {\bibfnamefont {Z.}~\bibnamefont {Zhu}}, \bibinfo
  {author} {\bibfnamefont {X.}~\bibnamefont {Cai}}, \bibinfo {author}
  {\bibfnamefont {Y.}~\bibnamefont {Jia}}, \ and\ \bibinfo {author}
  {\bibfnamefont {J.-H.}\ \bibnamefont {Cho}},\ }\href@noop {} {\bibfield
  {journal} {\bibinfo  {journal} {Inorganic chemistry}\ }\textbf {\bibinfo
  {volume} {57}},\ \bibinfo {pages} {5083} (\bibinfo {year}
  {2018}{\natexlab{b}})}\BibitemShut {NoStop}%
\bibitem [{\citenamefont {Solozhenko}\ \emph {et~al.}(1995)\citenamefont
  {Solozhenko}, \citenamefont {Will},\ and\ \citenamefont
  {Elf}}]{solozhenko1995isothermal}%
  \BibitemOpen
  \bibfield  {author} {\bibinfo {author} {\bibfnamefont {V.}~\bibnamefont
  {Solozhenko}}, \bibinfo {author} {\bibfnamefont {G.}~\bibnamefont {Will}}, \
  and\ \bibinfo {author} {\bibfnamefont {F.}~\bibnamefont {Elf}},\ }\href@noop
  {} {\bibfield  {journal} {\bibinfo  {journal} {Solid state communications}\
  }\textbf {\bibinfo {volume} {96}},\ \bibinfo {pages} {1} (\bibinfo {year}
  {1995})}\BibitemShut {NoStop}%
\bibitem [{\citenamefont {Greenaway}\ and\ \citenamefont
  {Nitsche}(1965)}]{tis2_hfse2_abc}%
  \BibitemOpen
  \bibfield  {author} {\bibinfo {author} {\bibfnamefont {D.~L.}\ \bibnamefont
  {Greenaway}}\ and\ \bibinfo {author} {\bibfnamefont {R.}~\bibnamefont
  {Nitsche}},\ }\href@noop {} {\bibfield  {journal} {\bibinfo  {journal}
  {Journal of Physics and Chemistry of Solids}\ }\textbf {\bibinfo {volume}
  {26}},\ \bibinfo {pages} {1445} (\bibinfo {year} {1965})}\BibitemShut
  {NoStop}%
\bibitem [{\citenamefont {Bronsema}\ \emph
  {et~al.}(1986{\natexlab{b}})\citenamefont {Bronsema}, \citenamefont
  {De~Boer},\ and\ \citenamefont {Jellinek}}]{mos2_mose2_abc}%
  \BibitemOpen
  \bibfield  {author} {\bibinfo {author} {\bibfnamefont {K.}~\bibnamefont
  {Bronsema}}, \bibinfo {author} {\bibfnamefont {J.}~\bibnamefont {De~Boer}}, \
  and\ \bibinfo {author} {\bibfnamefont {F.}~\bibnamefont {Jellinek}},\
  }\href@noop {} {\bibfield  {journal} {\bibinfo  {journal} {Zeitschrift
  f{\"u}r anorganische und allgemeine Chemie}\ }\textbf {\bibinfo {volume}
  {540}},\ \bibinfo {pages} {15} (\bibinfo {year}
  {1986}{\natexlab{b}})}\BibitemShut {NoStop}%
\bibitem [{\citenamefont {Lejaeghere}\ \emph {et~al.}(2016)\citenamefont
  {Lejaeghere}, \citenamefont {Bihlmayer}, \citenamefont {Bj{\"o}rkman},
  \citenamefont {Blaha}, \citenamefont {Bl{\"u}gel}, \citenamefont {Blum},
  \citenamefont {Caliste}, \citenamefont {Castelli}, \citenamefont {Clark},
  \citenamefont {Dal~Corso}, \citenamefont {de~Gironcoli}, \citenamefont
  {Deutsch}, \citenamefont {Dewhurst}, \citenamefont {Di~Marco}, \citenamefont
  {Draxl}, \citenamefont {Du{\l}ak}, \citenamefont {Eriksson}, \citenamefont
  {Flores-Livas}, \citenamefont {Garrity}, \citenamefont {Genovese},
  \citenamefont {Giannozzi}, \citenamefont {Giantomassi}, \citenamefont
  {Goedecker}, \citenamefont {Gonze}, \citenamefont {Gr{\r a}n{\"a}s},
  \citenamefont {Gross}, \citenamefont {Gulans}, \citenamefont {Gygi},
  \citenamefont {Hamann}, \citenamefont {Hasnip}, \citenamefont {Holzwarth},
  \citenamefont {Iu{\c s}an}, \citenamefont {Jochym}, \citenamefont {Jollet},
  \citenamefont {Jones}, \citenamefont {Kresse}, \citenamefont {Koepernik},
  \citenamefont {K{\"u}{\c c}{\"u}kbenli}, \citenamefont {Kvashnin},
  \citenamefont {Locht}, \citenamefont {Lubeck}, \citenamefont {Marsman},
  \citenamefont {Marzari}, \citenamefont {Nitzsche}, \citenamefont
  {Nordstr{\"o}m}, \citenamefont {Ozaki}, \citenamefont {Paulatto},
  \citenamefont {Pickard}, \citenamefont {Poelmans}, \citenamefont {Probert},
  \citenamefont {Refson}, \citenamefont {Richter}, \citenamefont {Rignanese},
  \citenamefont {Saha}, \citenamefont {Scheffler}, \citenamefont {Schlipf},
  \citenamefont {Schwarz}, \citenamefont {Sharma}, \citenamefont {Tavazza},
  \citenamefont {Thunstr{\"o}m}, \citenamefont {Tkatchenko}, \citenamefont
  {Torrent}, \citenamefont {Vanderbilt}, \citenamefont {van Setten},
  \citenamefont {Van~Speybroeck}, \citenamefont {Wills}, \citenamefont {Yates},
  \citenamefont {Zhang},\ and\ \citenamefont
  {Cottenier}}]{DFTReproducibility2016lejaeghere}%
  \BibitemOpen
  \bibfield  {author} {\bibinfo {author} {\bibfnamefont {K.}~\bibnamefont
  {Lejaeghere}}, \bibinfo {author} {\bibfnamefont {G.}~\bibnamefont
  {Bihlmayer}}, \bibinfo {author} {\bibfnamefont {T.}~\bibnamefont
  {Bj{\"o}rkman}}, \bibinfo {author} {\bibfnamefont {P.}~\bibnamefont {Blaha}},
  \bibinfo {author} {\bibfnamefont {S.}~\bibnamefont {Bl{\"u}gel}}, \bibinfo
  {author} {\bibfnamefont {V.}~\bibnamefont {Blum}}, \bibinfo {author}
  {\bibfnamefont {D.}~\bibnamefont {Caliste}}, \bibinfo {author} {\bibfnamefont
  {I.~E.}\ \bibnamefont {Castelli}}, \bibinfo {author} {\bibfnamefont {S.~J.}\
  \bibnamefont {Clark}}, \bibinfo {author} {\bibfnamefont {A.}~\bibnamefont
  {Dal~Corso}}, \bibinfo {author} {\bibfnamefont {S.}~\bibnamefont
  {de~Gironcoli}}, \bibinfo {author} {\bibfnamefont {T.}~\bibnamefont
  {Deutsch}}, \bibinfo {author} {\bibfnamefont {J.~K.}\ \bibnamefont
  {Dewhurst}}, \bibinfo {author} {\bibfnamefont {I.}~\bibnamefont {Di~Marco}},
  \bibinfo {author} {\bibfnamefont {C.}~\bibnamefont {Draxl}}, \bibinfo
  {author} {\bibfnamefont {M.}~\bibnamefont {Du{\l}ak}}, \bibinfo {author}
  {\bibfnamefont {O.}~\bibnamefont {Eriksson}}, \bibinfo {author}
  {\bibfnamefont {J.~A.}\ \bibnamefont {Flores-Livas}}, \bibinfo {author}
  {\bibfnamefont {K.~F.}\ \bibnamefont {Garrity}}, \bibinfo {author}
  {\bibfnamefont {L.}~\bibnamefont {Genovese}}, \bibinfo {author}
  {\bibfnamefont {P.}~\bibnamefont {Giannozzi}}, \bibinfo {author}
  {\bibfnamefont {M.}~\bibnamefont {Giantomassi}}, \bibinfo {author}
  {\bibfnamefont {S.}~\bibnamefont {Goedecker}}, \bibinfo {author}
  {\bibfnamefont {X.}~\bibnamefont {Gonze}}, \bibinfo {author} {\bibfnamefont
  {O.}~\bibnamefont {Gr{\r a}n{\"a}s}}, \bibinfo {author} {\bibfnamefont
  {E.~K.~U.}\ \bibnamefont {Gross}}, \bibinfo {author} {\bibfnamefont
  {A.}~\bibnamefont {Gulans}}, \bibinfo {author} {\bibfnamefont
  {F.}~\bibnamefont {Gygi}}, \bibinfo {author} {\bibfnamefont {D.~R.}\
  \bibnamefont {Hamann}}, \bibinfo {author} {\bibfnamefont {P.~J.}\
  \bibnamefont {Hasnip}}, \bibinfo {author} {\bibfnamefont {N.~A.~W.}\
  \bibnamefont {Holzwarth}}, \bibinfo {author} {\bibfnamefont {D.}~\bibnamefont
  {Iu{\c s}an}}, \bibinfo {author} {\bibfnamefont {D.~B.}\ \bibnamefont
  {Jochym}}, \bibinfo {author} {\bibfnamefont {F.}~\bibnamefont {Jollet}},
  \bibinfo {author} {\bibfnamefont {D.}~\bibnamefont {Jones}}, \bibinfo
  {author} {\bibfnamefont {G.}~\bibnamefont {Kresse}}, \bibinfo {author}
  {\bibfnamefont {K.}~\bibnamefont {Koepernik}}, \bibinfo {author}
  {\bibfnamefont {E.}~\bibnamefont {K{\"u}{\c c}{\"u}kbenli}}, \bibinfo
  {author} {\bibfnamefont {Y.~O.}\ \bibnamefont {Kvashnin}}, \bibinfo {author}
  {\bibfnamefont {I.~L.~M.}\ \bibnamefont {Locht}}, \bibinfo {author}
  {\bibfnamefont {S.}~\bibnamefont {Lubeck}}, \bibinfo {author} {\bibfnamefont
  {M.}~\bibnamefont {Marsman}}, \bibinfo {author} {\bibfnamefont
  {N.}~\bibnamefont {Marzari}}, \bibinfo {author} {\bibfnamefont
  {U.}~\bibnamefont {Nitzsche}}, \bibinfo {author} {\bibfnamefont
  {L.}~\bibnamefont {Nordstr{\"o}m}}, \bibinfo {author} {\bibfnamefont
  {T.}~\bibnamefont {Ozaki}}, \bibinfo {author} {\bibfnamefont
  {L.}~\bibnamefont {Paulatto}}, \bibinfo {author} {\bibfnamefont {C.~J.}\
  \bibnamefont {Pickard}}, \bibinfo {author} {\bibfnamefont {W.}~\bibnamefont
  {Poelmans}}, \bibinfo {author} {\bibfnamefont {M.~I.~J.}\ \bibnamefont
  {Probert}}, \bibinfo {author} {\bibfnamefont {K.}~\bibnamefont {Refson}},
  \bibinfo {author} {\bibfnamefont {M.}~\bibnamefont {Richter}}, \bibinfo
  {author} {\bibfnamefont {G.-M.}\ \bibnamefont {Rignanese}}, \bibinfo {author}
  {\bibfnamefont {S.}~\bibnamefont {Saha}}, \bibinfo {author} {\bibfnamefont
  {M.}~\bibnamefont {Scheffler}}, \bibinfo {author} {\bibfnamefont
  {M.}~\bibnamefont {Schlipf}}, \bibinfo {author} {\bibfnamefont
  {K.}~\bibnamefont {Schwarz}}, \bibinfo {author} {\bibfnamefont
  {S.}~\bibnamefont {Sharma}}, \bibinfo {author} {\bibfnamefont
  {F.}~\bibnamefont {Tavazza}}, \bibinfo {author} {\bibfnamefont
  {P.}~\bibnamefont {Thunstr{\"o}m}}, \bibinfo {author} {\bibfnamefont
  {A.}~\bibnamefont {Tkatchenko}}, \bibinfo {author} {\bibfnamefont
  {M.}~\bibnamefont {Torrent}}, \bibinfo {author} {\bibfnamefont
  {D.}~\bibnamefont {Vanderbilt}}, \bibinfo {author} {\bibfnamefont {M.~J.}\
  \bibnamefont {van Setten}}, \bibinfo {author} {\bibfnamefont
  {V.}~\bibnamefont {Van~Speybroeck}}, \bibinfo {author} {\bibfnamefont
  {J.~M.}\ \bibnamefont {Wills}}, \bibinfo {author} {\bibfnamefont {J.~R.}\
  \bibnamefont {Yates}}, \bibinfo {author} {\bibfnamefont {G.-X.}\ \bibnamefont
  {Zhang}}, \ and\ \bibinfo {author} {\bibfnamefont {S.}~\bibnamefont
  {Cottenier}},\ }\href {\doibase 10.1126/science.aad3000} {\bibfield
  {journal} {\bibinfo  {journal} {Science}\ }\textbf {\bibinfo {volume} {351}}
  (\bibinfo {year} {2016}),\ 10.1126/science.aad3000}\BibitemShut {NoStop}%
\bibitem [{\citenamefont {Jackson}\ \emph {et~al.}(2010)\citenamefont
  {Jackson}, \citenamefont {Ramakrishnan}, \citenamefont {Muriki},
  \citenamefont {Canon}, \citenamefont {Cholia}, \citenamefont {Shalf},
  \citenamefont {Wasserman},\ and\ \citenamefont
  {Wright}}]{2010-jackson-cholia-lbl-cloud-con}%
  \BibitemOpen
  \bibfield  {author} {\bibinfo {author} {\bibfnamefont {K.}~\bibnamefont
  {Jackson}}, \bibinfo {author} {\bibfnamefont {L.}~\bibnamefont
  {Ramakrishnan}}, \bibinfo {author} {\bibfnamefont {K.}~\bibnamefont
  {Muriki}}, \bibinfo {author} {\bibfnamefont {S.}~\bibnamefont {Canon}},
  \bibinfo {author} {\bibfnamefont {S.}~\bibnamefont {Cholia}}, \bibinfo
  {author} {\bibfnamefont {J.}~\bibnamefont {Shalf}}, \bibinfo {author}
  {\bibfnamefont {H.}~\bibnamefont {Wasserman}}, \ and\ \bibinfo {author}
  {\bibfnamefont {N.}~\bibnamefont {Wright}},\ }\href {\doibase 10.1109/????}
  {\bibfield  {journal} {\bibinfo  {journal} {Proceedings of the IEEE Second
  International Conference on Cloud Computing Technology and Science (CloudCom
  2010)}\ ,\ \bibinfo {pages} {159}} (\bibinfo {year} {2010})}\BibitemShut
  {NoStop}%
\bibitem [{\citenamefont {Crowley}\ \emph {et~al.}(2016)\citenamefont
  {Crowley}, \citenamefont {Tahir-Kheli},\ and\ \citenamefont
  {Goddard}}]{goddard2016JPCL}%
  \BibitemOpen
  \bibfield  {author} {\bibinfo {author} {\bibfnamefont {J.~M.}\ \bibnamefont
  {Crowley}}, \bibinfo {author} {\bibfnamefont {J.}~\bibnamefont
  {Tahir-Kheli}}, \ and\ \bibinfo {author} {\bibfnamefont {W.~A.}\ \bibnamefont
  {Goddard}},\ }\href {\doibase 10.1021/acs.jpclett.5b02870} {\bibfield
  {journal} {\bibinfo  {journal} {The Journal of Physical Chemistry Letters}\
  }\textbf {\bibinfo {volume} {7}},\ \bibinfo {pages} {1198} (\bibinfo {year}
  {2016})},\ \bibinfo {note} {pMID: 26944092},\ \Eprint
  {http://arxiv.org/abs/https://doi.org/10.1021/acs.jpclett.5b02870}
  {https://doi.org/10.1021/acs.jpclett.5b02870} \BibitemShut {NoStop}%
\bibitem [{\citenamefont {Harl}\ \emph {et~al.}(2010)\citenamefont {Harl},
  \citenamefont {Schimka},\ and\ \citenamefont
  {Kresse}}]{kresse2010vdWCorrection}%
  \BibitemOpen
  \bibfield  {author} {\bibinfo {author} {\bibfnamefont {J.}~\bibnamefont
  {Harl}}, \bibinfo {author} {\bibfnamefont {L.}~\bibnamefont {Schimka}}, \
  and\ \bibinfo {author} {\bibfnamefont {G.}~\bibnamefont {Kresse}},\ }\href
  {\doibase 10.1103/PhysRevB.81.115126} {\bibfield  {journal} {\bibinfo
  {journal} {Phys. Rev. B}\ }\textbf {\bibinfo {volume} {81}},\ \bibinfo
  {pages} {115126} (\bibinfo {year} {2010})}\BibitemShut {NoStop}%
\bibitem [{\citenamefont {Shishkin}\ \emph {et~al.}(2007)\citenamefont
  {Shishkin}, \citenamefont {Marsman},\ and\ \citenamefont
  {Kresse}}]{shishkin2007accurate}%
  \BibitemOpen
  \bibfield  {author} {\bibinfo {author} {\bibfnamefont {M.}~\bibnamefont
  {Shishkin}}, \bibinfo {author} {\bibfnamefont {M.}~\bibnamefont {Marsman}}, \
  and\ \bibinfo {author} {\bibfnamefont {G.}~\bibnamefont {Kresse}},\
  }\href@noop {} {\bibfield  {journal} {\bibinfo  {journal} {Physical review
  letters}\ }\textbf {\bibinfo {volume} {99}},\ \bibinfo {pages} {246403}
  (\bibinfo {year} {2007})}\BibitemShut {NoStop}%
\bibitem [{\citenamefont {Skone}\ \emph {et~al.}(2014)\citenamefont {Skone},
  \citenamefont {Govoni},\ and\ \citenamefont {Galli}}]{scHSE2014galli}%
  \BibitemOpen
  \bibfield  {author} {\bibinfo {author} {\bibfnamefont {J.~H.}\ \bibnamefont
  {Skone}}, \bibinfo {author} {\bibfnamefont {M.}~\bibnamefont {Govoni}}, \
  and\ \bibinfo {author} {\bibfnamefont {G.}~\bibnamefont {Galli}},\ }\href
  {\doibase 10.1103/PhysRevB.89.195112} {\bibfield  {journal} {\bibinfo
  {journal} {Phys. Rev. B}\ }\textbf {\bibinfo {volume} {89}},\ \bibinfo
  {pages} {195112} (\bibinfo {year} {2014})}\BibitemShut {NoStop}%
\end{thebibliography}%

\end{document}